\documentclass[fleqn,usenatbib]{mnras}
\usepackage{graphicx,url}
\usepackage{newtxtext}
\usepackage[varg]{newtxmath}
\usepackage[T1]{fontenc}
\usepackage{ae,aecompl}
\usepackage{microtype}
\usepackage{amsmath}	
\usepackage{amssymb}	
\begin{document}

\label{firstpage}

\title[Source Finding for LOFAR using Faraday Moments]{Source Finding in Linear Polarization for LOFAR, and SKA Predecessor Surveys, using Faraday Moments}

\author[J.~S.\ Farnes, et al.]{J.~S.\ Farnes$^{1,2}$\thanks{email:\texttt{j.farnes@astro.ru.nl}}, G.\ Heald$^{3,4,5}$, H.\ Junklewitz$^{6}$, D.~D.\ Mulcahy$^{7}$, M.\ Haverkorn$^{1}$, \newauthor C.~L.\ Van Eck$^{1}$, C.~J.\ Riseley$^{3,7,8}$, M.\ Brentjens$^{4}$, C.\ Horellou$^{9}$, V.\ Vacca$^{10}$, D.~I.\ Jones$^{1}$, \newauthor A.\ Horneffer$^{6}$, R.\ Paladino$^{11}$\\
  $^{1}$Department of Astrophysics/IMAPP, Radboud University, PO Box 9010, NL-6500 GL Nijmegen, the Netherlands.\\
  $^{2}$Oxford e-Research Centre (OeRC), Keble Road, Oxford OX1 3QG, England.\\ 
  $^{3}$CSIRO Astronomy and Space Science, 26 Dick Perry Avenue, Kensington WA 6151, Australia.\\
  $^{4}$ASTRON, the Netherlands Institute for Radio Astronomy, Postbus 2, NL-7990 AA Dwingeloo, the Netherlands.\\ 
  $^{5}$Kapteyn Astronomical Institute, University of Groningen, PO Box 800, 9700 AV, Groningen, The Netherlands.\\
  $^{6}$Max-Planck-Institut für Radioastronomie, Auf dem Hügel 69, 53121, Bonn, Germany.\\ 
  $^{7}$Jodrell Bank Center for Astrophysics, School of Physics and Astronomy,
The University of Manchester, Manchester, M13 9PL, UK.\\
  $^{8}$School of Physics and Astronomy, University of Southampton, Highfield, SO17 1BJ, Southampton, UK.\\
  $^{9}$Chalmers University of Technology, Dept. of Earth and Space Sciences, Onsala Space Observatory, SE-439 92 Onsala, Sweden.\\
   $^{10}$INAF - Osservatorio Astronomico di Cagliari, Via della Scienza 5 - I-09047 Selargius (CA), Italy.\\
  $^{11}$INAF-Osservatorio di Radioastronomia, Via P. Gobetti, 101
40129 Bologna, Italy.\\ }

\date{Accepted ---}

\pagerange{\pageref{firstpage}--\pageref{lastpage}}

\pubyear{2015}

\maketitle

\begin{abstract}
The optimal source-finding strategy for linear polarization data is an unsolved problem, with many inhibitive factors imposed by the technically-challenging nature of polarization observations. Such an algorithm is essential for Square Kilometre Array (SKA) pathfinder surveys, such as the Multifrequency Snapshot Sky Survey (MSSS) with the LOw Frequency ARray (LOFAR), as data volumes are significant enough to prohibit manual inspection. We present a new strategy of `Faraday Moments' for source-finding in linear polarization with LOFAR, using the moments of the frequency-dependent full-Stokes data (i.e.\ the mean, standard deviation, skewness, and excess kurtosis). Through simulations of the sky, we find that moments can identify polarized sources with a high completeness: 98.5\% at a signal--to--noise of 5. While the method has low reliability, Rotation Measure (RM) Synthesis can be applied per candidate source to filter out instrumental and spurious detections. This combined strategy will result in a complete and reliable catalogue of polarized sources that includes the full sensitivity of the observational bandwidth. We find that the technique can reduce the number of pixels on which RM Synthesis needs to be performed by a factor of $\approx1\times10^{5}$ for source distributions anticipated with modern radio telescopes. Through tests on LOFAR data, we find that the technique works effectively in the presence of diffuse emission. Extensions of this method are directly applicable to other upcoming radio surveys such as the POlarization Sky Survey of the Universe's Magnetism (POSSUM) with the Australia Square Kilometre Array Pathfinder (ASKAP), and the SKA itself.
\end{abstract}



\section{Introduction}\label{intro}
Magnetic fields are ubiquitous throughout the Universe, and these cosmic magnetic fields are best studied through spectropolarimetric radio observations. Spectropolarimetry with modern correlators on interferometers such as e.g.\ the LOw Frequency ARray \citep[LOFAR;][]{2013A&A...556A...2V}, allows for measurements of the Stokes parameters, $I$, $Q$, $U$, and $V$, using a large number of channels across an observational bandwidth. These channels allow for measurements of the linearly polarized fraction, and of the Faraday rotation. The Faraday rotation occurs as linearly polarized radiation travelling through a magnetised plasma undergoes a phenomenon that can be modelled as birefringence. The linear polarization can be considered as two counter-rotating circularly polarized components which experience different refractive indices. Upon exiting the plasma, Faraday rotation will have caused the electric vector of the incoming linearly polarized wave to rotate. In a simple model with just one emitting source along a line of sight, with no internal Faraday rotation, and only a single slab of plasma between the observer and the source, the electric vector polarization angle (EVPA) will be rotated by an amount proportional to the squared wavelength of the radiation as described by,
\begin{equation}
\chi_{\textrm{EVPA}} = \chi_{0} + \textrm{RM}\lambda^2 \\,
\end{equation}
where $\chi_{\textrm{EVPA}}$ is the observed EVPA, $\chi_{0}$ is the intrinsic EVPA at the source, and $\lambda$ is the wavelength of the radiation. The factor of proportionality is known as the rotation measure (RM), which is related to the integral of the magnetic field component along the line-of-sight, which is here defined as,
\begin{equation}
\textrm{RM} = -\frac{e^3}{2\pi m_{e}^2 c^4} \int_{0}^{d} n_e B_{\parallel} ds \approx 0.812 \int_{d}^{0} n_e B_{\parallel} ds \\,
\end{equation}
where $n_{e}$ is generally the electron number density of the plasma in cm$^{-3}$, $B_{\parallel}$ is the strength of the component of the magnetic field that is parallel to the line-of-sight in $\muup$G, and $ds$ is a finite element of the path length in pc. The constants $e$, $m_{e}$, and $c$ are the electronic charge, the mass of the electron, and the speed of electromagnetic radiation in a vacuum respectively. The integral from $0$ to $d$ represents the distance along the line of sight between the observer and the source. Experimentally, the measured RM is retrieved by fitting a straight line to $\chi_{\textrm{EVPA}}(\lambda^2)$ \citep[e.g.][]{1983A&AS...52..317R}, by using RM Synthesis \citep[e.g.][]{2005A&A...441.1217B}, or by $QU$-fitting \citep[e.g.][]{2012MNRAS.421.3300O}. 

The ability to retrieve the polarized quantities of a radio source is entirely dependent on the ability to find radio sources within noisy images. It is of importance to planned future surveys to investigate suitable strategies for source-finding in linear polarization with interferometers such as LOFAR, which are very well suited for deep radio surveys \citep[e.g.][]{2016arXiv160609437H,2017A&A...601A..25C}, but are particularly technically challenging due to operation at low-radio frequencies \citep{2015A&A...574A.114V}, with potentially subarcsecond angular resolution \citep{2015A&A...574A..73M}, at high sensitivity \citep{2016MNRAS.459..277S}, and with the ability to make precise Faraday rotation measurements \citep{2013A&A...552A..58S}. Nevertheless, finding linearly polarized sources faces many hurdles: (i) at sub-arcminute resolution, the peak in linearly polarized intensity can be offset from the peak in total intensity \citep[see for example Fig.~1 in][]{2015ApJ...806...83O}, (ii) the statistics in polarized intensity, $P=\sqrt{Q^2+U^2}$, are Rician, rather than Gaussian, while all publically available source-finders are geared towards Gaussian noise statistics, (iii) the full sensitivity is not provided in any single channel of $Q$, $U$, or $P$, and RM Synthesis is therefore required to retrieve the full point-source sensitivity from the data, (iv) sources detected in $Q$ and $U$ can have both positive and negative brightness, and these values oscillate and mix across the observing bandwidth due to Faraday rotation, and (v) in some cases, $Q$ and $U$ images can be more sensitive than $I$ images, which in principle could lead to sources that can be found in $P$ but not in $I$. Source-finding in circular polarization, Stokes $V$, is beyond the scope of this paper in which we focus on linear polarization, but also faces similar challenges due to the full-sensitivity not being provided in a single channel and the process of Faraday conversion across the observing band. Moreover, the linear feeds used for observations at low radio frequencies with instruments such as LOFAR are more suitable for measuring circular rather than linear polarization, which further increases the difficulty of detecting faint linearly polarized sources.

Furthermore, the ideal source-finder is also both highly complete and reliable. The definitions of `completeness' and `reliability' are rigourously detailed in \citet{2012MNRAS.422.1812H}. The completeness is measured as the number of sources with a \emph{measured} flux $S \ge S_{0}$ that are contained within the catalogue, while the reliability is related to the false-detection rate (as False-detection rate + Reliability = 100\%), which at a flux $S_{0}$ is defined as the fraction of catalogued sources with $S \ge S_{0}$ which are not identified with a real source.

There are two proposed ``ideal'' strategies, although neither have yet been addressed in the literature: (a) develop a three-dimensional source-finder to identify 3D blobs in Faraday cubes that have right ascension, declination, and Faraday depth axes, or (b) develop an astronomical source-finder that accounts for Rician noise statistics. However, strategy (a) of finding 3D structures (e.g.\ Gaussians) would be affected by sidelobes from the rotation measure spread function \citep[RMSF; equivalent to the point spread function in Faraday space, see][]{2005A&A...441.1217B}. In addition, strategy (b) is a significantly complex issue that has been addressed by functional Magnetic Resonance Imaging studies (that also operate in Rician noise), but with no clear optimal solution \citep[e.g.][]{fmripaper}. It is possible to make a critical assumption that Rician noise can be parameterised by a Gaussian, however independent studies by \citet{2012PASA...29..214G} and \citet{2012ApJ...750..139M} find that polarized intensity is more strongly biased than Rician statistics suggest. In combination with typical interferometric imaging artefacts, \citet{2012PASA...29..214G} found that the false-detection rates at $8\sigma_{QU}$ are similar to Rician false detection rates at $4.9\sigma_{QU}$, suggesting that an underlying assumption of normality is not appropriate. In addition, for both (a) and (b) it is not clear how the full sensitivity of the band could be used for such a source-finder, i.e.\ source-finding in $P$ would need to take place on either a per-channel or per-Faraday-depth basis. This naturally limits the sensitivity at which the source-finder can operate, and thereby restricts the completeness of the source-finding. Furthermore, both of these strategies would be computationally challenging, as they would require RM Synthesis of the entire sky area that has been observed, which mostly consists of noisy and empty pixels. 

Source-finding in linear polarization is therefore clearly a non-trivial issue, with no current optimal solution for the next generation of radio surveys. An optimised source-finding strategy would be of use to surveys such as the Multifrequency Snapshot Sky Survey (MSSS) with LOFAR \citep{2015A&A...582A.123H}, the POlarization Sky Survey of the Universe's Magnetism (POSSUM) with the Australia Square Kilometre Array Pathfinder \citep[ASKAP;][]{2007PASA...24..174J,2010AAS...21547013G}, the GaLactic and Extragalactic All-sky Murchison Widefield Array (GLEAM) survey with the MWA \citep{2017MNRAS.464.1146H}, the Very Large Array Sky Survey (VLASS) with the Karl G. Jansky Very Large Array \citep[VLA;][]{2016AAS...22732409L}, and for surveys with the Square Kilometre Array (SKA) itself \citep[e.g.][]{2015aska.confE..92J}. This paper is structured as follows: in Section~\ref{simulatedSEDs} we discuss the new Faraday Moments technique for source-finding and test the properties of the method at high signal--to--noise ratios, in Section~\ref{simulations} we test our new source-finding technique on simulated LOFAR observations, in Section~\ref{realdata} we test the technique on real LOFAR observations, in Section~\ref{prescription} we devise a full-formalism for Faraday Moment source-finding and test the method across a substantial range of signal--to--noise ratios between 3 to 500, and in Section~\ref{conclusions} we provide conclusions on our findings. We only provide pseudo-colour scales and coordinate grids for images when these are necessary for the image interpretation.


\section{Faraday Moments}\label{simulatedSEDs} 

\subsection{Calculating Faraday Moments}
\label{calcmoments}
We present a new technique for source-finding in linear polarization, and have developed a new strategy that uses moment images derived using data across an observational bandwidth. As an example, a source that is bright in Stokes $Q$ and not in Stokes $U$, and with little Faraday rotation, will appear as a peak in an image of the mean value of $Q$ across the band. In a similar way, a source with significant Faraday rotation will appear as peaks in images of the standard deviation of $Q$ and $U$ across the band. Moments therefore provide unique ways to identify sources in linear polarization based upon their Faraday properties, and we therefore call the technique ``Faraday Rotation Moments'' or ``Faraday Moments''.

For this technique, Faraday Moment images must be generated at the location of every pixel in each $Q$, $U$, and $P=\sqrt{Q^2+U^2}$ datacube. All observed wavelengths are used. We calculate moments using the following equations,
\begin{equation}
\mu_{Q} = \frac{1}{n}\sum_{i=1}^n Q\left( \lambda_{i} \right) \\,
\label{eqn1}
\end{equation}

\begin{equation}
\sigma_{Q} = \sqrt{\frac{1}{n-1} \sum_{i=1}^n \left( Q\left( \lambda_{i} \right) - \mu_{Q} \right) ^2} \\,
\label{eqn2}
\end{equation}

\begin{equation}
\psi_{Q} = \frac{\tfrac{1}{n} \sum_{i=1}^n \left(Q\left( \lambda_{i} \right)-\mu_{Q}\right)^3}{\left[\tfrac{1}{n-1} \sum_{i=1}^n \left(Q\left( \lambda_{i} \right)-\mu_{Q}\right)^2\right]^{3/2}} \\,
\label{eqn3}
\end{equation}

\begin{equation}
\kappa_{Q} = \frac{\tfrac{1}{n} \sum_{i=1}^n \left(Q\left( \lambda_{i} \right)-\mu_{Q}\right)^4}{\left[\tfrac{1}{n-1} \sum_{i=1}^n \left(Q\left( \lambda_{i} \right)-\mu_{Q}\right)^2\right]^2} - 3 \\,
\label{eqn4}
\end{equation}    
where $n$ is the number of samples at different wavelengths, $\lambda$, where $\mu_{Q}$ is the mean of Stokes $Q$, $\sigma_{Q}$ is the standard deviation of Stokes $Q$, $\psi_{Q}$ is the skewness of Stokes $Q$, and $\kappa_{Q}$ is the excess kurtosis of Stokes $Q$ \citep[e.g.][]{statisticspaper}.\footnote{Note that conventionally, the skew is denoted via $\gamma^{1}$ and the excess kurtosis via $\gamma^{2}$. For clarity, we instead use the alternative notation of $\psi$ and $\kappa$. Similarly, the second moment is conventionally the variance, $\sigma^{2}$, although we here define the second moment as the standard deviation, $\sigma$.} Note that the denominator of the $\psi_{Q}$ and $\kappa_{Q}$ equations can be further simplified to $\sigma_{Q}^3$ and $\sigma_{Q}^4$ respectively. We use the excess kurtosis ($\kappa=$kurtosis$-3$), rather than the kurtosis, in order to make the moment directly comparable to the normal distribution. We also use \emph{unbiased estimators} for each moment, which for raw sample moments is $\propto 1/n$, and for central moments (in which calculation uses up a degree of freedom by using the sample mean) is $\propto 1/(n-1)$. While equations~\ref{eqn1} to \ref{eqn4} are defined for Stokes $Q$, similar images can also be made for both Stokes $U$ and for $P$. This provides images of ($\mu_{Q}$,  $\sigma_{Q}$, $\psi_{Q}$, $\kappa_{Q}$), ($\mu_{U}$,  $\sigma_{U}$, $\psi_{U}$, $\kappa_{U}$), ($\mu_{P}$,  $\sigma_{P}$, $\psi_{P}$, $\kappa_{P}$). The means and standard deviations derived from a radio astronomy image will have the same units as the datacubes themselves -- in Jy~beam$^{-1}$, while skew and excess kurtosis are always dimensionless. In practice, we want to calculate these moments in a fast way, so that it is computationally inexpensive, and we are able to do so using tools in commonly available packages such as \textsc{numpy/scipy}.
     
\subsection{The Properties of Faraday Moments}
\label{comparisonofmoments}
  \begin{figure*}
   \centering
   \includegraphics[trim=0.0cm 1.4cm 0cm 0.0cm,clip=true,angle=0,origin=c,width=\hsize]{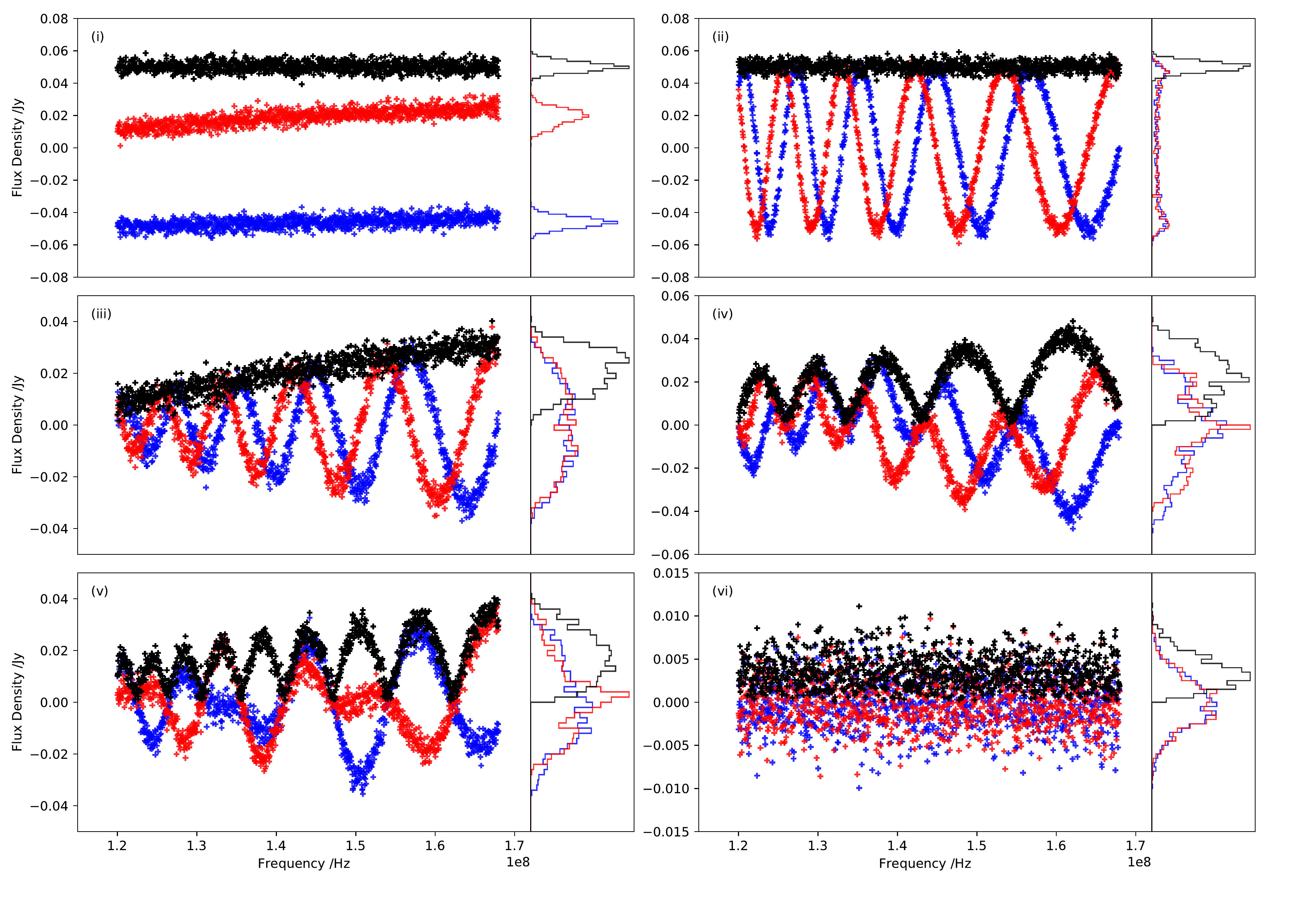}
   \caption{Simulated SEDs of $Q$ (blue), $U$ (red), and $P$ (black) for (i: top left) a purely Faraday rotating screen with low RM, (ii: top right) a purely Faraday rotating screen with high RM, (iii: middle left) a Faraday rotating and depolarizing screen, (iv: middle right) a Burn slab, (v: bottom left) two interfering Faraday components, (vi: bottom right) no signal, other than Gaussian (in $Q$/$U$) and Rayleigh (in $P$) noise. The effects of a spectral index have not been included into the simulated SEDs, but would only serve to increase the detectable moments of $P$. The noise level is 3~mJy. The panels on the right of each SED show the corresponding histograms, which are more clearly presented in Fig.~\ref{Dists}.}
              \label{SEDs}%
    \end{figure*}
    
      \begin{figure*}
   \centering
   \includegraphics[trim=0.0cm 17.5cm 0.0cm 0.0cm,clip=true,angle=0,origin=c,width=\hsize]{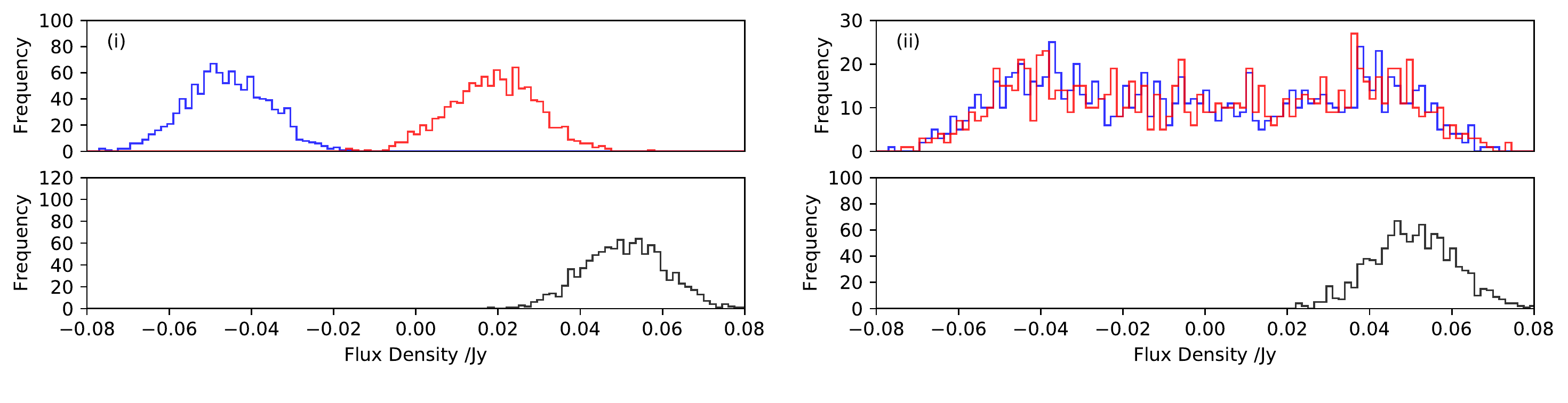}
   \includegraphics[trim=0.0cm 8.75cm 0.0cm 8.75cm,clip=true,angle=0,origin=c,width=\hsize]{Distributions4.pdf}
      \includegraphics[trim=0.0cm 0.0cm 0.0cm 17.5cm,clip=true,angle=0,origin=c,width=\hsize]{Distributions4.pdf}
   \caption{Histograms showing the distributions of $Q$ (blue), $U$ (red), and $P$ (black) with observational frequency for (i: top left) a purely Faraday rotating screen with low RM, (ii: top right) a purely Faraday rotating screen with high RM, (iii: middle left) a Faraday rotating and depolarizing screen, (iv: middle right) a Burn slab, (v: bottom left) two interfering Faraday components, (vi: bottom right) no signal, other than Gaussian (in $Q$/$U$) and Rayleigh (in $P$) noise. These histograms are similar to the panels shown on the right of each SED in Fig.~\ref{SEDs}, but are here expanded to allow for closer inspection. Unlike Fig.~\ref{SEDs}, the SEDs used for these histograms were created with a s/n of 5, rather than 16. At lower s/n, each moment behaves as if it were increasingly convolved with the underlying noise distribution.}
              \label{Dists}%
    \end{figure*}
The key to being able to detect polarized sources using Faraday Moments is being able to distinguish the moments of real sources from the moments expected due to noise. We therefore need to understand the properties of Faraday Moments, when applied to typical polarized sources. In order to explore this, simulated spectral energy distributions (SEDs) for $Q$, $U$, and $P$ are shown in Fig.~\ref{SEDs}. Note that this Figure demonstrates the behaviour of our method at high signal--to--noise (s/n) ratios. For an exploration of the method across a range of s/n, please see Section~\ref{prescription}. Each SED is simulated across the LOFAR HBA band, with a lowest frequency of 120~MHz and a bandwidth of 48~MHz separated into 1024 channels. The SEDs shown correspond to:
\begin{enumerate}
\item a purely Faraday rotating screen as in eqn.~\ref{standardrotation} (with low RM$=0.05$~rad~m$^{-2}$),
\item a purely Faraday rotating screen as in eqn.~\ref{standardrotation} (with high RM$=5.0$~rad~m$^{-2}$), 
\item both a Faraday rotating and depolarizing screen as in eqn.~\ref{burndepol} (with RM$=5.0$~rad~m$^{-2}$ and a Burn-style depolarization with $\sigma_{\textrm{RM}}=0.15$~rad~m$^{-2}$), 
\item a Burn slab as in eqn.~\ref{burnslab} (with the front edge of the screen at a Faraday depth of $1.0$~rad~m$^{-2}$ and with an extent of $5.0$~rad~m$^{-2}$), 
\item two interfering depolarizing Faraday components as in eqn.~\ref{twocomponents} (with RMs of $5.0$ and $-3.5$~rad~m$^{-2}$ respectively, and $\sigma_{\textrm{RM}}$ of $0.15$ and $0.1$~rad~m$^{-2}$ respectively), 
\item no signal, other than Gaussian (in $Q$ and $U$) and Rayleigh (in $P$) noise\footnote{The noise in polarized intensity follows a Rician distribution, although only in cases where there is a signal. For images from radio telescopes, a signal essentially fills the entire sky. In the complete absence of signal (such as in our simulations), the noise follows a Rayleigh distribution, which can be considered as a special case of the Rician distribution.}. 
\end{enumerate}
All signals shown have a maximum s/n of 16, with a noise level of 3~mJy. As all the moments constitute some type of ``average'' that uses the entire bandwidth, the s/n is solely dependent on the band-averaged noise properties rather than those in a single channel. This is a standard scenario for polarization data, and is frequently encountered in techniques such as RM Synthesis \citep[e.g.][]{2005A&A...441.1217B}. The case of diffuse polarized extended emission is considered during the application to real data in Section~\ref{realdata}. The histograms corresponding to the distributions, which will be parameterised using the moment equations, are shown to the right of each plot. These same distributions are shown in further detail in Fig.~\ref{Dists}. The equations that describe each of the shown SEDs are given by,
\begin{equation}
\tilde{P} = p_{0} e^{2j\left(\phi\lambda^2+\chi_{0}\right) },
\label{standardrotation}
\end{equation}
\begin{equation}
\tilde{P} = p_{0} e^{-2\sigma_{\textrm{RM}}^2 \lambda^4} e^{ 2j\left(\phi\lambda^2+\chi_{0}\right) },
\label{burndepol}
\end{equation}
\begin{equation}
\tilde{P} = p_{0} \frac{\sin{\phi_{s} \lambda^2}}{\phi_{s} \lambda^2} e^{ 2j\left(\chi_{0}+\phi_{f}\lambda^2+0.5\phi_{s}\lambda^2\right) },
\label{burnslab}
\end{equation}
\begin{equation}
\begin{split}
\tilde{P} &= p_{1} e^{-2\sigma_{\textrm{RM1}}^2 \lambda^4} e^{ 2j\left(\phi_{1}\lambda^2+\chi_{1}\right) } + \\ &+ p_{2} e^{-2\sigma_{\textrm{RM2}}^2 \lambda^4} e^{ 2j\left(\phi_{2}\lambda^2+\chi_{2}\right) } .
\end{split}
\label{twocomponents}
\end{equation}
where for the $x$-th polarized component: $p_{x}$ is the intrinsic polarization degree, $\chi_{x}$ is the polarization angle at infinite frequency, $\sigma_{\textrm{RMx}}$ is the standard deviation of RMs within the beam, $\phi_{x}$ is the Faraday depth, $\phi_{f}$ is the Faraday depth of the front edge of a Burn-slab, $\phi_{s}$ is the extent in Faraday depth of a Burn-slab, $\tilde{P}$ is the complex polarization vector, and $j$ is the imaginary unit. Further extensive descriptions of each polarized model are provided in \citet{1966MNRAS.133...67B}, \citet{1998MNRAS.299..189S}, \citet{2012MNRAS.421.3300O}, \citet{2014ApJS..212...15F}, and \citet{2015AJ....149...60S}. The effects of a spectral index have not been included into the simulated SEDs, but would only serve to increase the detectable moments of $P$.

\begin{figure}
   \centering
   \includegraphics[trim=0.0cm 16.7cm 18cm 0.0cm,clip=true,angle=0,origin=c,width=\hsize]{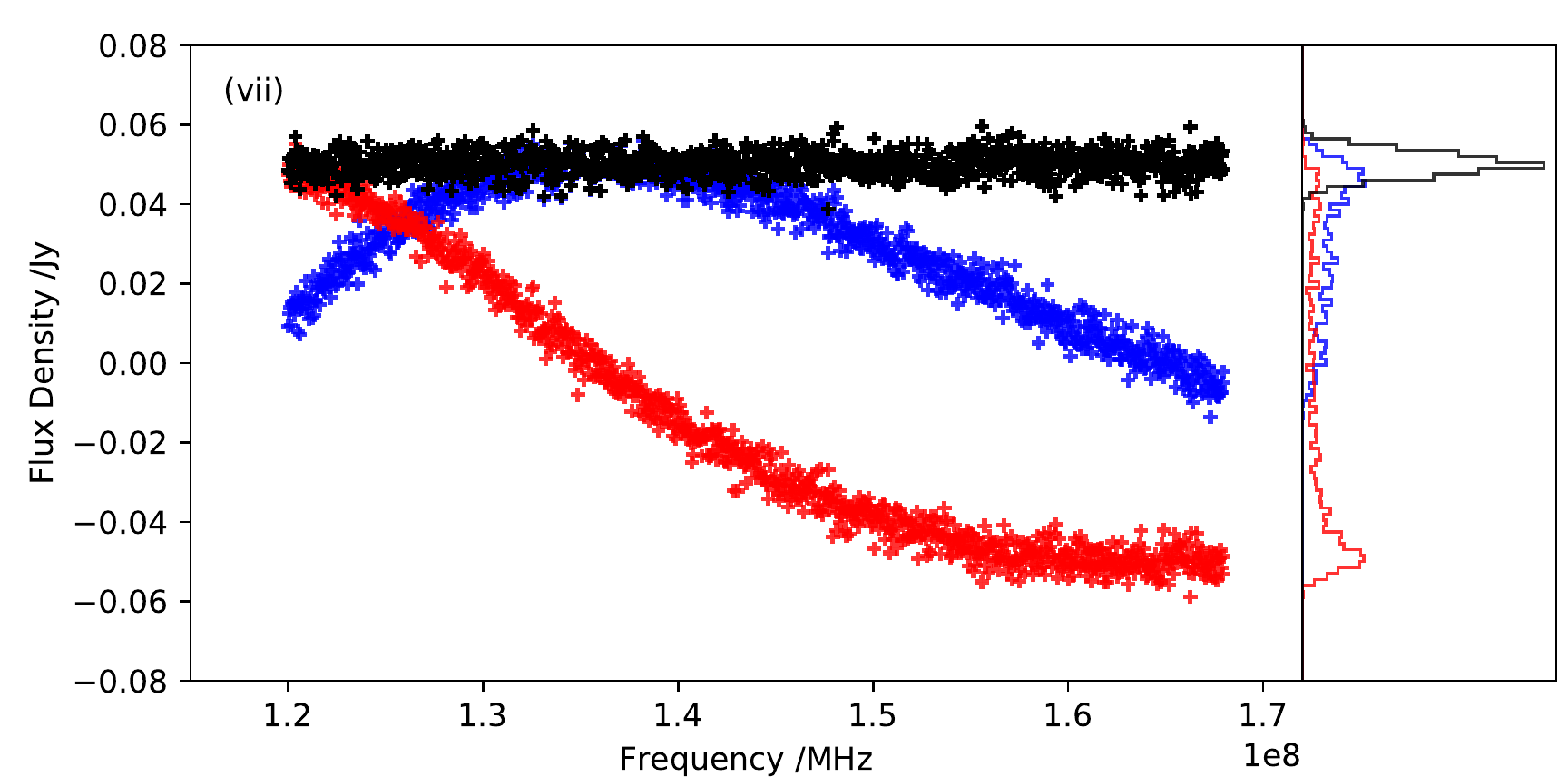}
   \vspace{-10pt}
   \caption{A simulated SED for the extra case of a purely Faraday rotating screen with an `intermediate' RM of 0.5~rad~m$^{-2}$, such that the polarization angle rotates by $1.5$~radians across the observing bandwidth. Stokes $Q$ (blue), $U$ (red), and $P$ (black) are shown. The panels on the right of the SED show the corresponding histograms. Such an intermediate RM value replicates the same structure in its moments as a high RM such as case (ii). However, this intermediate RM SED also has substantial skew in $Q$ and $U$.}
              \label{SEDs_ref}%
    \end{figure}

In Fig.~\ref{Dists} it is clear that case (i) provides approximately normal distributions, albeit possibly slightly peaked, in $Q$, $U$, and $P$, case (ii) provides non-normal distributions in $Q$ and $U$ due to the turning points in frequency space, but is approximately normal in $P$, cases (iii, iv, v) provide non-normal distributions in $Q$, $U$, and also $P$ because of the broadband depolarization, and case (vi) provides a normal distribution in $Q$ and $U$, and is non-normal in $P$. Please note that case (ii) for ``high'' RM is for a relatively low value of 5~rad~m$^{-2}$. Intermediate RM values also replicate the same structure and we show this extra case in Fig.~\ref{SEDs_ref}.

The moments of each of these distributions are given in Table~\ref{table1} for 1000 realisations of the noise, which allows us to provide $1\sigma$ uncertainties for each moment. In all cases, one of the moments differs from those of the noise distribution of case (vi). In practice, we find that the skewness is a weak indicator of Faraday rotation effects in $Q$ and $U$, except for a very small number of sources, but is a reasonable indicator in $P$. However, in our simulations the excess kurtosis is another excellent indicator of associated polarization. Note that the Faraday rotation simulations tend to have negative excess kurtosis in $Q$ and $U$, as generally a leptokurtic distribution (with a high-peak, $\kappa>0$) does not occur. Faraday rotation distributions (in $Q$ and $U$) therefore tend to be platykurtic with a flat-topped curve ($\kappa<0$), or mesokurtic with a normal distribution ($\kappa=0$), particularly for the most extremely non-normal distributions. The same does not necessarily hold true for sources in $P$, particularly faint sources that are best described by the Rician distribution, which tend to have a leptokurtic moment.

There could be complicated selection effects based upon the different s/n ratios in each independent moment image. One could thereby envisage a scenario where sources of some physical type would be systematically excluded by the Faraday Moments technique. In principle, it should be possible to use detailed numerical simulations of some form to obtain a quantitative analysis of the s/n ratio in the moment images in relation to each other, and how this could possibly introduce biases that relate to  different RMs or polarization angles. Nevertheless, such systematic biases are unlikely given the many different polarization SEDs that have been considered, and the measured completeness at retrieving these sources. As we are only concerned here with source detection, rather than source parameter estimation, there is no effect expected for the applicability of the method. Such an analysis is therefore beyond the scope of this paper.

  \begin{table*}
      \caption[]{Moments of the simulated SEDs for 1000 realisations of the noise. The noise in each pixel of the image cube is drawn from a Gaussian with a mean of 0.0~mJy and a standard deviation of 3~mJy, and the moment of the 100 channels is then calculated for each pixel. The $1\sigma$ uncertainties are provided for each moment. Uncertainties for the means and standard deviations, $\mu$ and $\sigma$, are not shown and in all cases are $\le3\times10^{-6}$~Jy. In case (vi) for the simulated noise, the idealised mean, skew, and excess kurtosis of the normally-distributed $Q$ and $U$ should be exactly 0.0, while the standard deviation should be exactly 0.003. Note that analytical values cannot be calculated for these quantities. The exact distribution of each moment is not well-defined. If one naively made the poor assumption that the moments could be described by a Rician distribution, the estimate would be dominated by variations in Rician bias, rather than by the noise. In many cases, a Rician distribution does not even provide a reasonable model of the noise (see Section~\ref{conventional}). However, the distributions are analytically calculable for the noise distribution of case (vi). Integrating the analytical formulae for the Rician distribution in \textsc{matlab} yields $\mu_{P}=0.0037599$, $\sigma_{P}=0.0019654$, $\psi_{P}=0.63111$, and $\kappa_{P}=0.24509$.}
         \label{table1}
     $$ 
         \begin{array}{crrrrrr}
            \hline
            \noalign{\smallskip}
           \textrm{}      &  \textrm{Case} \\
           \textrm{Parameter}      &  \textrm{(i)} &  \textrm{(ii)}&  \textrm{(iii)}&  \textrm{(iv)} &  \textrm{(v)} &  \textrm{(vi)}\\
            \noalign{\smallskip}
            \hline
            \noalign{\smallskip}
\mu_{Q}                & -0.0460332   &  -0.002071   &  -0.002245   &  -0.002896  &  -0.000212   &  0.0000038 \\[2pt]
\sigma_{Q}             & 0.003401   &  0.035602  &  0.019506   &  0.017388   &  0.018943  &  0.002994 \\[2pt]
\psi_{Q}         & -0.0107 \pm 0.0023  &  0.10355 \pm 0.00018  &  -0.0127 \pm 0.0005  &  -0.3068 \pm 0.0006  &  0.0620 \pm 0.0005  &  -0.0003 \pm 0.0025 \\[2pt]
\kappa_{Q}         & -0.065 \pm 0.005  &  -1.47500 \pm 0.00017  &  -0.6581 \pm 0.0007  &  -0.4730 \pm 0.0009  &  -0.7542 \pm 0.0008  &  0.000 \pm 0.005 \\[2pt]
\hline
\mu_{U}                & 0.019023   &  -0.000668   &  0.000065   &  -0.002549   &  0.001565  &  -0.000003 \\[2pt]
\sigma_{U}             & 0.005030   &  0.035296   &  0.019654  &  0.016441   &  0.016809   &  0.002996 \\[2pt]
\psi_{U}         & -0.2046 \pm 0.0017  &  0.02420 \pm 0.00018  &  0.0286 \pm 0.0006  &  -0.1476 \pm 0.0006  &  0.2810 \pm 0.0007  &  -0.0015 \pm 0.0024 \\[2pt]
\kappa_{U}         & -0.427 \pm 0.003  &  -1.46731 \pm 0.00018  &  -0.6235 \pm 0.0008  &  -0.9183 \pm 0.0007  &  -0.3795 \pm 0.0011  &  -0.008 \pm 0.005 \\[2pt]
\hline
\mu_{P}                & 0.050088   &  0.0500908   &  0.025004   &  0.021735  &  0.022840   &  0.0037569  \\[2pt]
\sigma_{P}             & 0.0029967  &  0.0029984  &  0.0121080   &  0.0107313  &  0.0110556   &  0.0019617  \\[2pt]
\psi_{P}         & 0.0015 \pm 0.0024  &  0.0012 \pm 0.0025  &  -0.2153 \pm 0.0006  &  0.0009 \pm 0.0008  &  0.2690 \pm 0.0008  &  0.629 \pm 0.003 \\[2pt]
\kappa_{P}         & 0.001 \pm 0.005  &  -0.005 \pm 0.005  &  -1.1764 \pm 0.0007  &  -0.8535 \pm 0.0011  &  -0.8024 \pm 0.0012  &  0.238 \pm 0.009 \\[2pt]
            \noalign{\smallskip}
            \hline
         \end{array}
     $$ 
   \end{table*}


\section{Application to Simulated LOFAR Data}\label{simulations}
\subsection{The Simulations and Moment Images}
\label{makesimulations}
To test our proposed methodology, we have simulated datacubes that are similar to LOFAR observations with frequency coverage from 120~MHz to 168~MHz, and separated into 100 channels equally spaced in frequency. Each field of view consists of $7200^2$ pixels, with a pixel width of $5$~arcsec, and includes a certain number of polarized sources drawn from a reasonable source distribution in both Stokes $I$ ($dN/dS$, see \citealt{2003AJ....125..465H,2013PASA...30...20N}) and in fractional polarization ($dN/dp$, see \citealt{2004MNRAS.349.1267T, 2010ApJ...714.1689G, 2014MNRAS.440.3113H, 2014ApJ...785...45R, 2014ApJ...787...99S}). These sources were used to randomly populate the field of view. Stokes $I$, $Q$, and $U$ fields were all injected with independent Gaussian noise in each channel. The images were all smoothed to $30$~arcsec resolution. The number of sources per field is in all cases $\sim230$--$250$, and the simulation includes a reasonable estimate for the LOFAR primary beam. In order to represent extragalactic extended sources, approximately 25\% of the sources are extended Gaussians. Each source has a spectral index and rotation measure, with spectral indices drawn from a normal distribution with a mean of $\alpha=-0.8$ and standard deviation of 0.3, and RMs drawn from a normal distribution with a mean of $0.0$~rad~m$^{-2}$ and a standard deviation of $40.0$~rad~m$^{-2}$. This accounts for an extragalactic component of $\approx7$~rad~m$^{-2}$ \citep{2015A&A...575A.118O} and a significant additional Galactic component \citep{2014ApJS..212...15F}. No depolarization effects, or sources with more complicated moments were included, although these will only increase our ability to distinguish sources from noise, as shown in Fig.~\ref{Dists}. None of these described properties have any strong effect on the outcome of our tests. Bandwidth depolarization would normally be significant when using 100 channels across a 48~MHz bandwidth (normal LOFAR observations use $\sim$1000 channels), and would only require RMs$\ge20.0$~rad~m$^{-2}$ for a source to be depolarized by a multiplicative factor of 0.85. However, bandwidth depolarization occurs while averaging polarization vectors within an individual channel. As the sources are directly injected into each channel for our simulations, there is no rotation within an individual channel-width and hence no bandwidth depolarization whatsoever, which is useful for the purpose of these simulations. Our simulations are therefore unaffected by bandwidth depolarization.

 \begin{figure*}
   \centering
   \includegraphics[trim=0.0cm 0.0cm 0.0cm 0.0cm,clip=false,angle=0,origin=c,width=5.8cm]{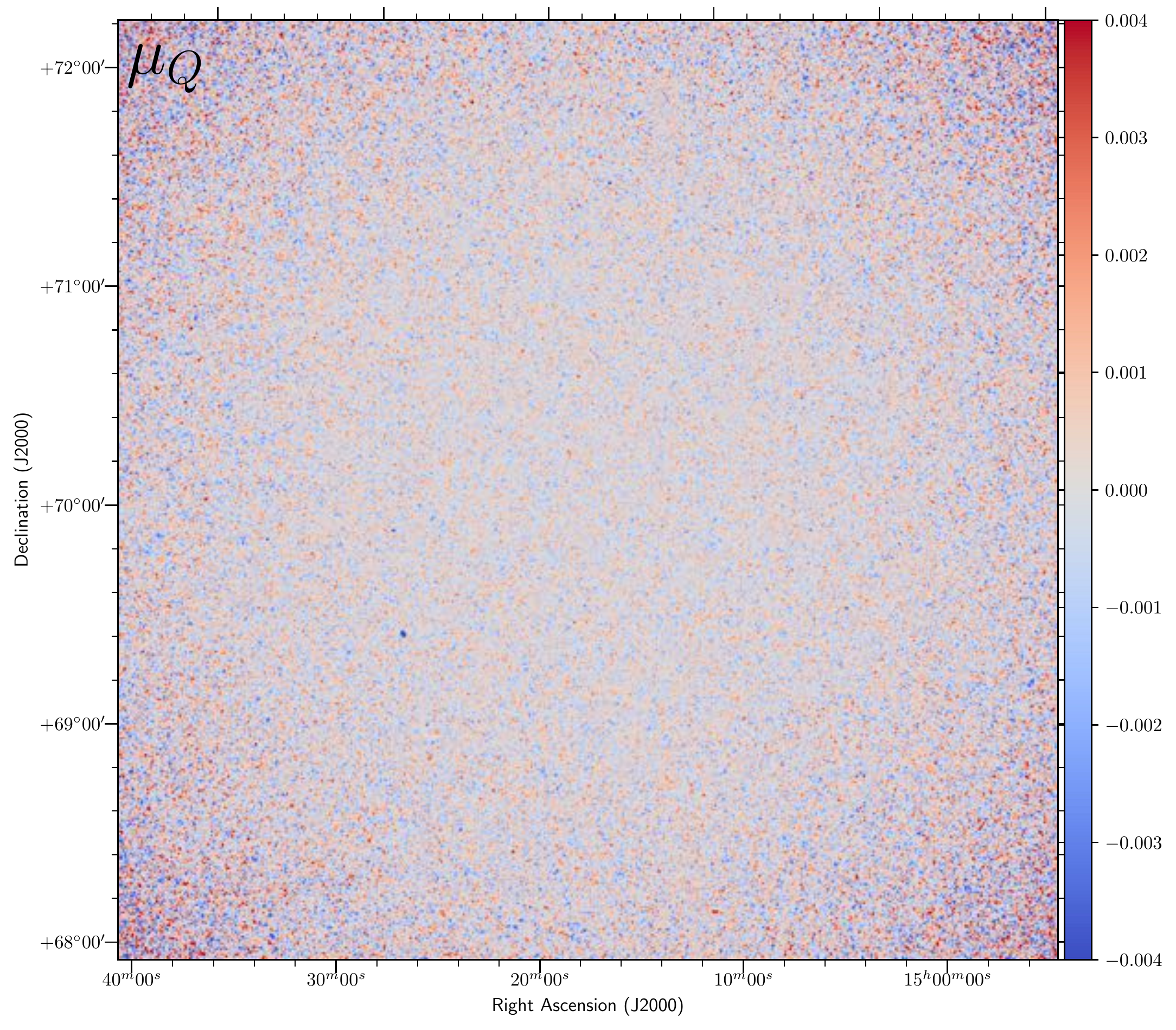}
   \includegraphics[trim=0.0cm 0.0cm 0.0cm 0.0cm,clip=false,angle=0,origin=c,width=5.8cm]{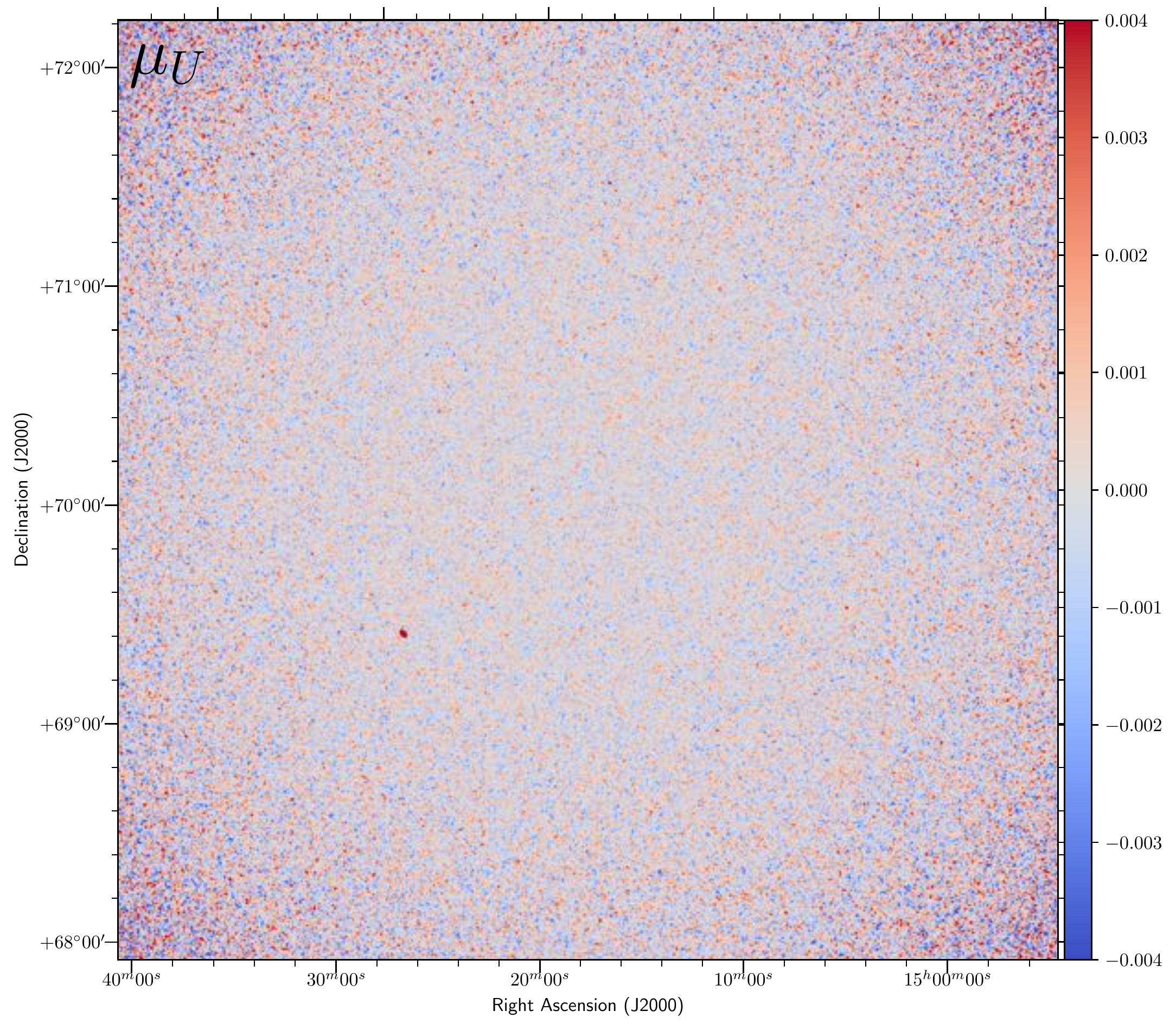}
   \includegraphics[trim=0.0cm 0.0cm 0.0cm 0.0cm,clip=false,angle=0,origin=c,width=5.8cm]{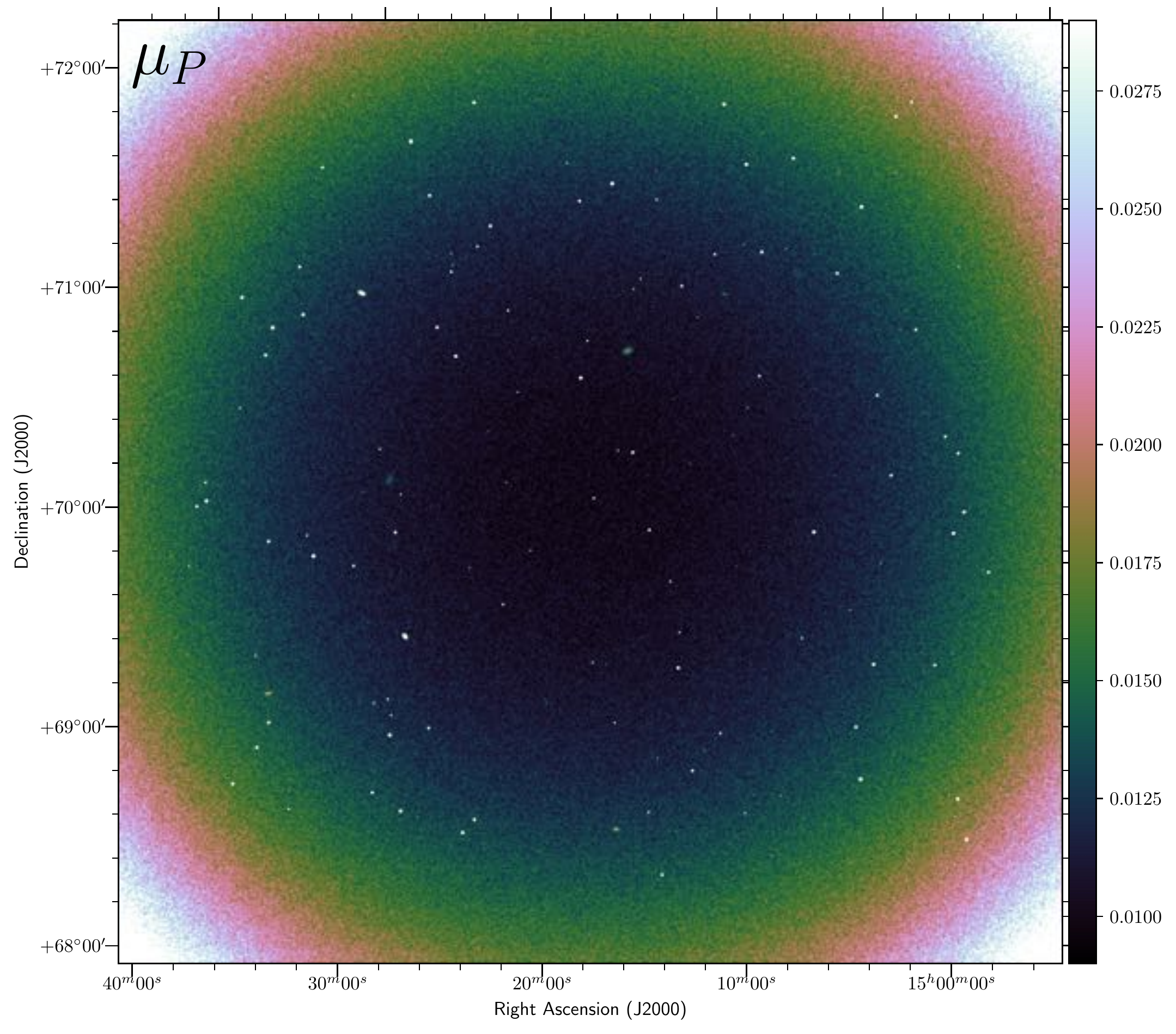}\\
      \includegraphics[trim=0.0cm 0.0cm 0.0cm 0.0cm,clip=false,angle=0,origin=c,width=5.8cm]{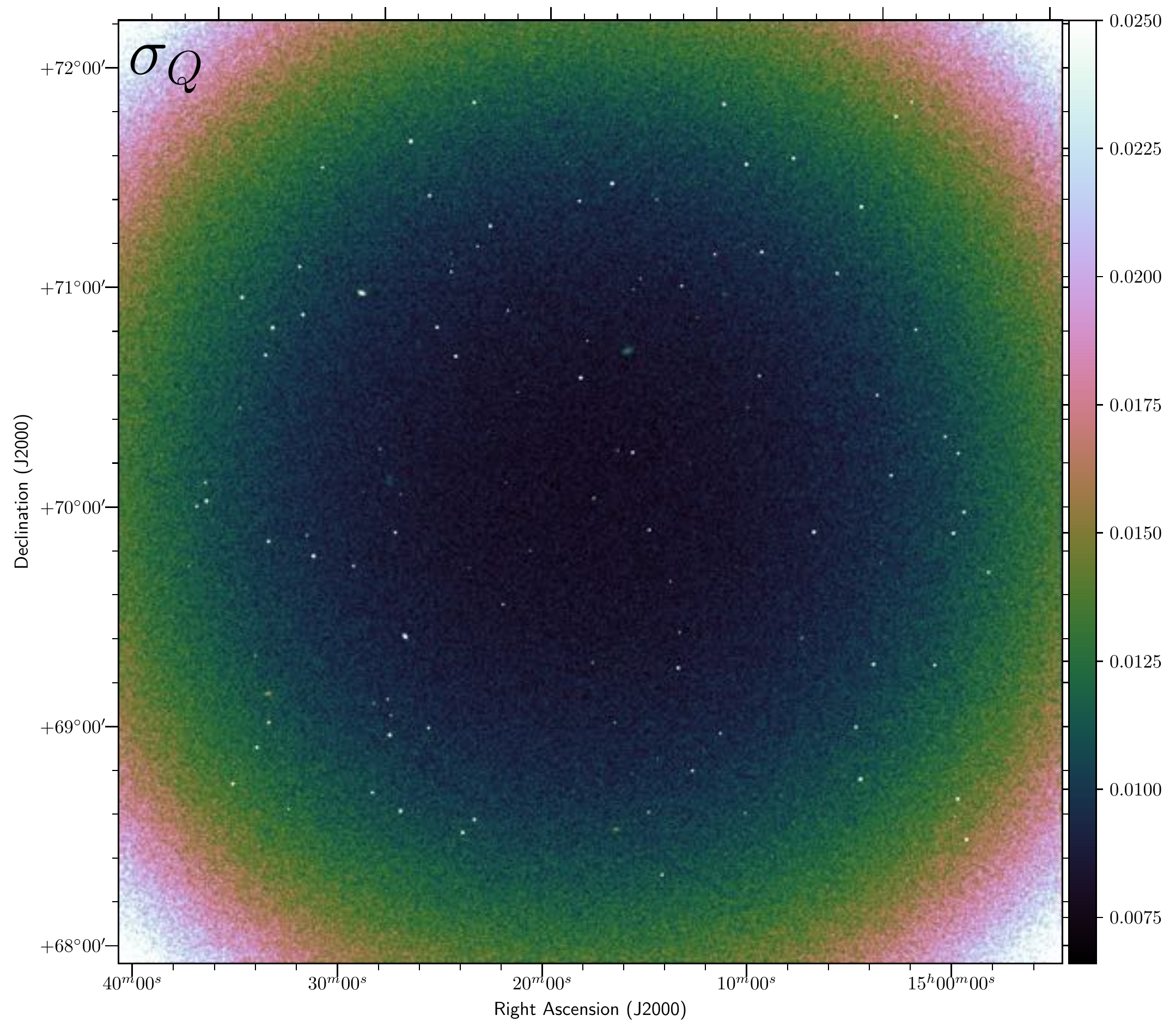}
   \includegraphics[trim=0.0cm 0.0cm 0.0cm 0.0cm,clip=false,angle=0,origin=c,width=5.8cm]{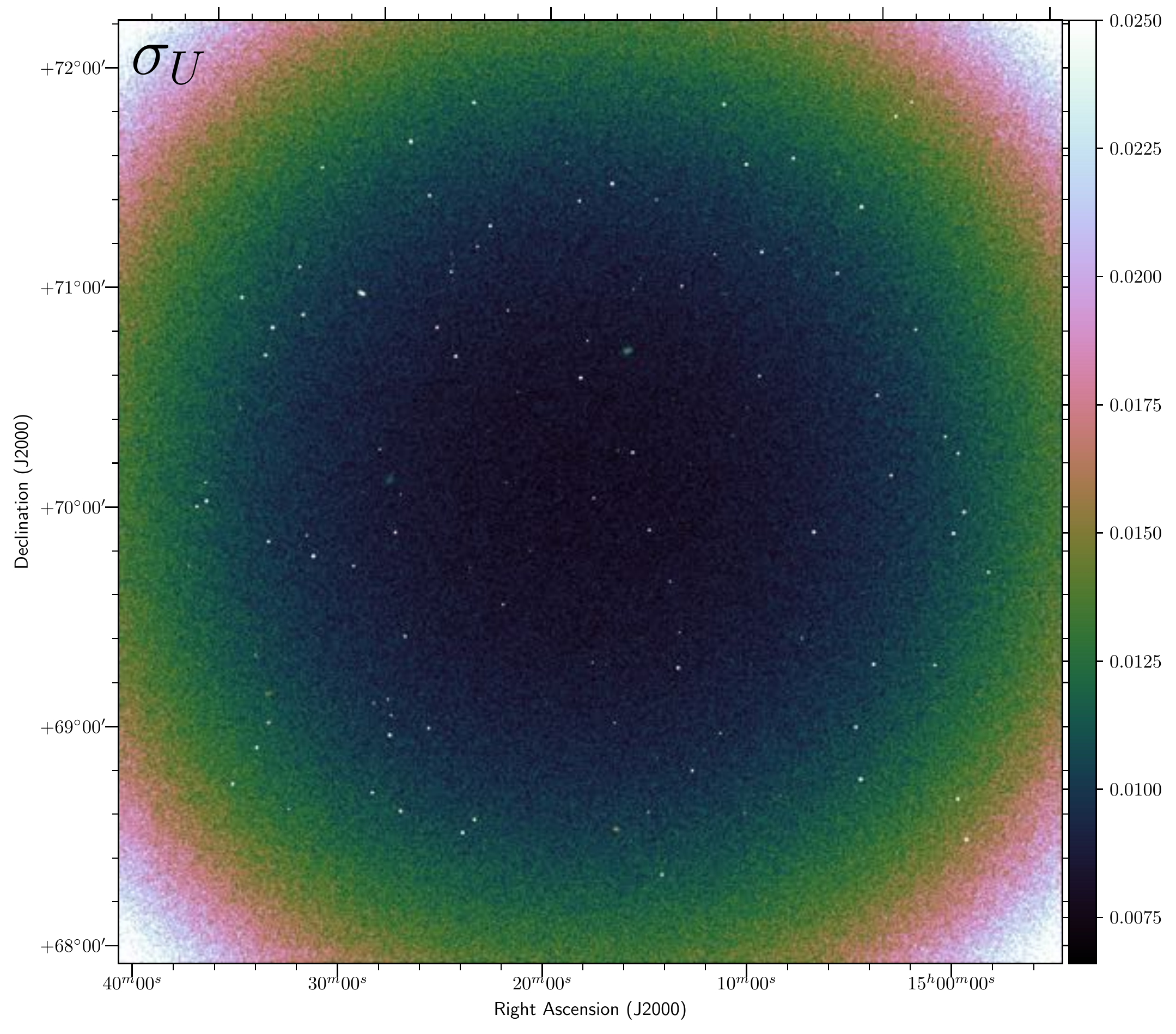}
   \includegraphics[trim=0.0cm 0.0cm 0.0cm 0.0cm,clip=false,angle=0,origin=c,width=5.8cm]{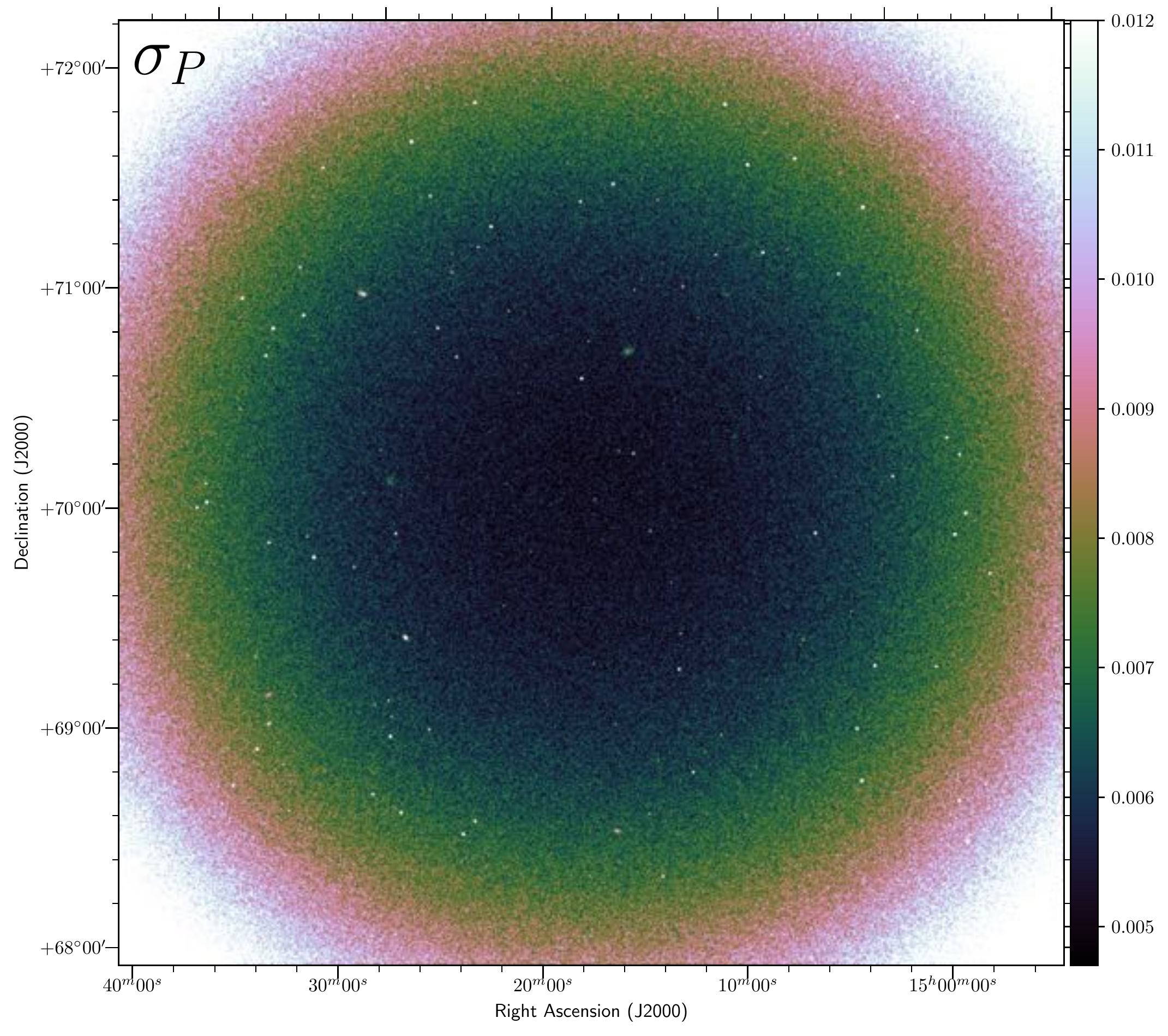}\\
      \includegraphics[trim=0.0cm 0.0cm 0.0cm 0.0cm,clip=false,angle=0,origin=c,width=5.8cm]{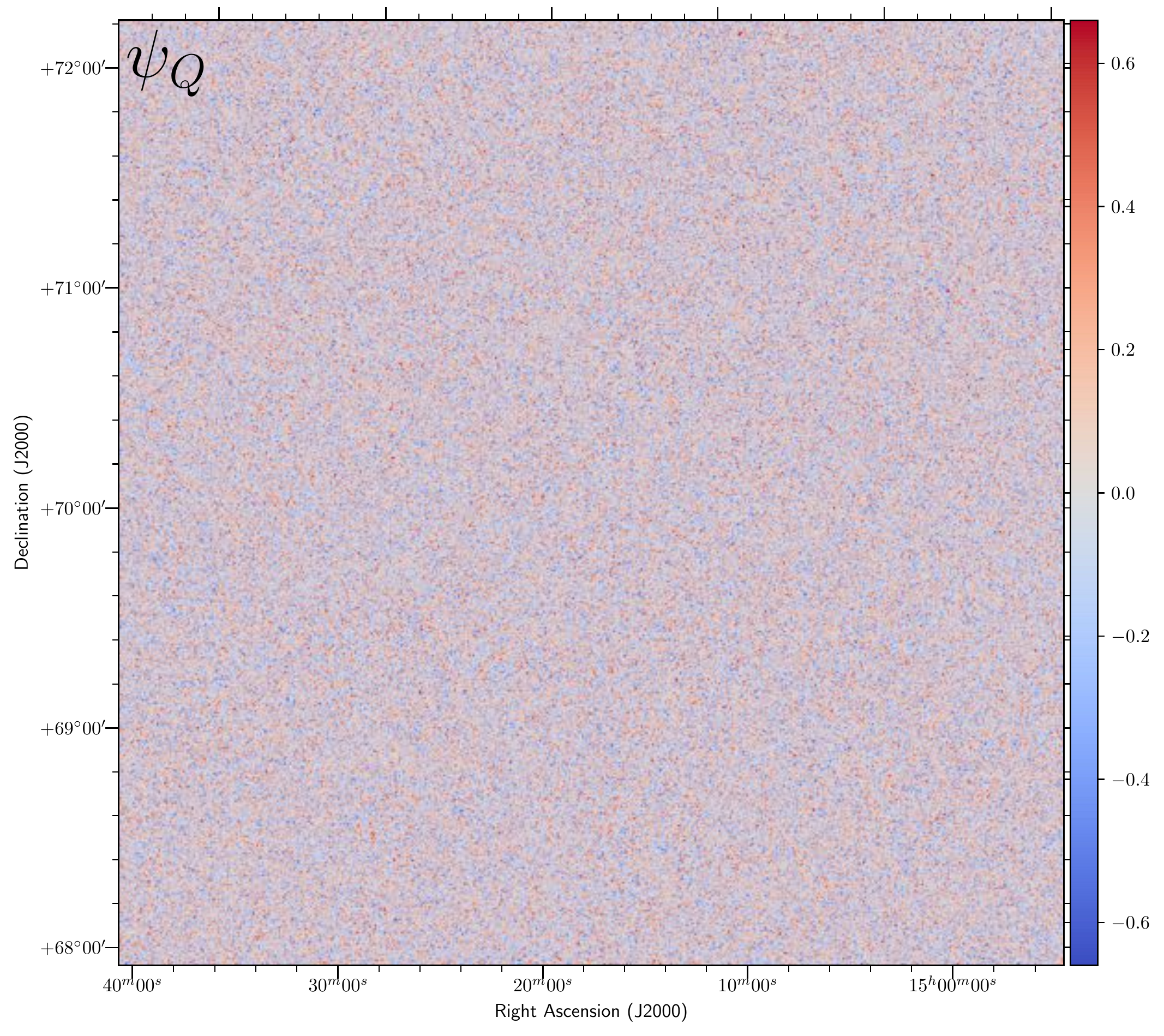}
   \includegraphics[trim=0.0cm 0.0cm 0.0cm 0.0cm,clip=false,angle=0,origin=c,width=5.8cm]{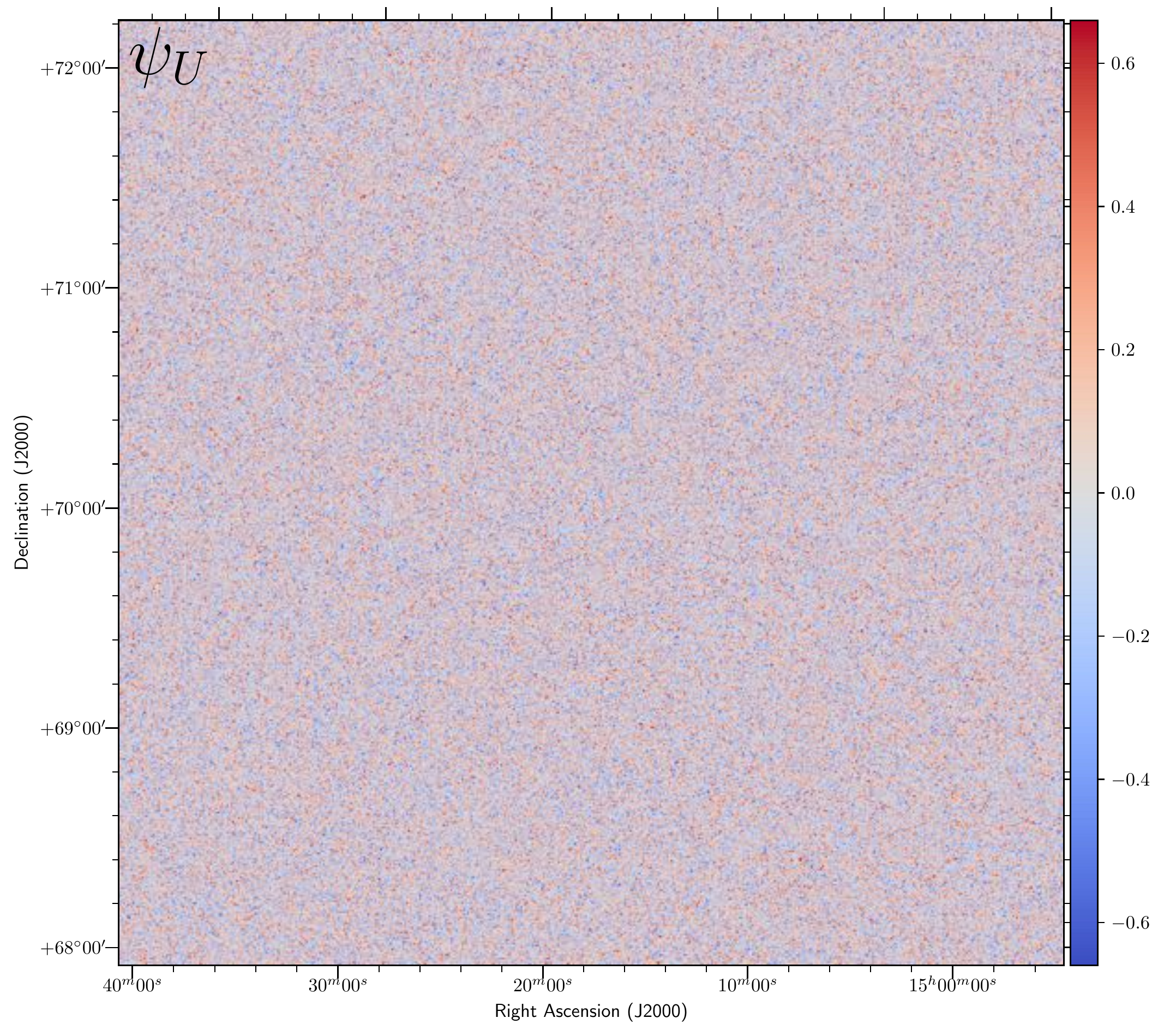}
   \includegraphics[trim=0.0cm 0.0cm 0.0cm 0.0cm,clip=false,angle=0,origin=c,width=5.8cm]{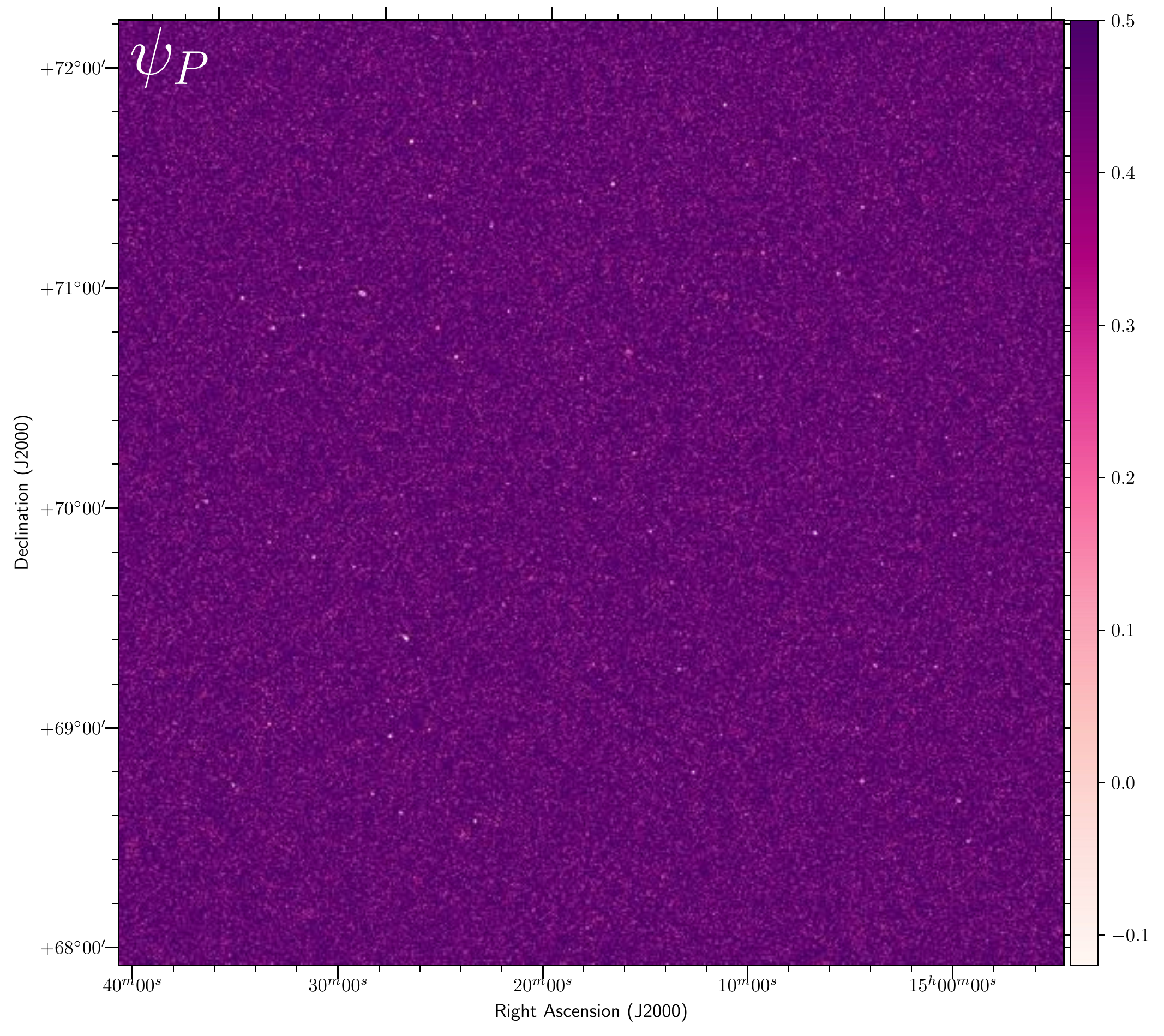}\\
      \includegraphics[trim=0.0cm 0.0cm 0.0cm 0.0cm,clip=false,angle=0,origin=c,width=5.8cm]{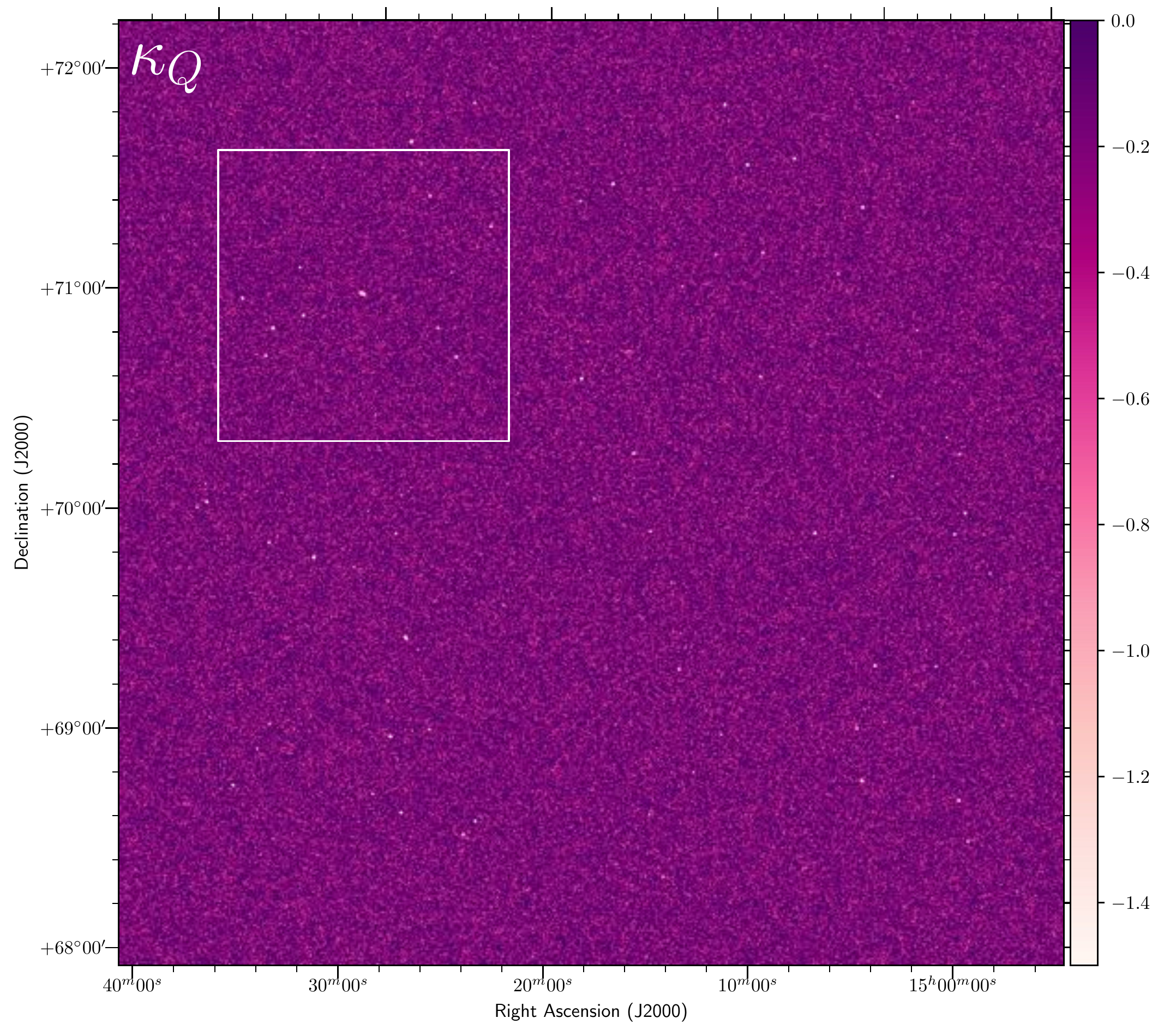}
   \includegraphics[trim=0.0cm 0.0cm 0.0cm 0.0cm,clip=false,angle=0,origin=c,width=5.8cm]{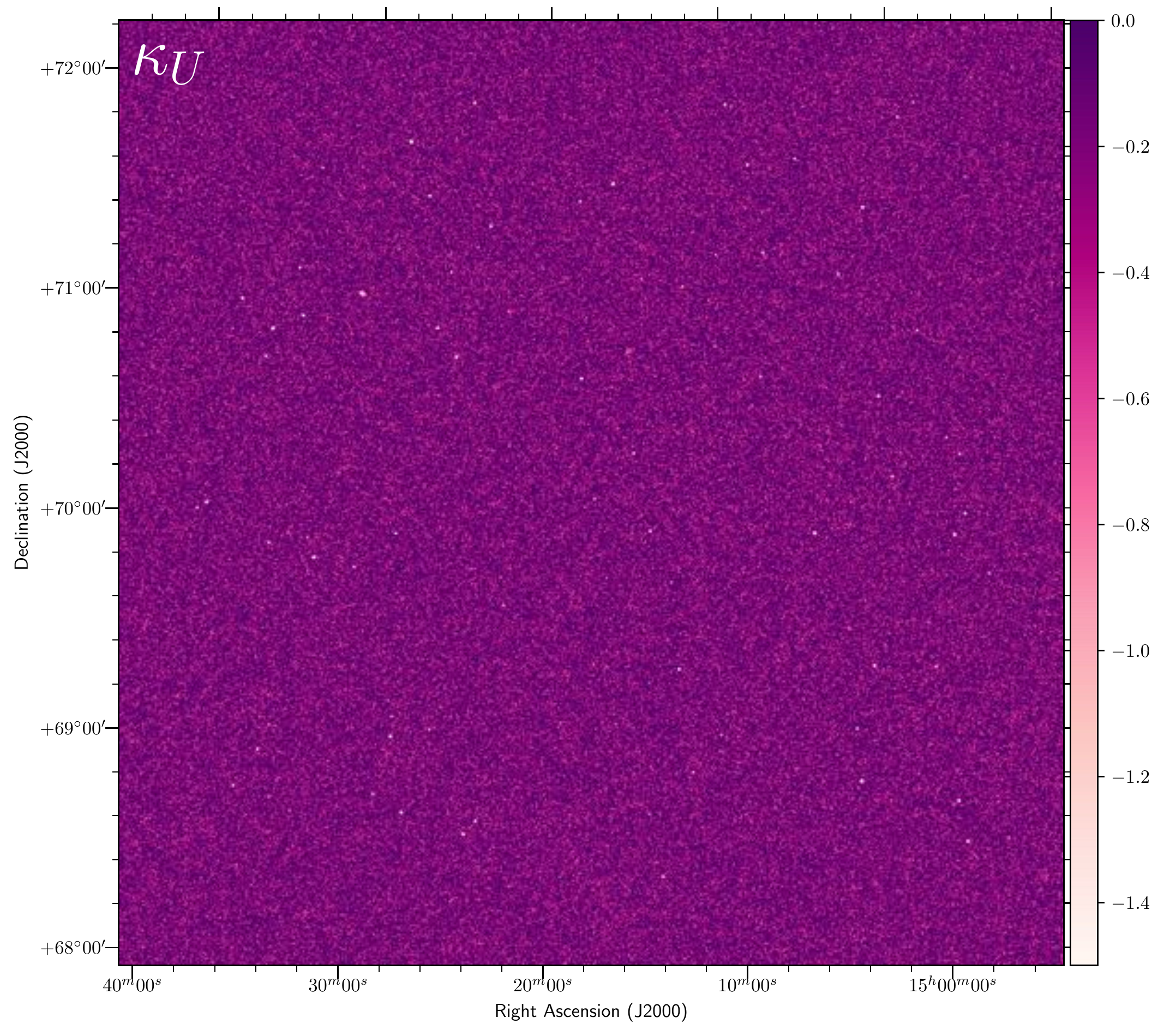}
   \includegraphics[trim=0.0cm 0.0cm 0.0cm 0.0cm,clip=false,angle=0,origin=c,width=5.8cm]{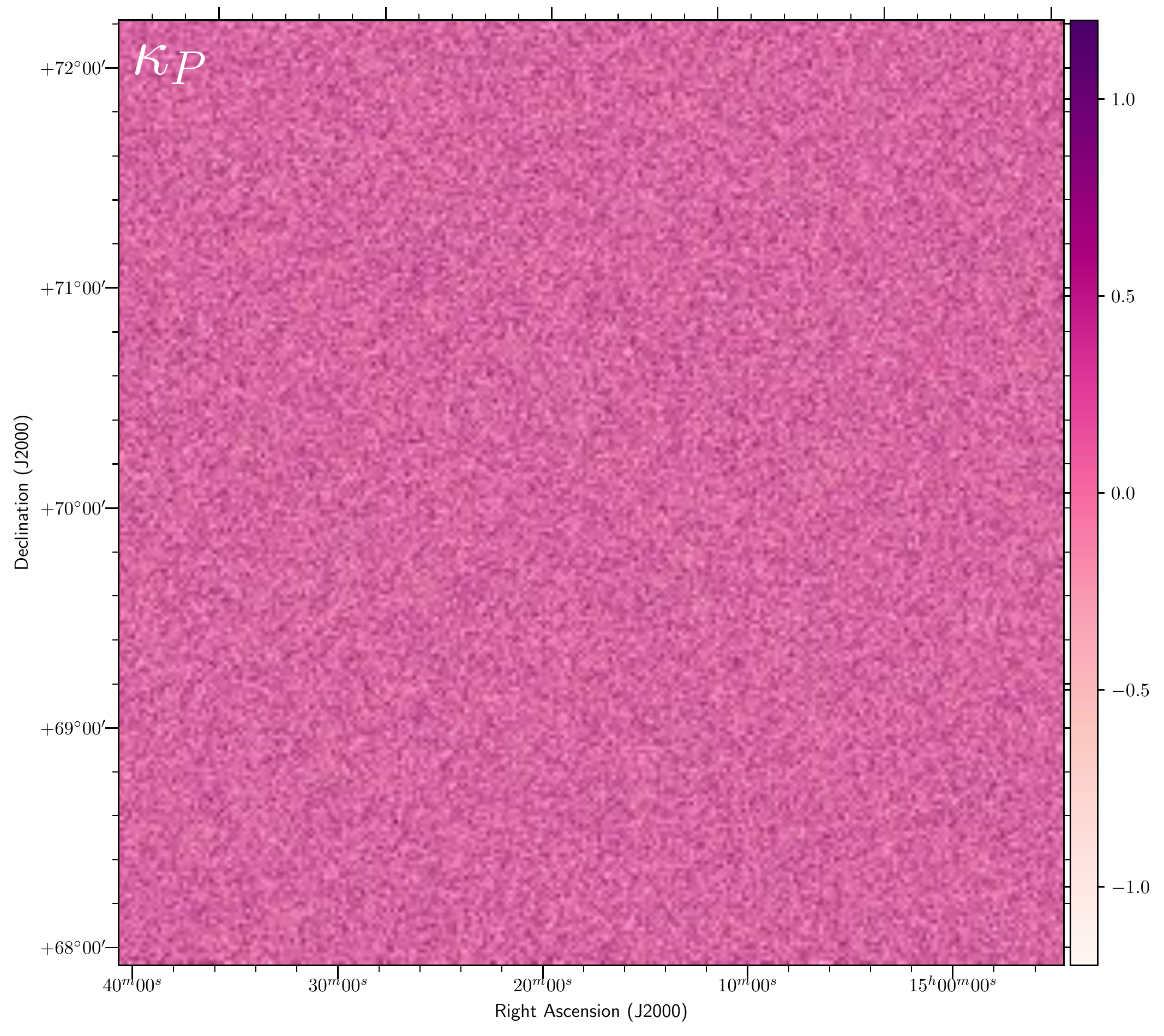}
   \caption{Examples of moment images, as derived from the simulated datacubes which contain $230$--$250$ sources. Further details on the simulations and image parameters are in Section~\ref{makesimulations}. The effect of the simulated LOFAR primary beam can be clearly seen in some moment images, as increased noise towards the periphery of the field. The moment images for the mean $\mu$ (top row), standard deviation $\sigma$ (2nd row), skewness $\psi$ (3rd row), and excess kurtosis $\kappa$ (bottom row), are all shown. The images shown are derived from the $Q$ data (left column), $U$ data (middle column), and $P$ data (right column). The pseudo-colour scales are chosen to provide contrast to each specific moment image. The $\sigma$ images all use the `cubehelix' colour-scheme \citep{2011BASI...39..289G}. The small subregion shown in Fig.~\ref{momentimagessmall} is indicated by the white box in the kurtosis of $Q$ image to the bottom left.}
              \label{momentimages}%
    \end{figure*}

 \begin{figure*}
   \centering
   \includegraphics[trim=0.0cm 0.0cm 0.0cm 0.0cm,clip=false,angle=0,origin=c,width=5.8cm]{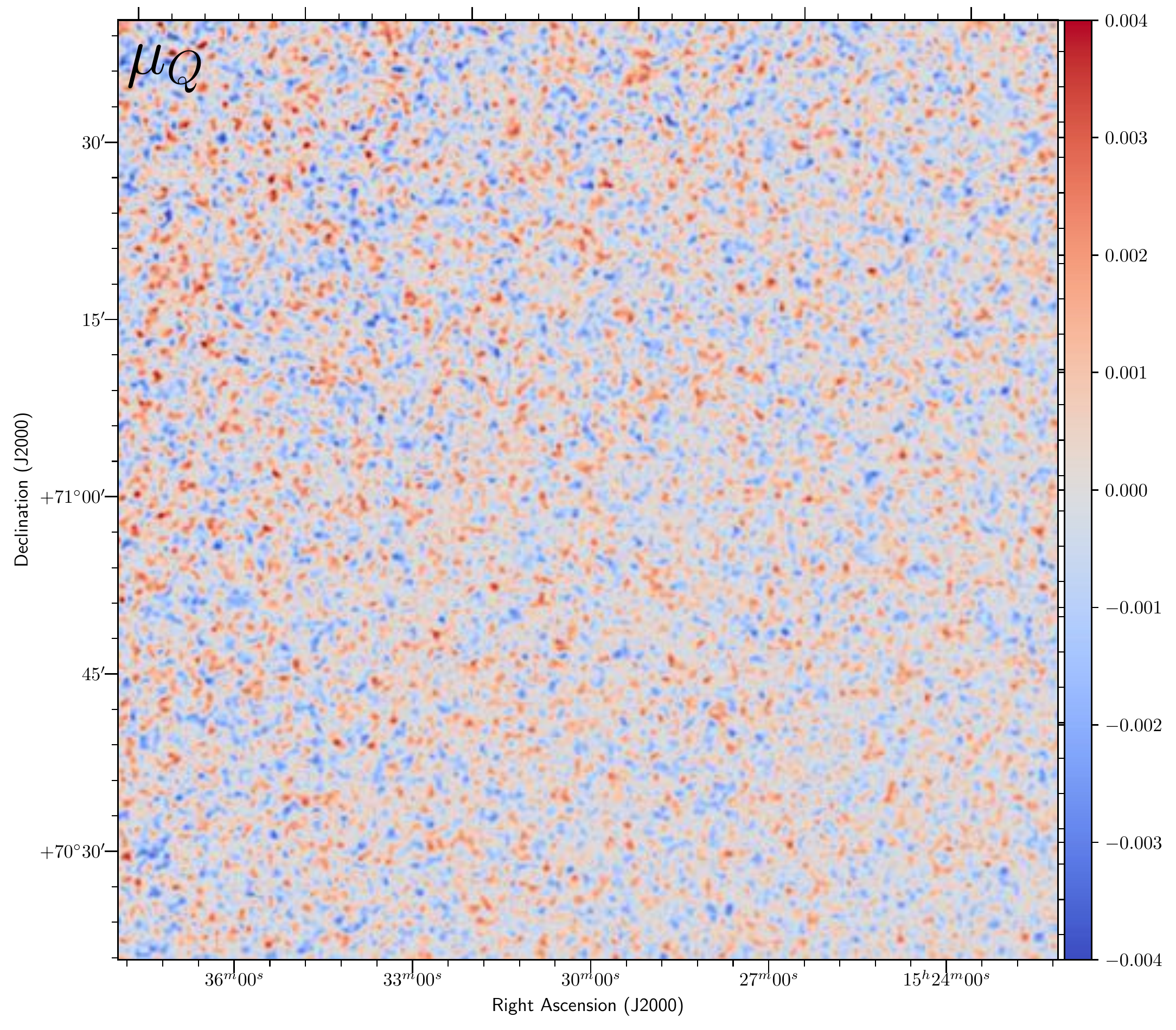}
   \includegraphics[trim=0.0cm 0.0cm 0.0cm 0.0cm,clip=false,angle=0,origin=c,width=5.8cm]{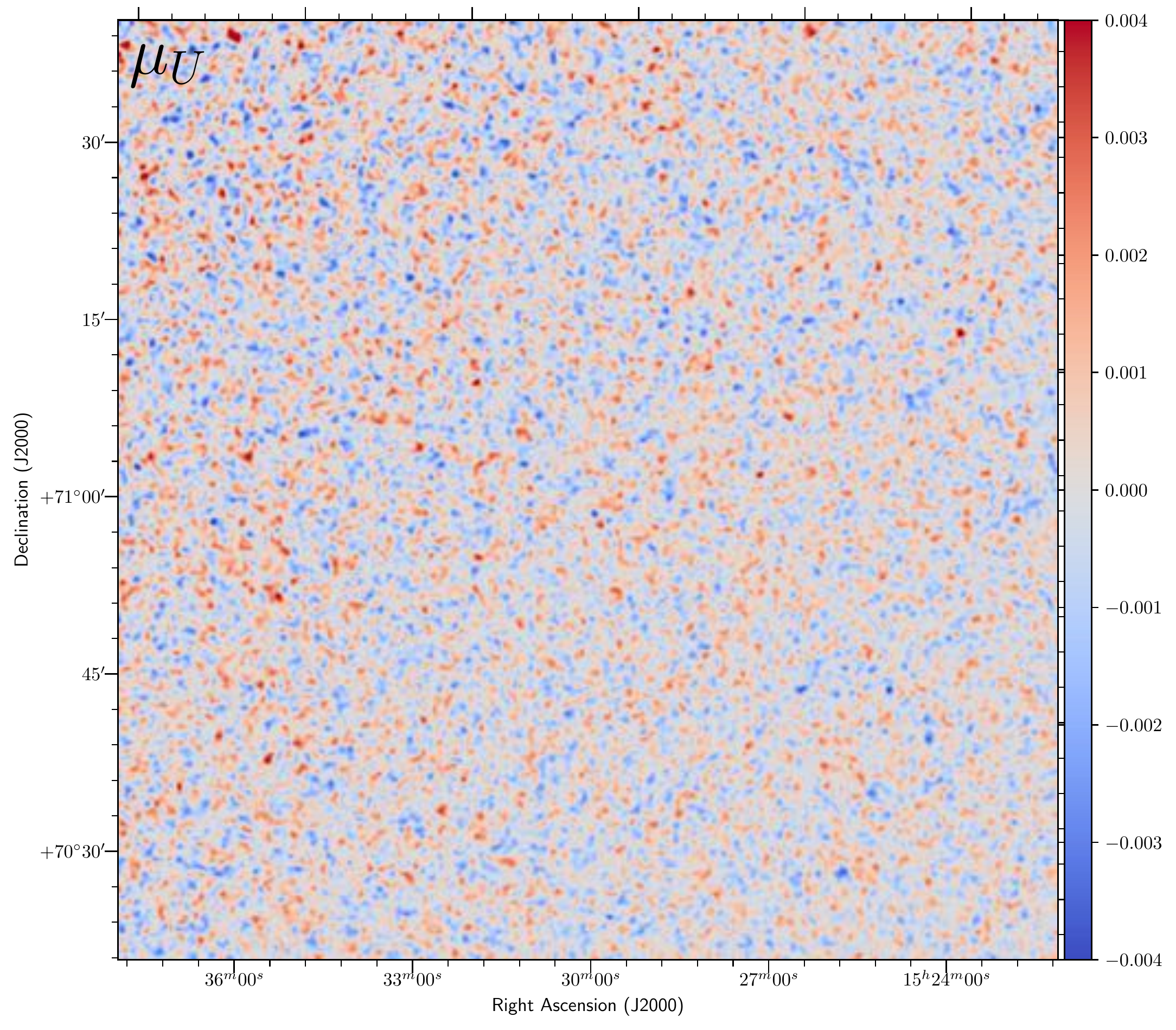}
   \includegraphics[trim=0.0cm 0.0cm 0.0cm 0.0cm,clip=false,angle=0,origin=c,width=5.8cm]{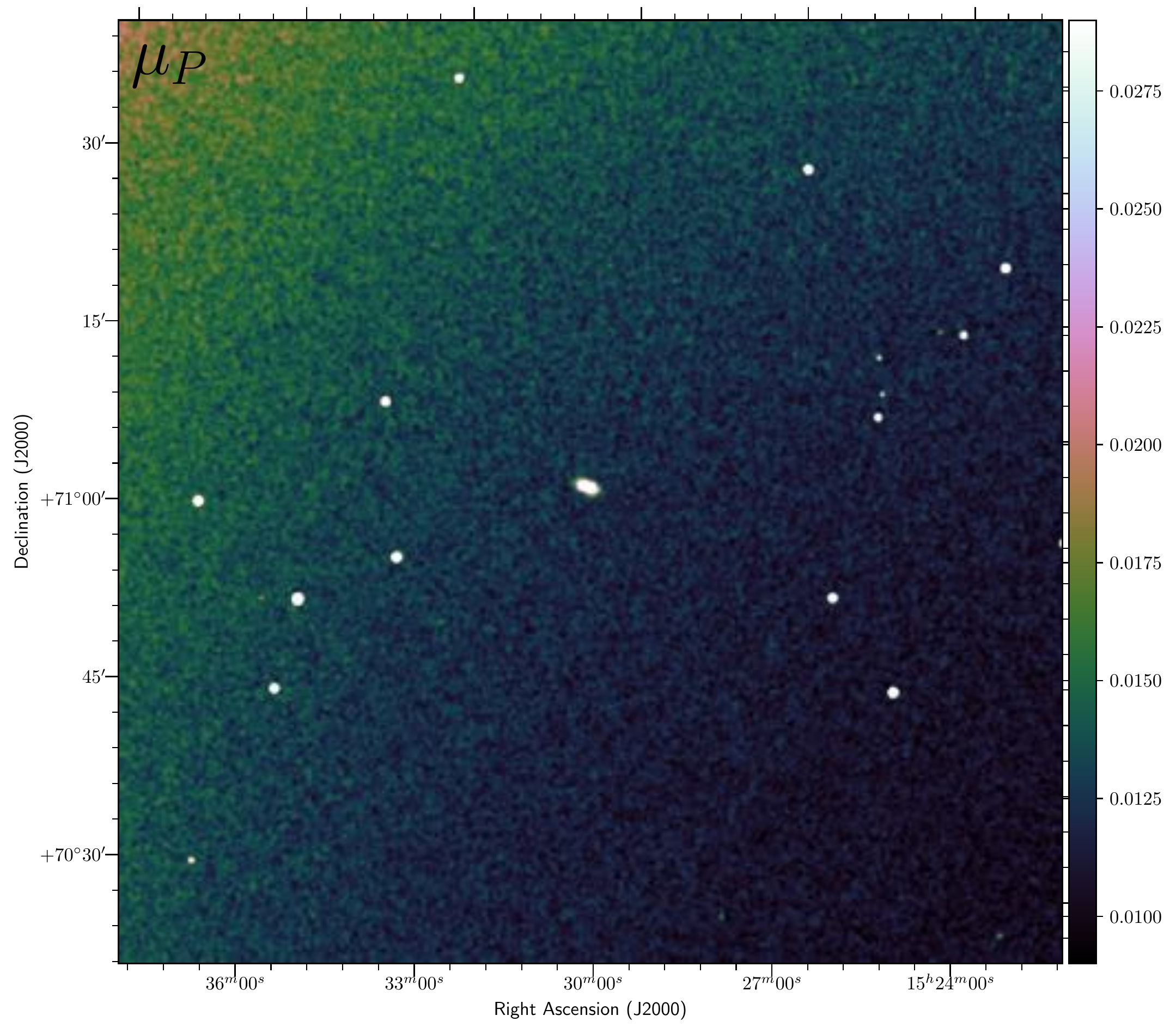}\\
      \includegraphics[trim=0.0cm 0.0cm 0.0cm 0.0cm,clip=false,angle=0,origin=c,width=5.8cm]{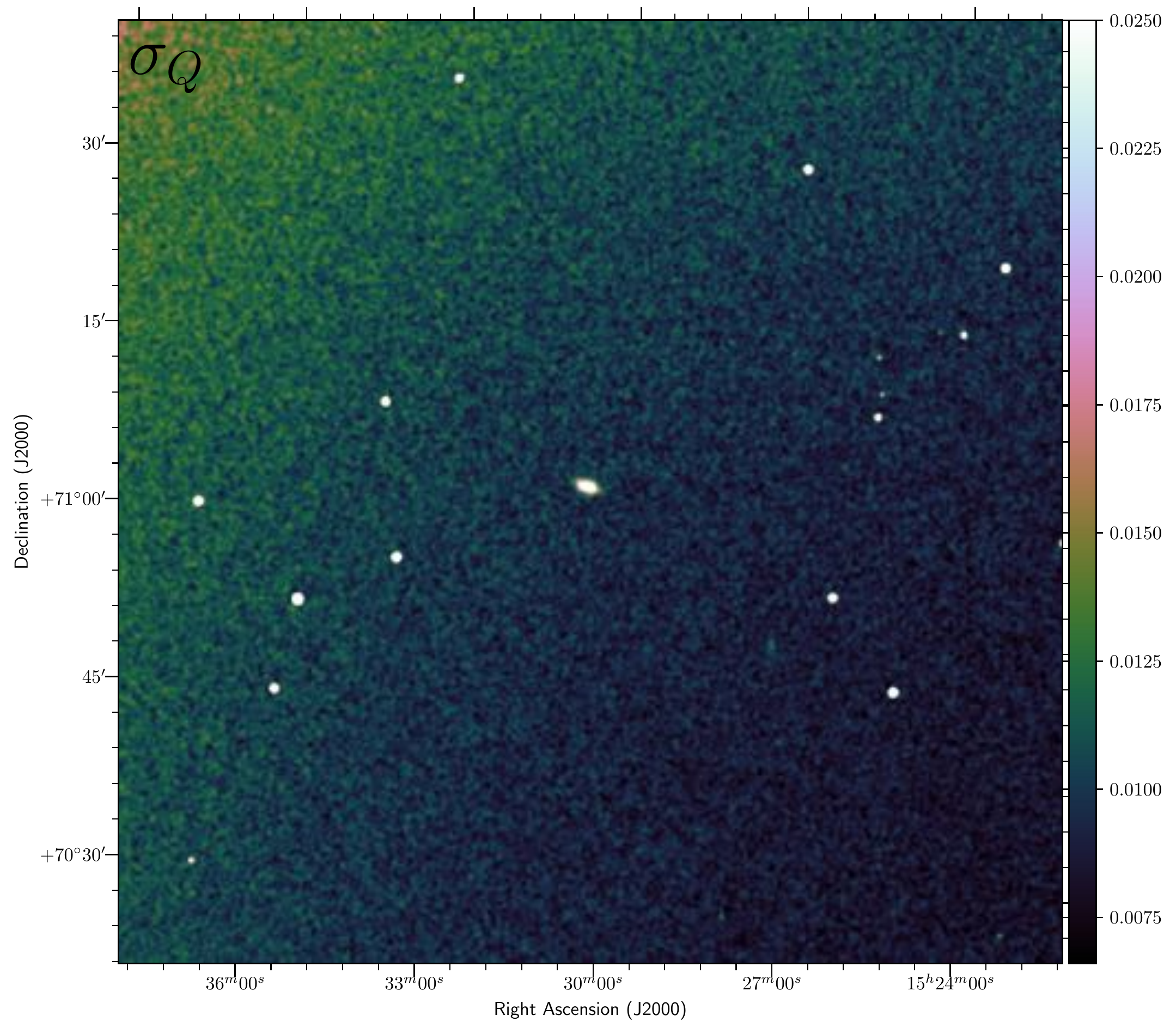}
   \includegraphics[trim=0.0cm 0.0cm 0.0cm 0.0cm,clip=false,angle=0,origin=c,width=5.8cm]{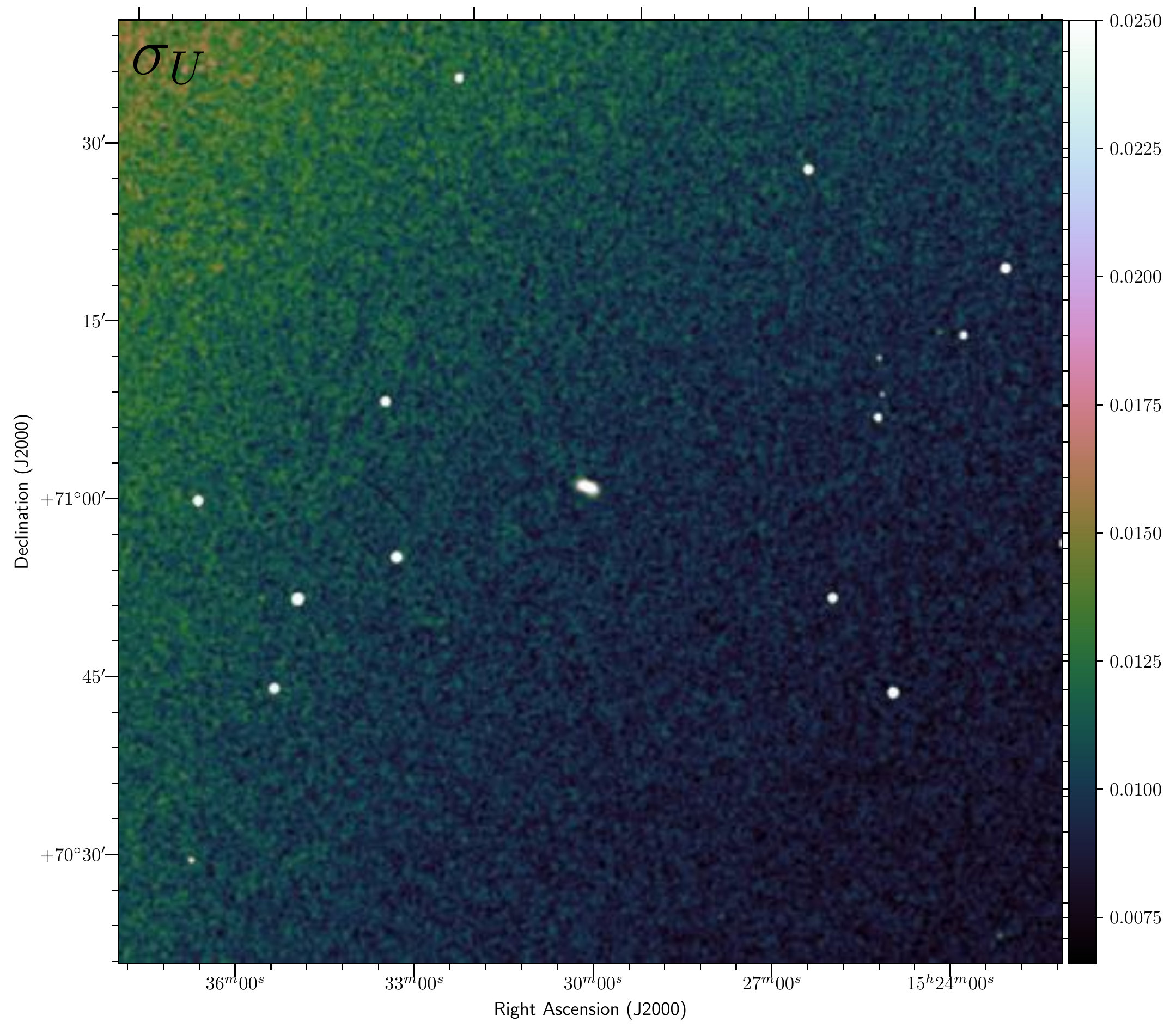}
   \includegraphics[trim=0.0cm 0.0cm 0.0cm 0.0cm,clip=false,angle=0,origin=c,width=5.8cm]{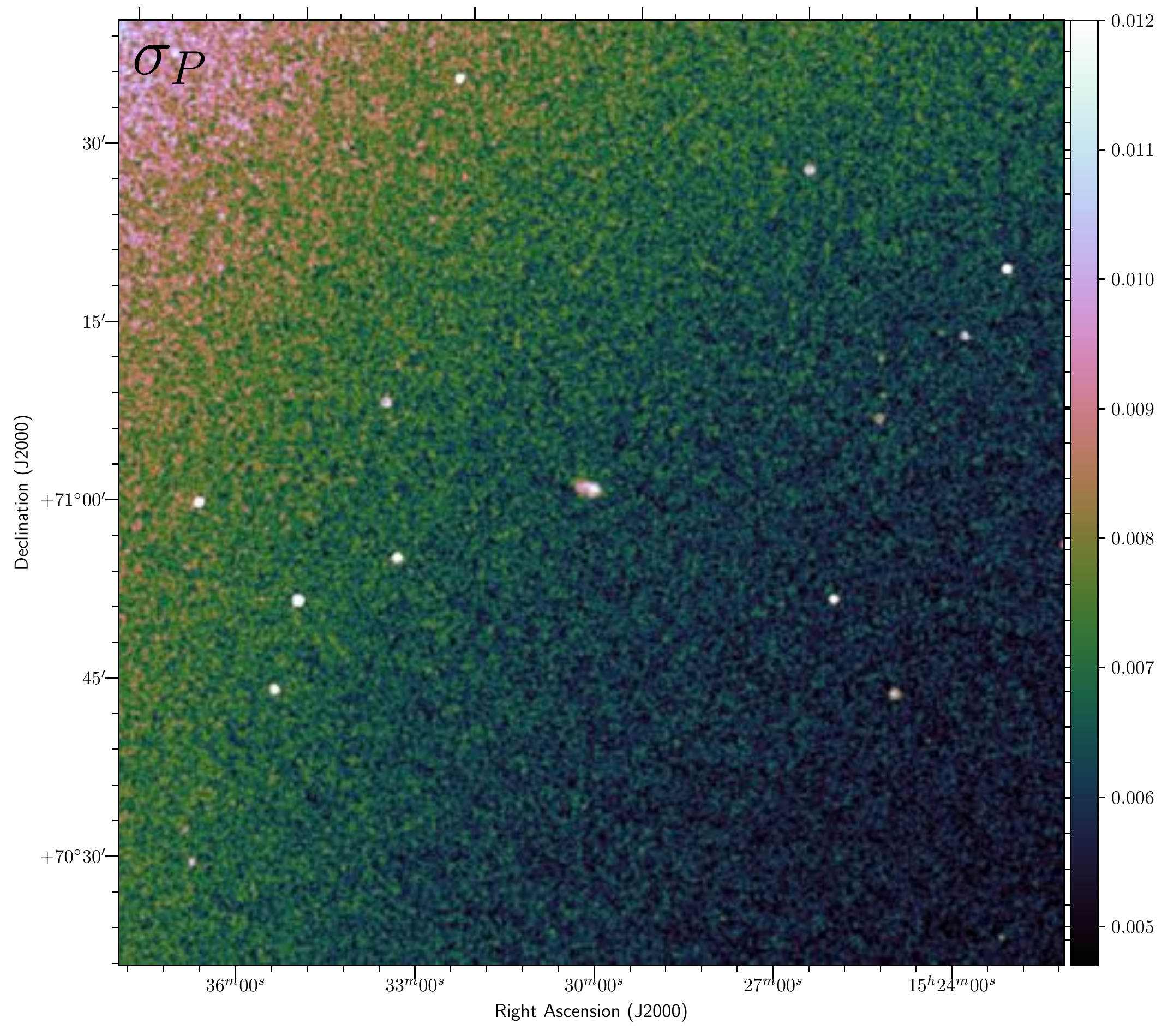}\\
      \includegraphics[trim=0.0cm 0.0cm 0.0cm 0.0cm,clip=false,angle=0,origin=c,width=5.8cm]{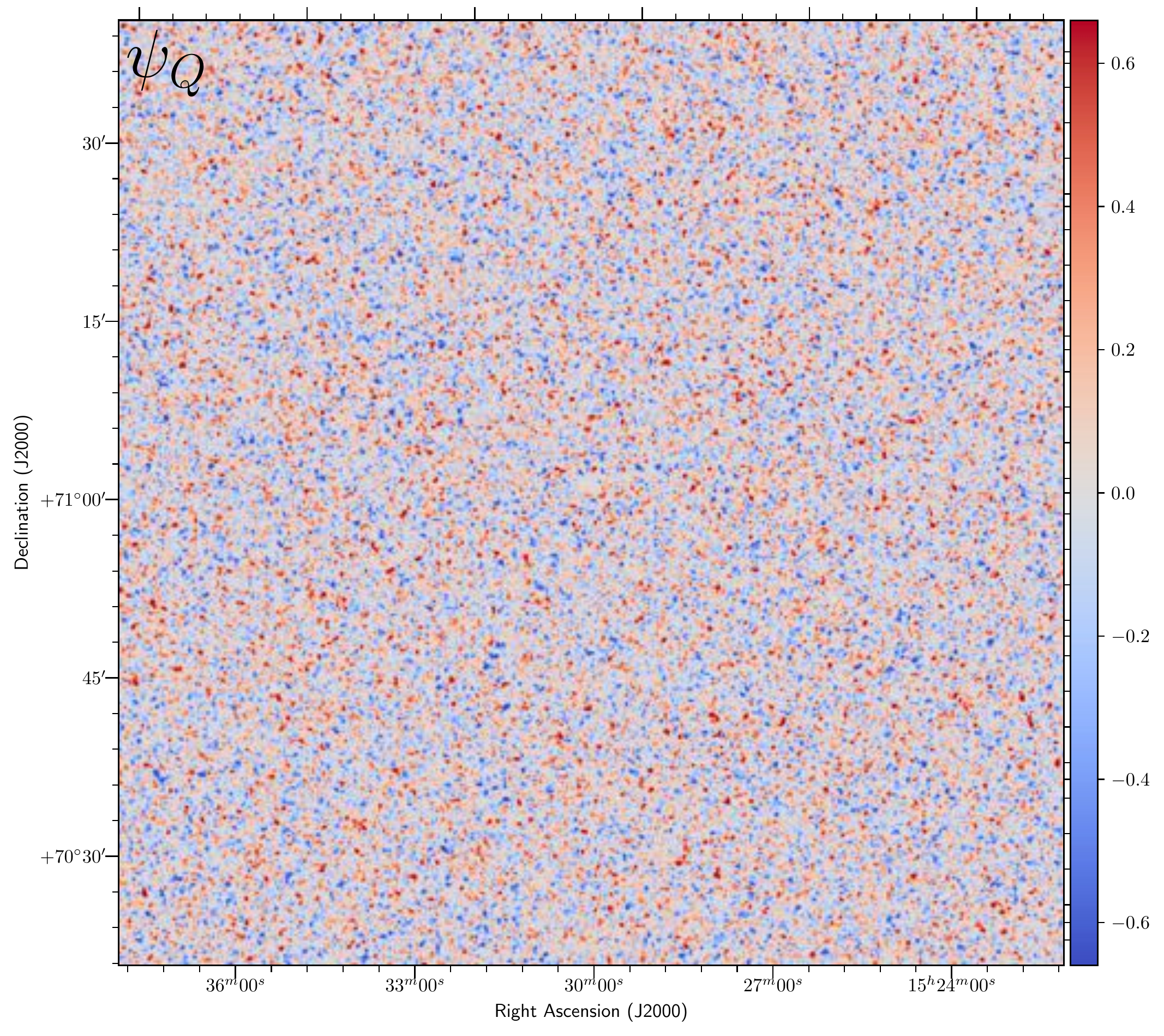}
   \includegraphics[trim=0.0cm 0.0cm 0.0cm 0.0cm,clip=false,angle=0,origin=c,width=5.8cm]{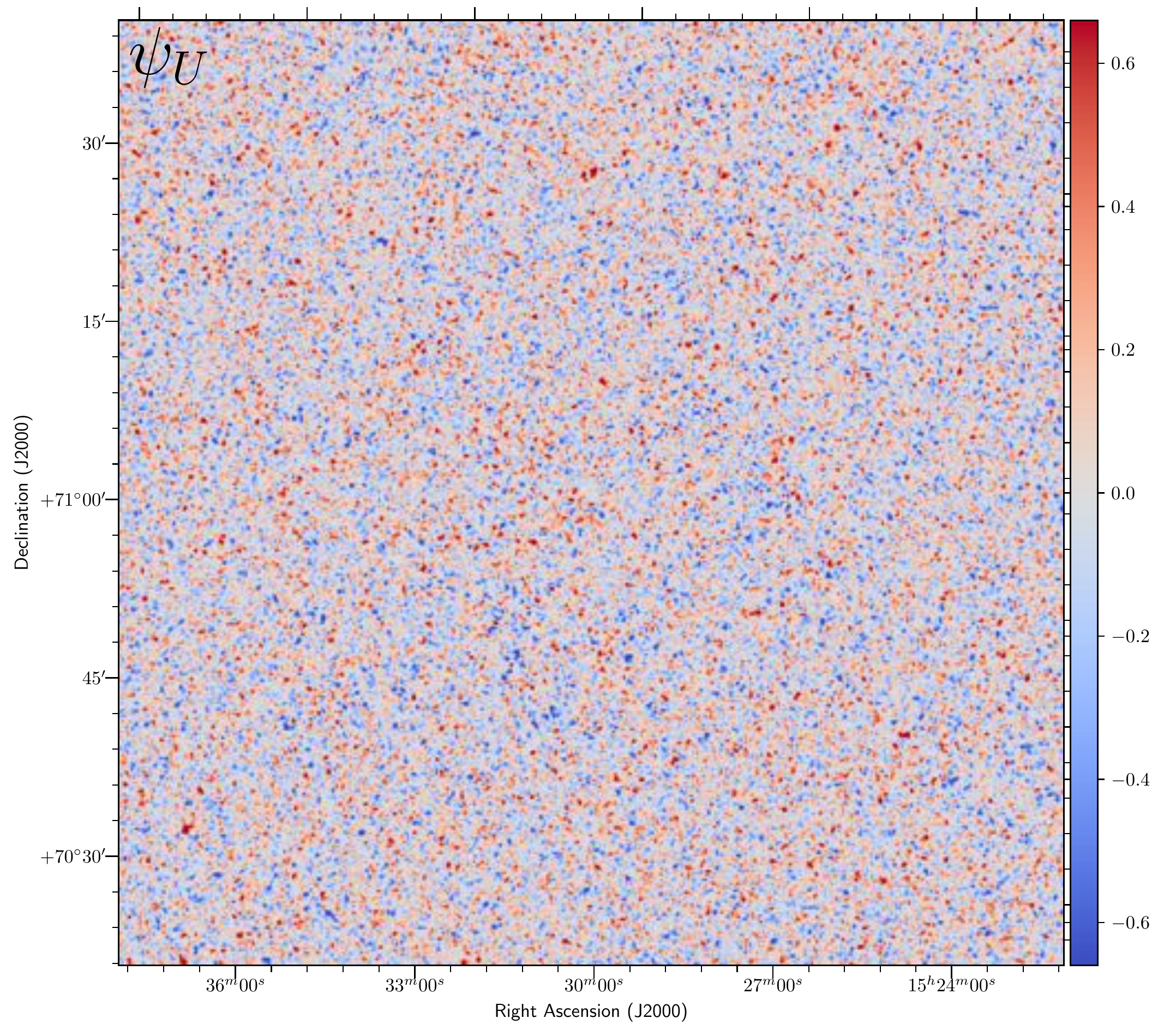}
   \includegraphics[trim=0.0cm 0.0cm 0.0cm 0.0cm,clip=false,angle=0,origin=c,width=5.8cm]{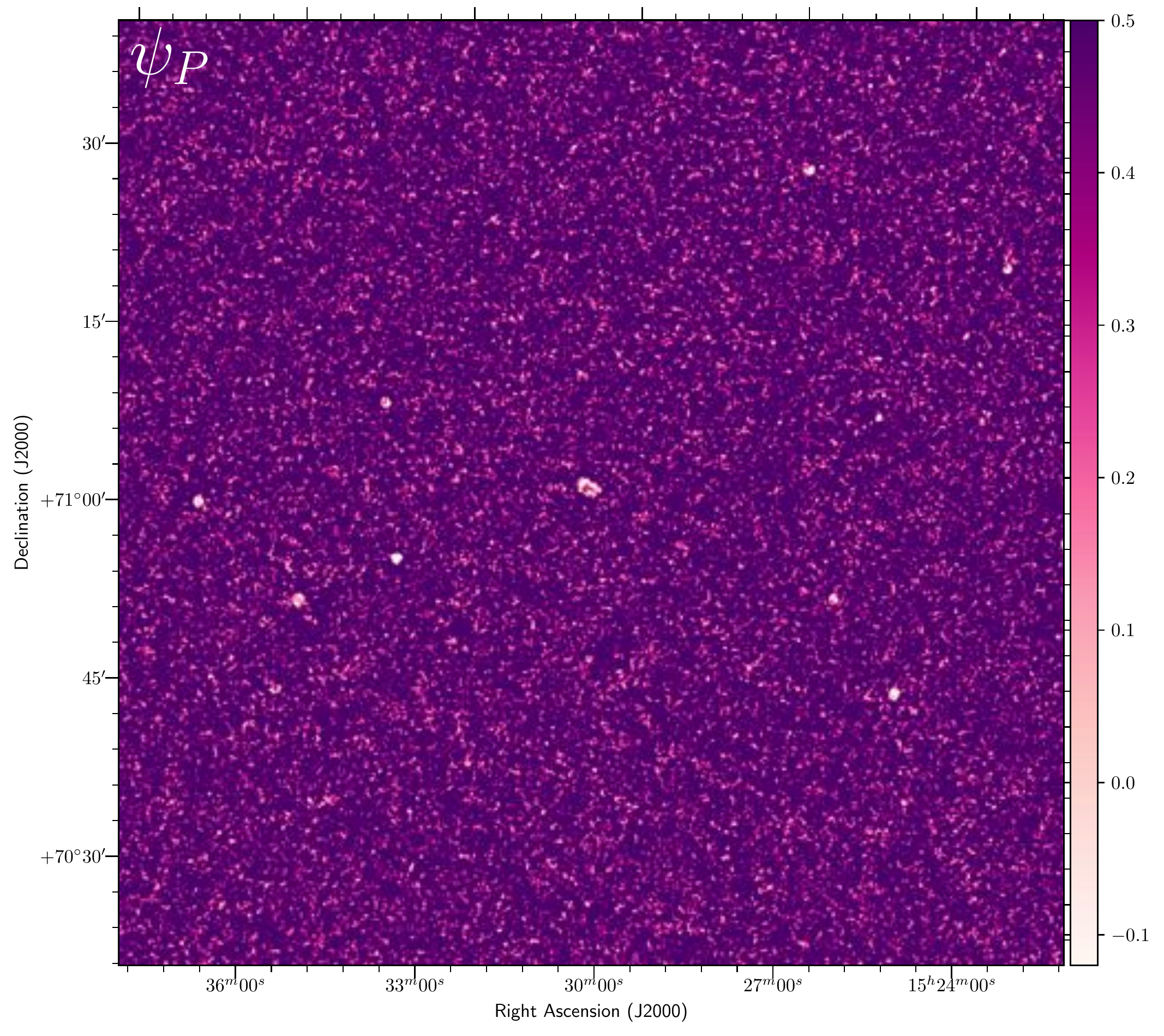}\\
      \includegraphics[trim=0.0cm 0.0cm 0.0cm 0.0cm,clip=false,angle=0,origin=c,width=5.8cm]{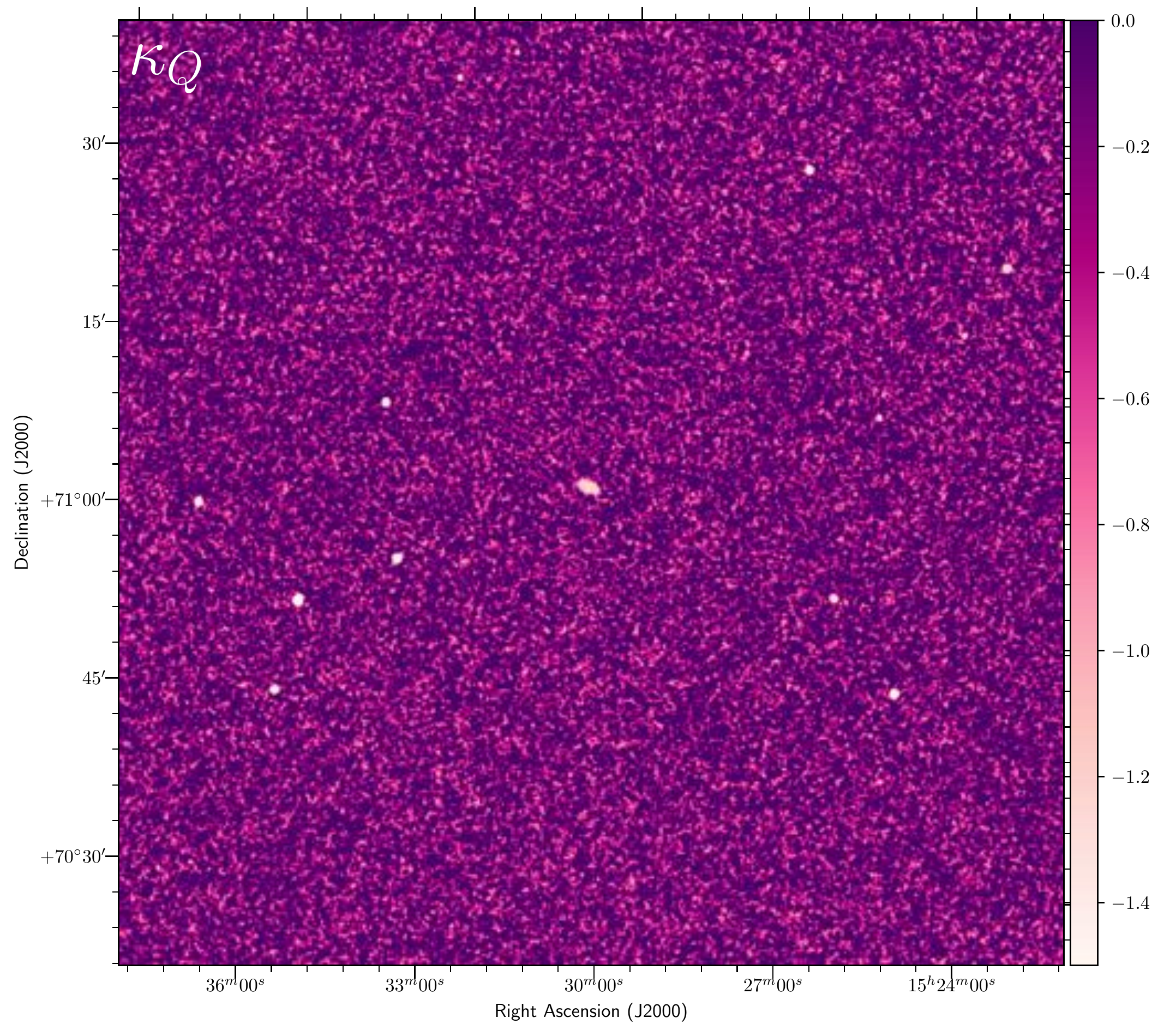}
   \includegraphics[trim=0.0cm 0.0cm 0.0cm 0.0cm,clip=false,angle=0,origin=c,width=5.8cm]{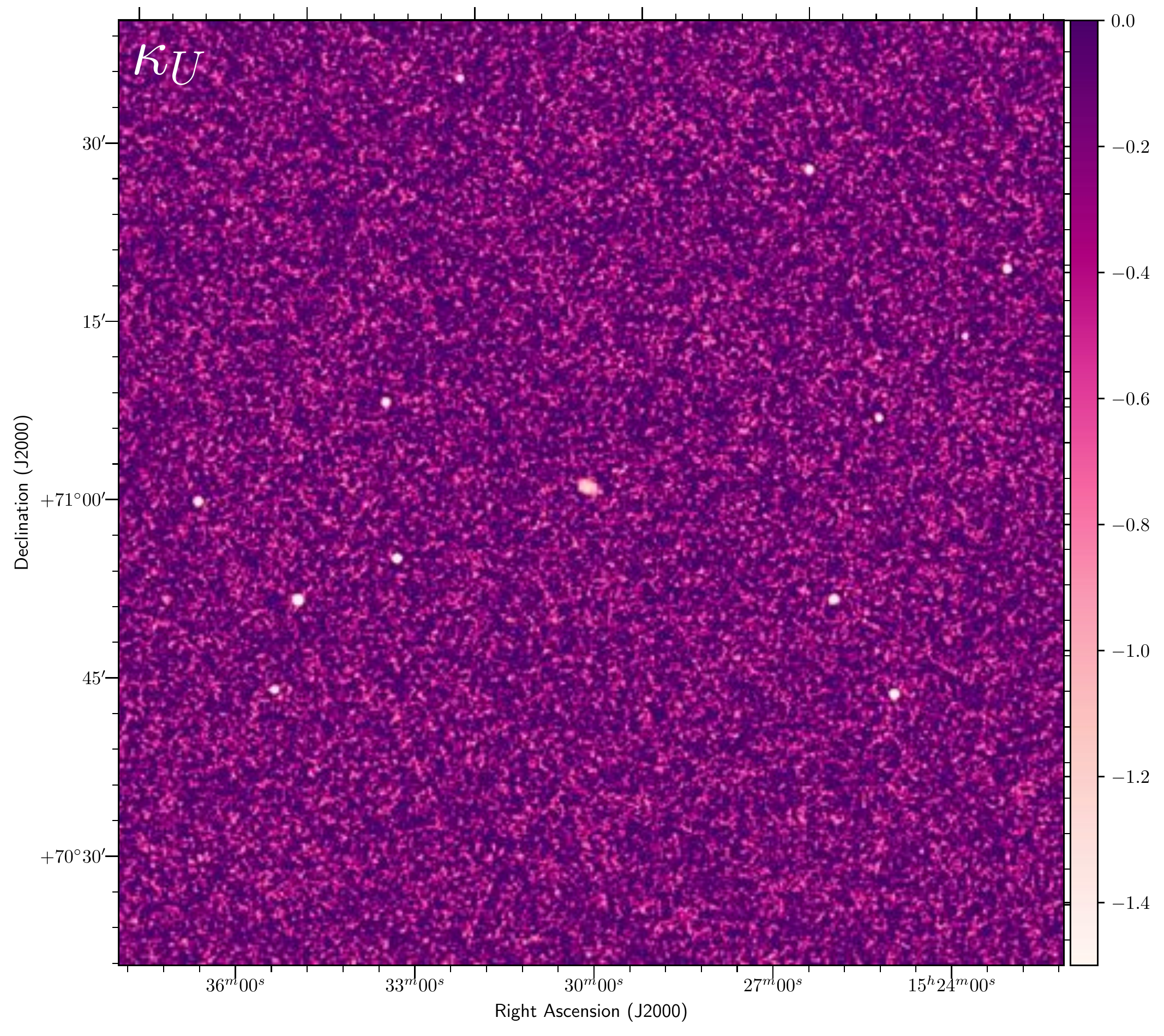}
   \includegraphics[trim=0.0cm 0.0cm 0.0cm 0.0cm,clip=false,angle=0,origin=c,width=5.8cm]{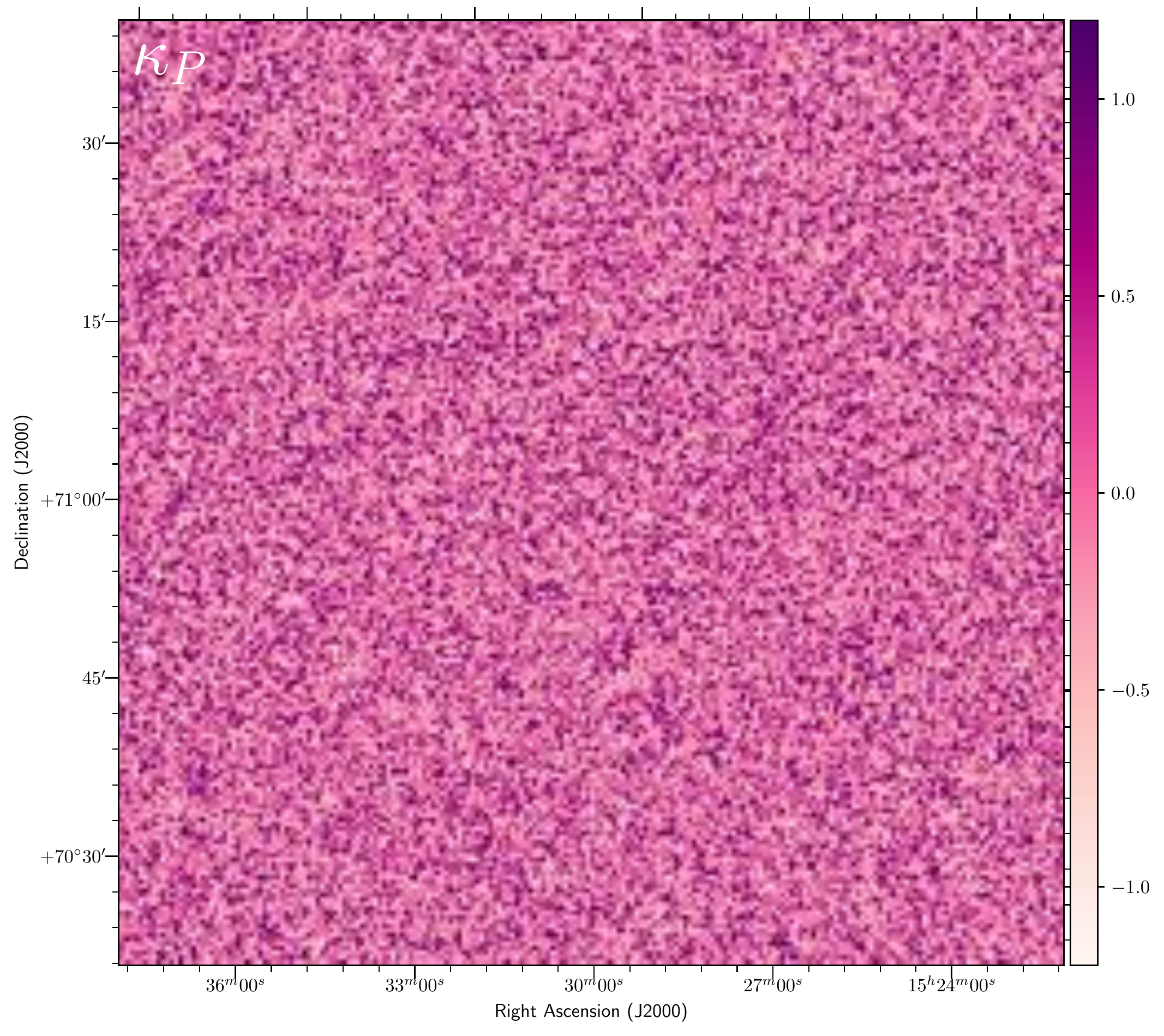}
   \caption{The same simulated images as shown in Fig.~\ref{momentimages}, but zoomed-in towards the small subregion. Further details on the simulations and image parameters are in Section~\ref{makesimulations}.}
              \label{momentimagessmall}%
    \end{figure*}

The moment images were generated for each pixel, $i$, using all observational wavelengths, $\lambda$, using the equations given in Section~\ref{calcmoments}. The derived moment images are shown in Fig.~\ref{momentimages}, and the same moment images zoomed-in towards a subregion are shown in Fig.~\ref{momentimagessmall}. These Faraday Moment images make clear which moments are most useful for source-finding. In particular, the $\mu_{P}$ and all $\sigma$ images are particularly useful. The $\mu_{Q}$ and $\mu_{U}$ images work well at detecting those sources with particularly low RMs. In this sense, one can consider the $\mu_{Q}$ and $\mu_{U}$ images as being equivalent to RM Synthesis with RM$=0$~rad~m$^{-2}$. In this way, these mean images can also add to the overall completeness of the method. However, the $\psi_{Q}$ and $\psi_{U}$ images do not work well at detecting sources, and are dominated by noise. However, $\psi_{P}$ provides negatively-valued dips at the location of some sources. Meanwhile, the $\kappa_{Q}$ and $\kappa_{U}$ images also provide negatively-valued dips at the location of Faraday-rotating sources, while $\kappa_{P}$ is also dominated by noise -- however, our simulations do not include the effects of depolarization, for which $\kappa_{P}$ can be a useful tracer (see Section~\ref{comparisonofmoments}). Our simulations do include a spectral index, however strictly speaking this cannot mimic the effects of depolarization across the observing band as we apply the Faraday Moments to the polarized intensity, rather than the polarized fraction. However, this does give rise to a change in polarized intensity across the band, which to some extent simulates changing polarization properties with frequency.

\subsection{Using Conventional Source-Finding on Mean and Standard Deviation Images}
\label{conventional}
\begin{figure}
   \centering
   \includegraphics[trim=1.6cm 3.65cm 3.0cm 4.1cm,clip=true,angle=0,origin=c,width=\hsize]{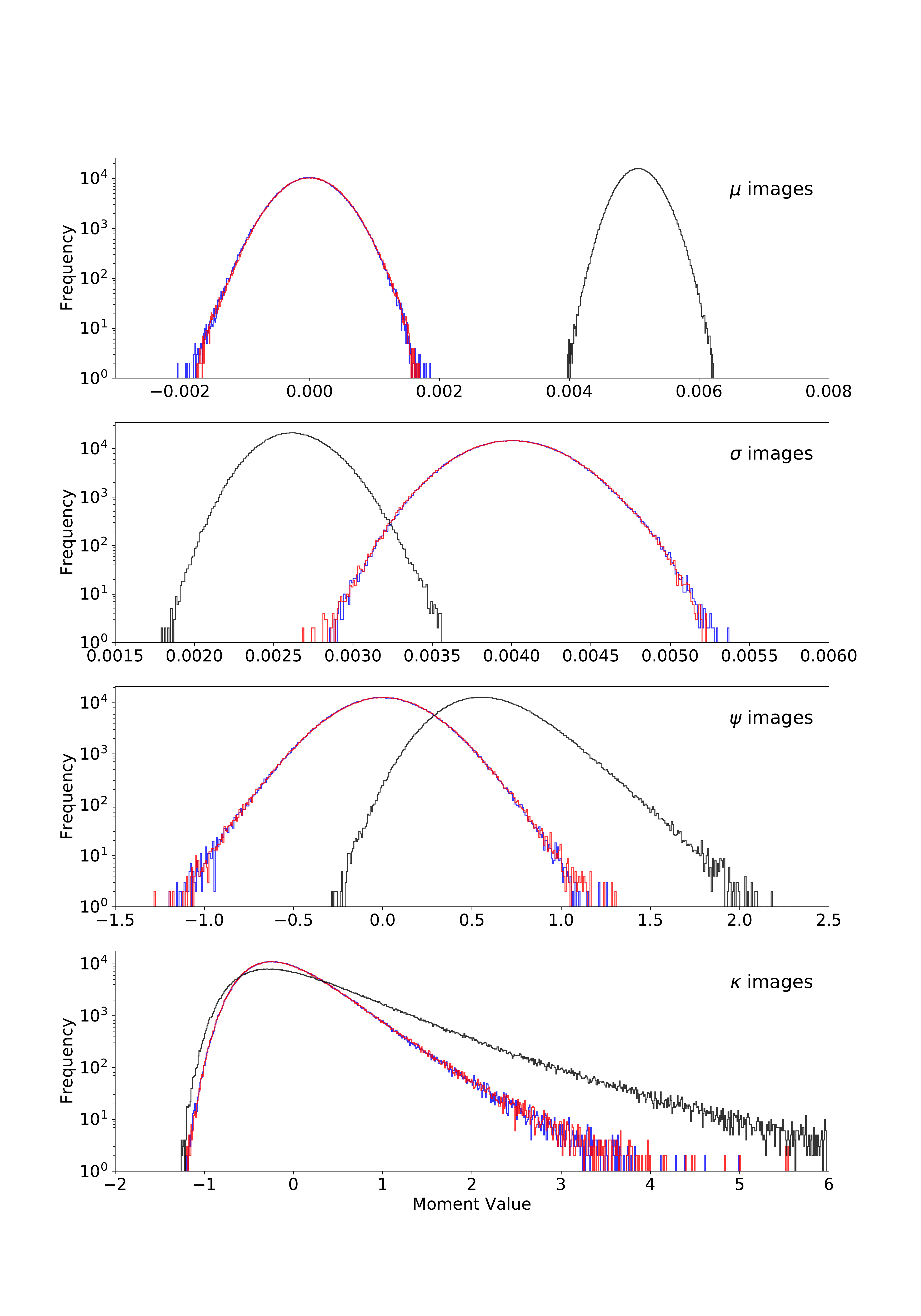}
   \caption{The distributions of noise in the central region of $1024^2$ pixels in the moment images, as derived from a simulated datacube that contains no sources. The distributions are shown for $Q$ (blue), $U$ (red), and $P$ (black). The noise distributions of the moment images for the mean $\mu$ (top panel), standard deviation $\sigma$ (2nd panel), skewness $\psi$ (3rd panel), and excess kurtosis $\kappa$ (bottom panel), are all shown. Note that the $y$-axis is shown as a log-scale in order to accentuate deviations from a normal distribution.}
              \label{NoiseDists}%
    \end{figure}

A critical condition for being able to use Faraday Moments for polarized source-finding, is being able to distinguish the moments of a real source from those that originate due to noise (as described in Section~\ref{comparisonofmoments}). All publicly available astronomical source-finding algorithms have all been designed and optimised to find islands of positive and negative values surrounded by Gaussian noise (see the useful review and test of various source-finders in \citealt{2012MNRAS.422.1812H}). The effect of non-Gaussian noise statistics can yield additional false positives, as is the case when source-finding in $P$ images \citep[e.g.][]{2012MNRAS.425..979H}. Whether a conventional source-finding algorithm can be successfully applied to the images is therefore dependent on the noise statistics. In some cases, such as for Rician statistics, source-finding can be applied as long as the algorithm is not pushed too deeply, however for other distributions typical source-finders may provide substantial numbers of false-positives. Histograms showing the noise in the central region of each $\mu$, $\sigma$, $\psi$, and $\kappa$ image are shown in Fig.~\ref{NoiseDists}. The $\mu$ moment images all appear to have approximately Gaussian noise in $Q$, $U$, and approximately Rician noise in $P$. The $\sigma$ images are all positive-definite, but still appear to have approximately Gaussian noise. Conventional source-finding can therefore be carried out on the $\mu$ and $\sigma$ moment images, although parameterisations of the s/n should only be measured using known normally-distributed noise. In this paper, we have performed all the source-finding using the \textsc{aegean} software \citep{2012MNRAS.422.1812H}. Although beyond the scope of this paper, the PyBDSF software \citep[the Python Blob Detector and SourceFinder, formerly PyBDSM;][]{2015ascl.soft02007M}, the Transients Project source extraction and measurement code \citep[PySE;][]{2014ascl.soft12011T}, and other source-finding packages \citep[e.g.][]{2015PASA...32...37H} are all viable alternatives.

\subsection{Using Source-Finding Alternatives on Skew and Excess Kurtosis Images}
\label{alternatives}
The skew, $\psi$, images have \emph{approximately} Gaussian noise, as shown in Fig.~\ref{NoiseDists}. This may lead one to believe that it is possible to apply a conventional source-finder. However, the situation is more complicated: a closer look reveals that the noise distribution of the $\psi$ images appears fat-tailed relative to the $\mu$ and $\sigma$ images. An assumption of normality is therefore not sufficient. Indeed, in the $\psi_{Q}$ and $\psi_{U}$ images the noise pixels all tend to have skew, while the majority of Faraday-rotating sources tend to zero-skew (with the notable exception of a source with a $\sim1.5$~radian rotation in polarization angle across the observing band). Only the $\psi_{P}$ image is therefore really useful, as sources appear as dips in an otherwise positive image (as the skew of the Rician distribution is positively valued). Furthermore, these dips are not Gaussian, but rather closer to a smoothed top-hat function. Typical source-finding algorithms are not well suited to handling this situation, even if one attempted to search for negative sources or to flip the sign of the images. In addition to this, while our simulations do not include any depolarizing sources, we do include the effect of a spectral index. Such sources also have $\psi_{P}$ skew, which in the case of depolarization can be both positive and negative (and therefore in some cases with similar skew to the noise pixels). This mix of complexities suggests that skew is not well suited to finding depolarizing sources, and that other moment images should be used for this purpose. However, skew is still a useful quantity for locating Faraday rotating sources, only without the use of a typical source-finder.

The noise statistics in the $\kappa$ images are strongly non-Gaussian, as shown in Fig.~\ref{NoiseDists}. These kurtosis images also show Faraday rotating sources appearing as dips in an otherwise positive image. This combination of factors again suggests that, similarly to the skew images, the kurtosis images cannot be used for source-finding with an off-the-shelf algorithm. This challenge is accentuated as the difference between the skew and kurtosis of a polarized pixel versus a noise pixel is small. We have attempted to search the skew and kurtosis images using the conventional \textsc{aegean} source-finder \citep{2012MNRAS.422.1812H}, and find that this is both a very unreliable and incomplete method -- generating a large number of false positives, with very few of the real sources being detected.

As an alternative to applying conventional source-finding algorithms, we have implemented another technique to determine if the skew and kurtosis images indicate the presence of a polarized source. In addition to calculating the moments, it is also possible to apply a statistical test of the null hypothesis that the skewness or excess kurtosis of the population from which the sample was drawn is that of the normal distribution. For tests of the skewness, further details are provided in \citet{jarquebera} and particularly \citet{dagostino}, while for tests of the kurtosis, further details are provided in \citet{anscombeglynn}. In practice, these tests were carried out using functions available in the \textsc{scipy} package. These tests provide a two-sided $p$-value, and allow the user to define a threshold at which they accept the skewness or kurtosis as non-normal. In combination with conventional source-finding on the $Q$, $U$, and $P$ images of the mean and standard deviation, this allows for selection of each pixel in which there is non-Gaussian skew or excess kurtosis. Note that although the distribution of $P$ is always non-normal, the magnitude of the $p$-value varies based upon the degree of non-normality at a given pixel, and the deviation from normality is greater for real sources with measurable Faraday Moments than it is for noise alone.

\begin{figure*}
   \centering
      \includegraphics[trim=0.0cm 0.0cm 0.0cm 0.0cm,clip=false,angle=0,origin=c,width=5.8cm]{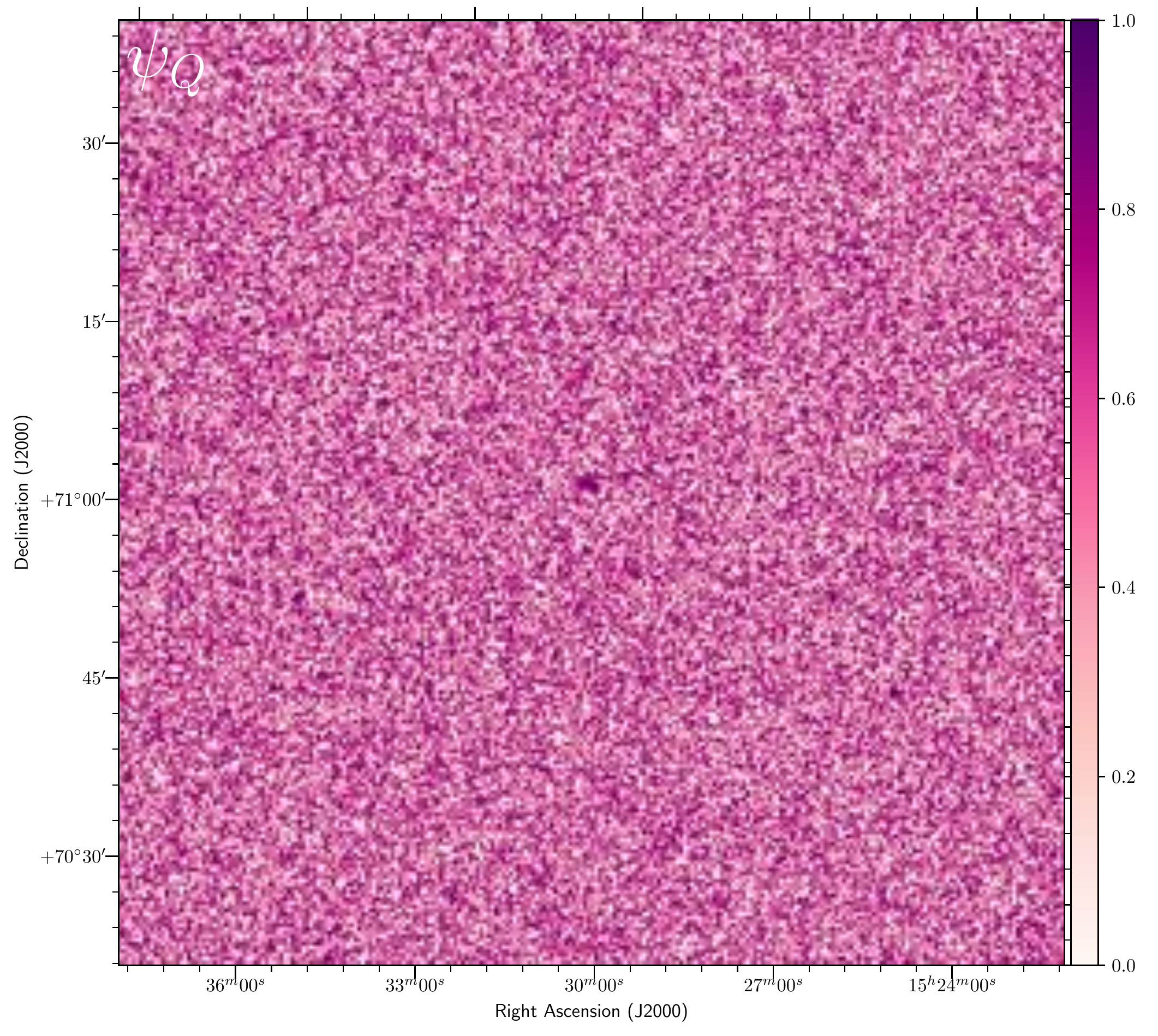}
   \includegraphics[trim=0.0cm 0.0cm 0.0cm 0.0cm,clip=false,angle=0,origin=c,width=5.8cm]{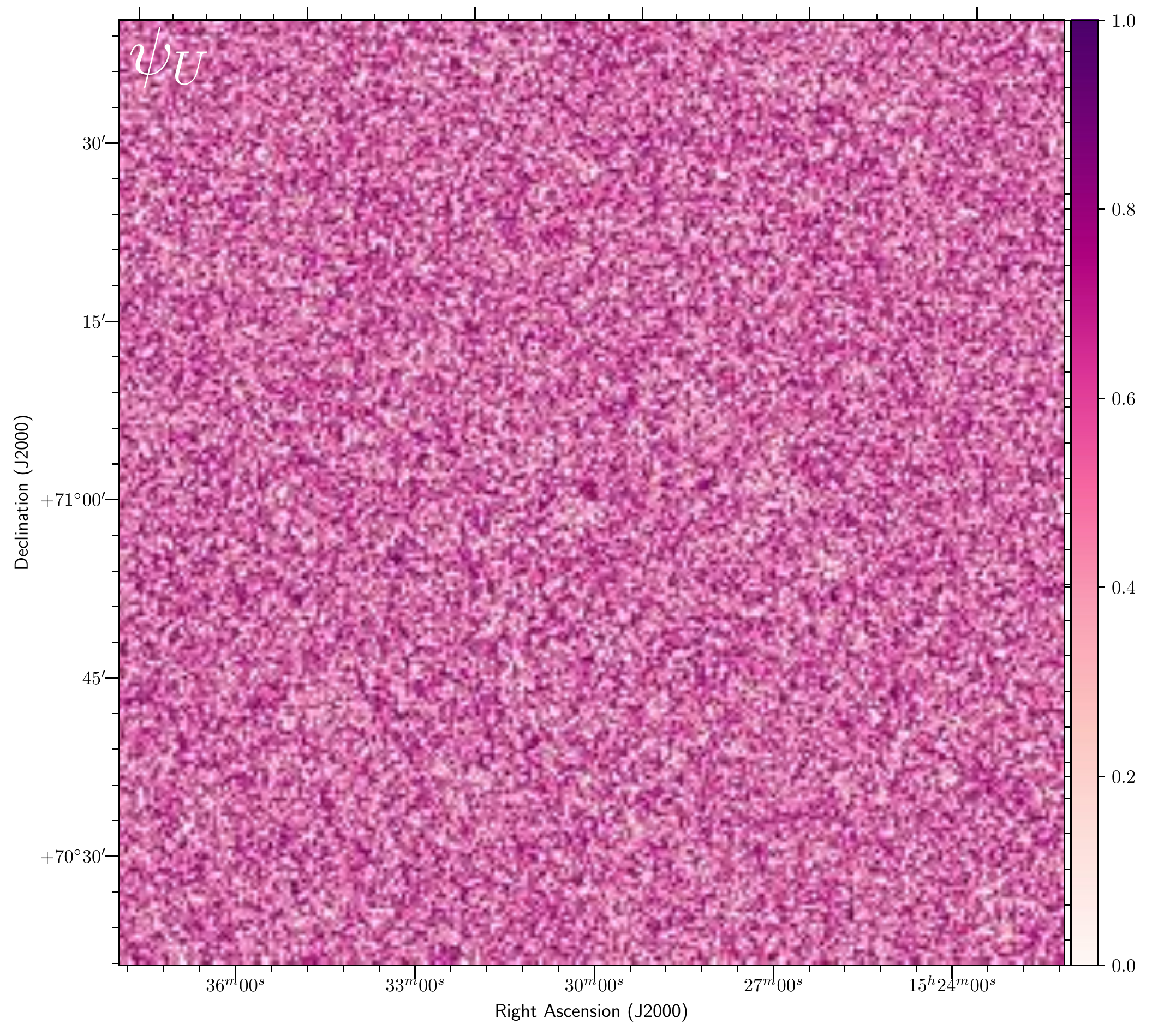}
   \includegraphics[trim=0.0cm 0.0cm 0.0cm 0.0cm,clip=false,angle=0,origin=c,width=5.8cm]{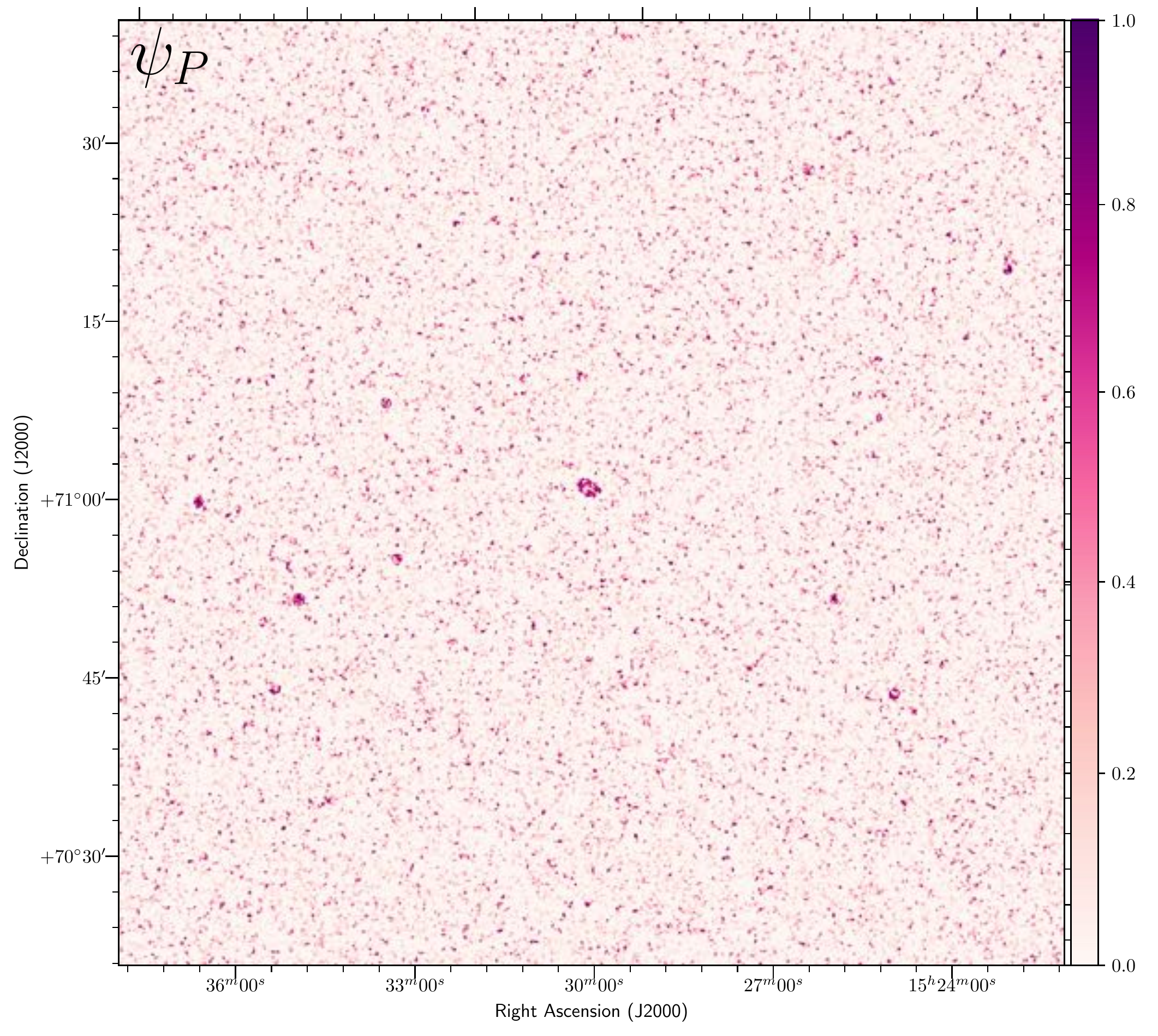}\\
      \includegraphics[trim=0.0cm 0.0cm 0.0cm 0.0cm,clip=false,angle=0,origin=c,width=5.8cm]{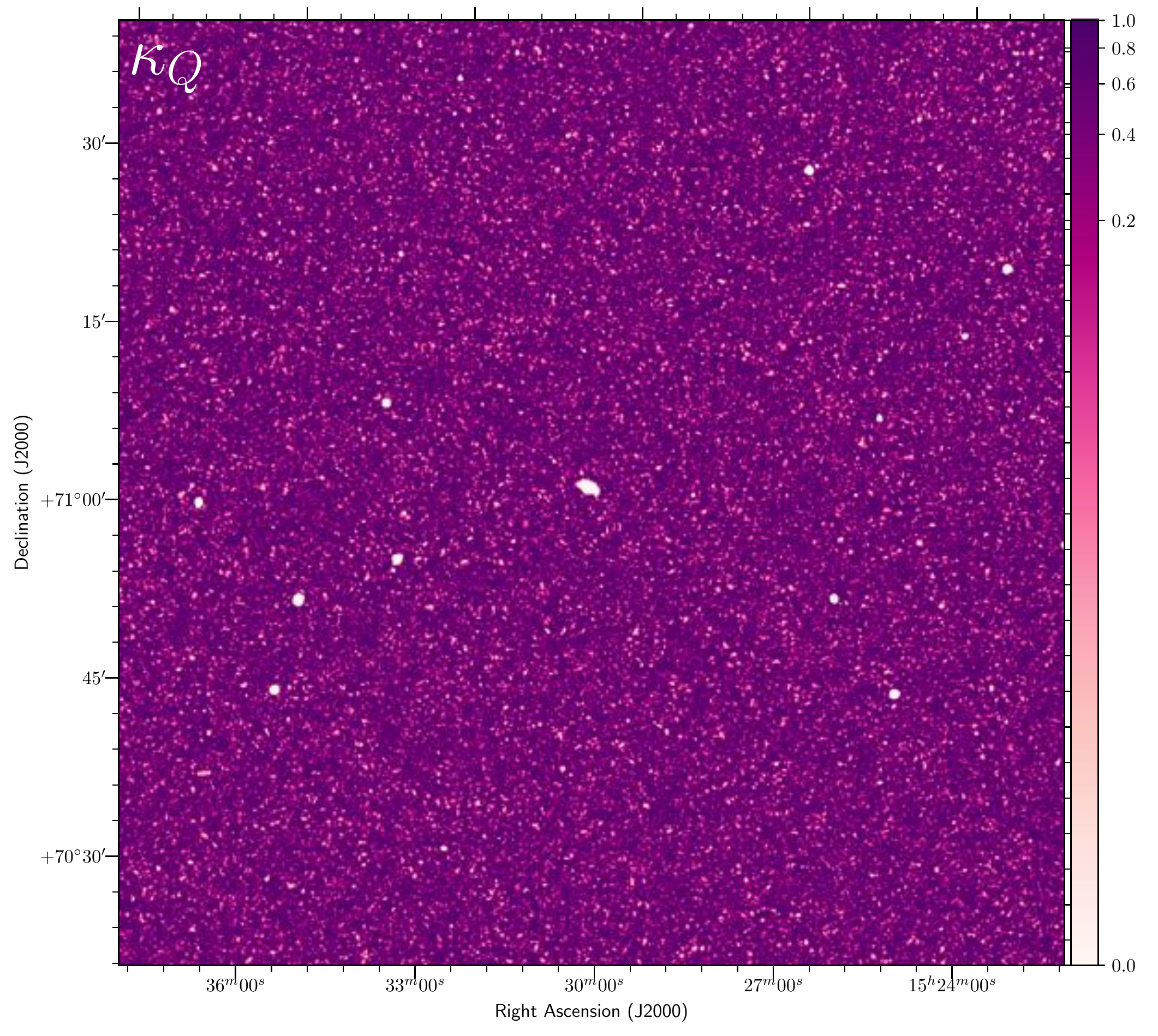}
   \includegraphics[trim=0.0cm 0.0cm 0.0cm 0.0cm,clip=false,angle=0,origin=c,width=5.8cm]{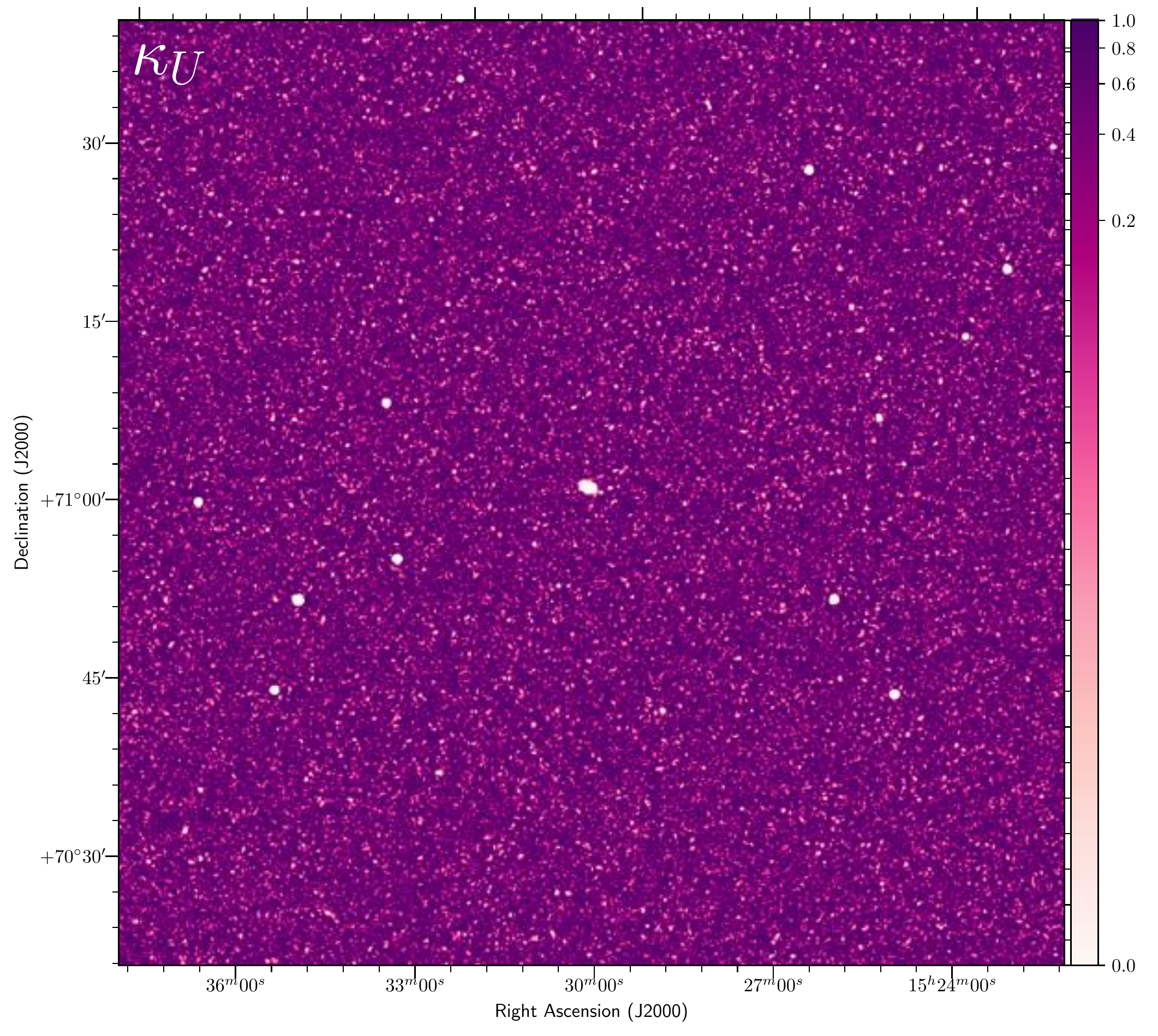}
   \includegraphics[trim=0.0cm 0.0cm 0.0cm 0.0cm,clip=false,angle=0,origin=c,width=5.8cm]{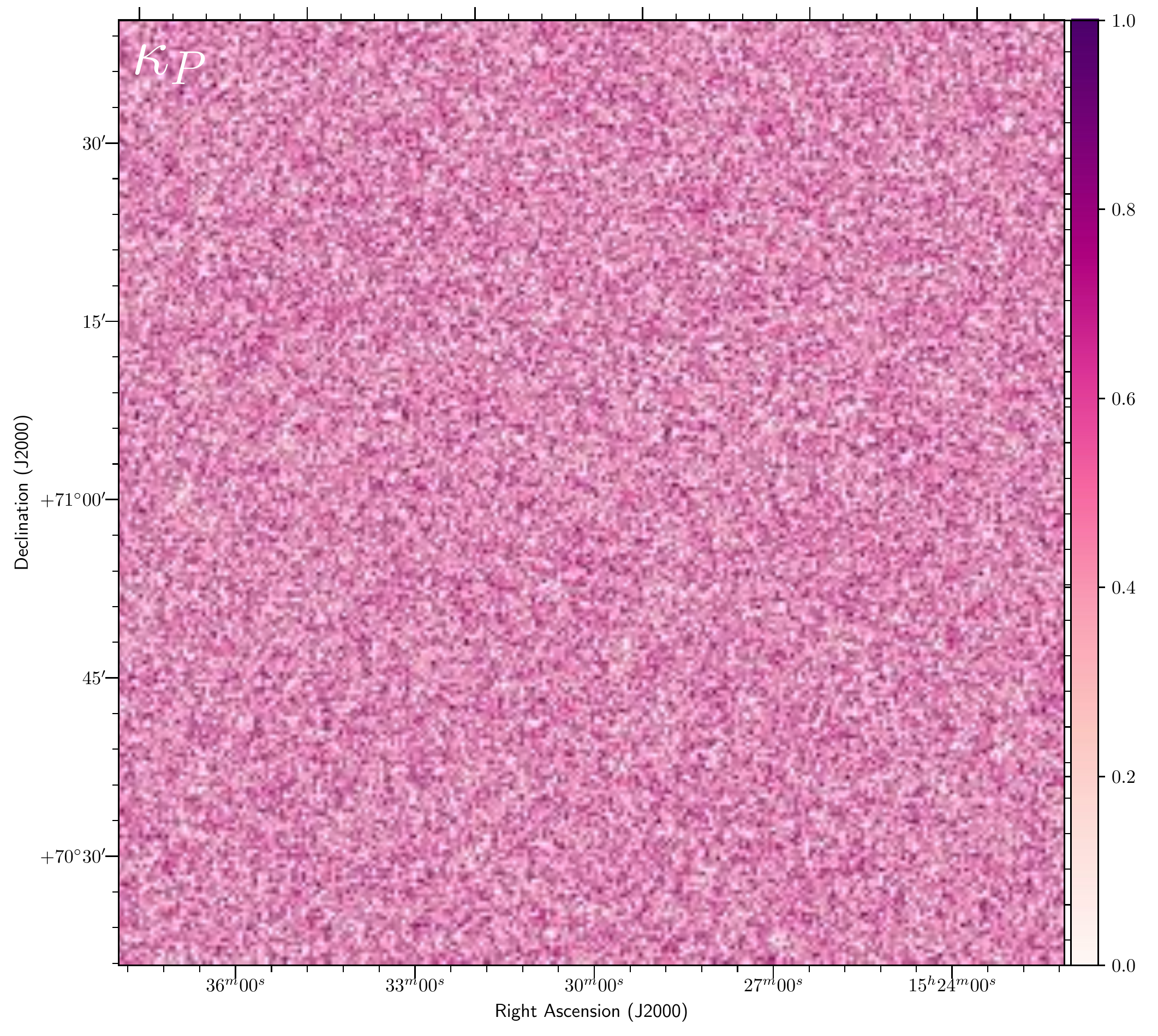}
   \caption{Similar simulated images as those shown in Fig.~\ref{momentimagessmall}, but showing the outputted $p$-values from the skewness (top row) and excess kurtosis (bottom row) tests. Images are shown for the $Q$ (left column), $U$ (middle column), and $P$ (right column) data. The excess kurtosis $p$-value images are particularly useful indicators of the presence of polarized sources.}
              \label{pvalueimagessmall}%
    \end{figure*}

The $p$-values provided for the simulated SEDs shown in Figs.~\ref{SEDs} and \ref{Dists}, are stated in Table~\ref{table2}. The $\psi$ skewness measurements are again not strong indicators of our simulated sources, with the property that noise pixels have low $\psi_{P}$ $p$-values of the order $p\approx10^{-15}$. However, the $\kappa$ measurements in $Q$ and $U$ are especially useful, with low $p$-values for cases with low and high-RMs (of the order $p\approx10^{-4}$), and for depolarizing sources (of the order $p\approx10^{-10}$). These properties can be seen in the $p$-value images, which are shown in Fig.~\ref{pvalueimagessmall}. Based on our simulations, we therefore recommend an excess kurtosis $p$-value cut-off of $\le0.001$ in order to ensure reliability. However, this parameter may vary under other observational circumstances. In this way, the $p$-value images can be used to identify pixels in which the excess kurtosis is not believed to be that of the normal distribution. All pixels with a value meeting this cut-off could in principle be listed as a source candidate. It may be possible to use these statistics, or similar alternatives, in order to isolate sources in the skew and kurtosis images from the noise. However, we will later show (see the caveats in Sections~\ref{realdata} and~\ref{prescription}) that in real data, the skew and kurtosis do not appear to provide any extra benefit to what is possible using the lower-order moments. In the future, it may be possible to use the D'Agostino--Pearson $K^2$ test or a similar test, which combines the skew and kurtosis statistics together in order to test for departures from normality.

\begin{table}
      \caption[]{The $p$-values for the skewness and excess kurtosis of the Simulated SEDs. As the moments for cases (iii), (iv), and (v) are all similar to one another, only cases (i), (ii), (iii), and (vi) are shown. As these are $p$-values, there are no uncertainties. The $p$-values for each case, should be contrasted against case (vi) for the noise, which fills the majority of the image pixels.}
         \label{table2}
     $$ 
         \begin{array}{crrrr}
            \hline
            \noalign{\smallskip}
           \textrm{}      &  \textrm{Case} \\
           \textrm{Parameter}      &  \textrm{(i)} &  \textrm{(ii)}&  \textrm{(iii)}&  \textrm{(vi)}\\
            \noalign{\smallskip}
            \hline
            \noalign{\smallskip}
            \psi_{Q}         & 0.19107 & 0.52982 & 0.67684 & 0.27792    \\[2pt]
            \kappa_{Q}         & 0.87519 & 0.00000 & 1.3377\times10^{-10} & 0.87873    \\[2pt]
            \hline
            \psi_{U}         & 0.11093 & 0.34314 & 0.65713 & 0.74846    \\[2pt]
            \kappa_{U}         & 0.00044 & 0.00000 & 4.2606\times10^{-13} & 0.45144    \\[2pt]
            \hline
            \psi_{P}         & 0.27408 & 0.10176 & 0.01167 & 5.2754\times10^{-15}    \\[2pt]
            \kappa_{P}         & 0.69830 & 0.67808 & 3.5049\times10^{-159} & 0.03779    \\[2pt]
            \noalign{\smallskip}
            \hline
         \end{array}
     $$ 
   \end{table}

\section{Application to real LOFAR Data}\label{realdata}
Real data can differ substantially from simulated data. Artefacts resulting from incomplete $uv$-coverage, instrumental polarization leakage, and diffuse polarized emission from the Galactic foreground -- all of which are present in both LOFAR data, and radio data more generally -- could affect the veracity of our method.

\begin{figure*}
   \centering
   \includegraphics[trim=0.0cm 0.0cm 0.0cm 0.0cm,clip=false,angle=0,origin=c,width=5.8cm]{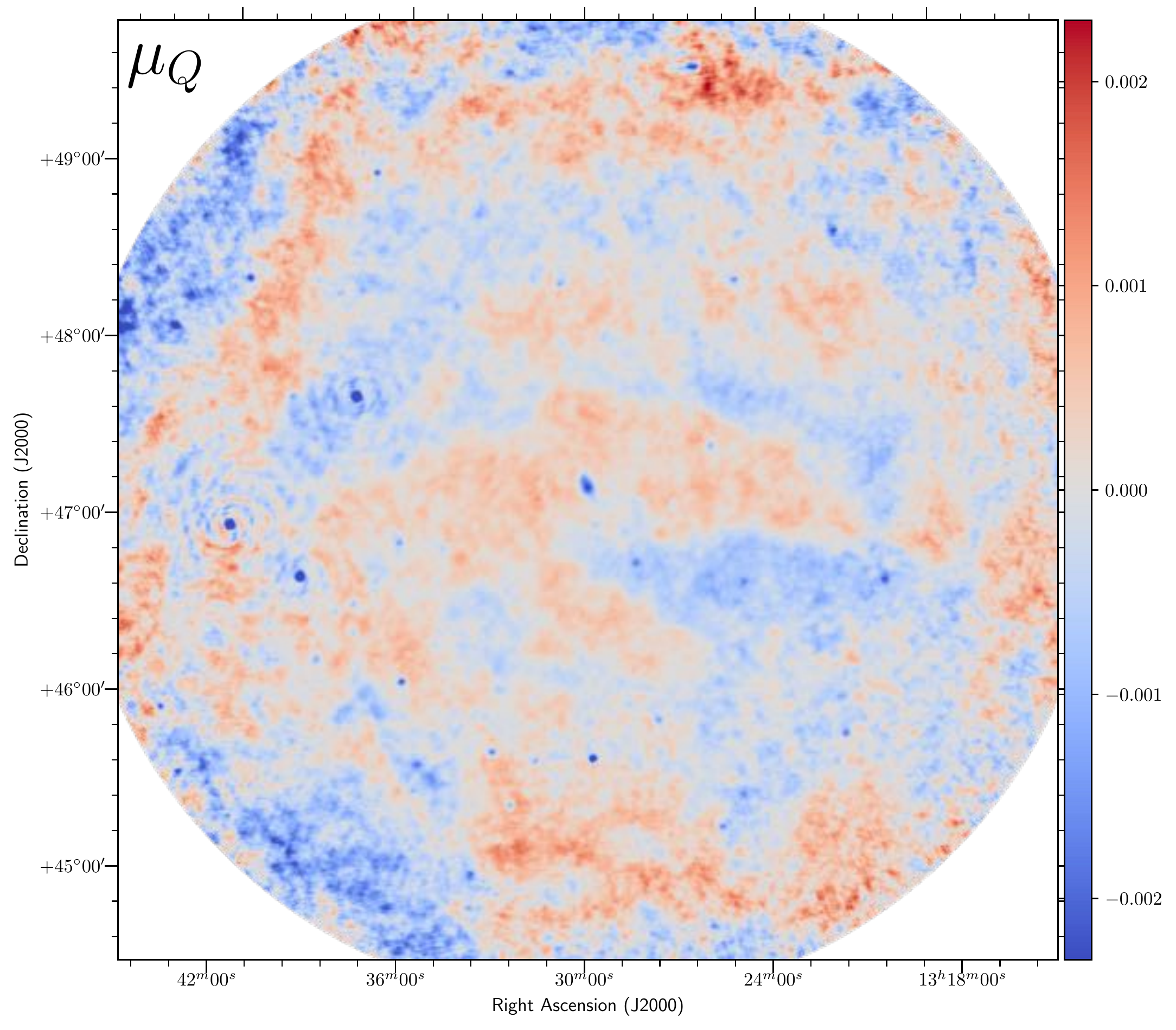}
   \includegraphics[trim=0.0cm 0.0cm 0.0cm 0.0cm,clip=false,angle=0,origin=c,width=5.8cm]{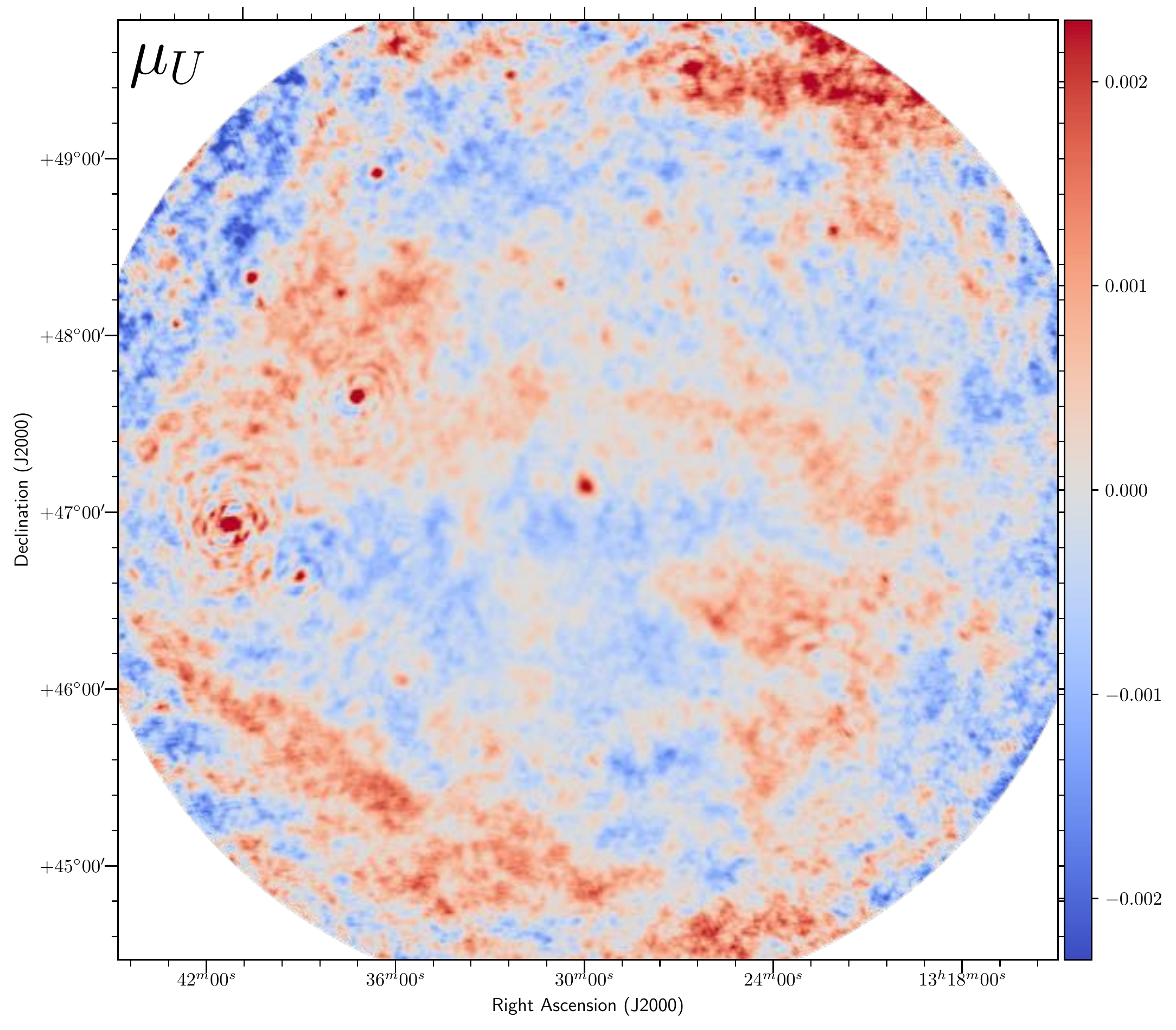}
   \includegraphics[trim=0.0cm 0.0cm 0.0cm 0.0cm,clip=false,angle=0,origin=c,width=5.8cm]{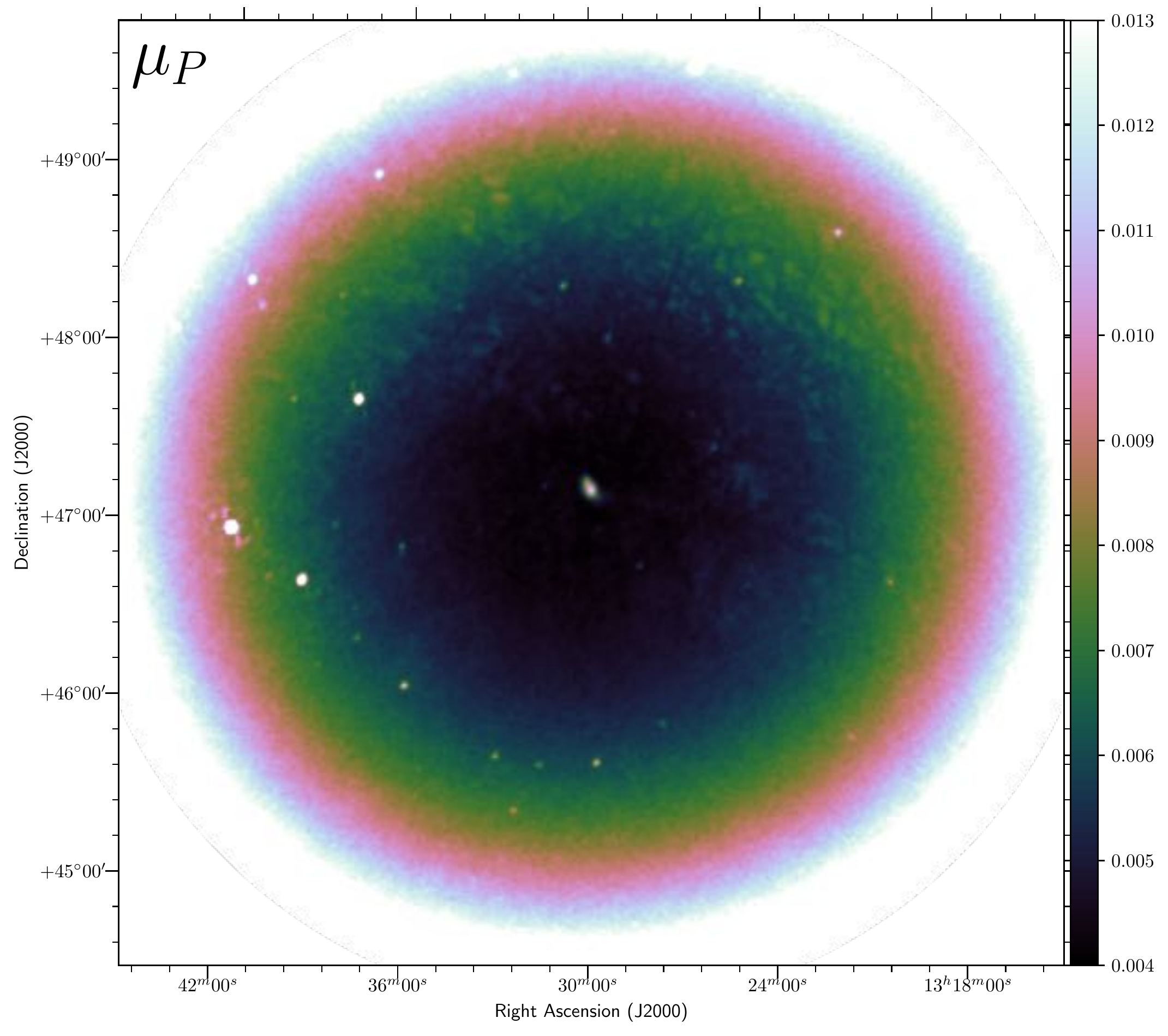}\\
      \includegraphics[trim=0.0cm 0.0cm 0.0cm 0.0cm,clip=false,angle=0,origin=c,width=5.8cm]{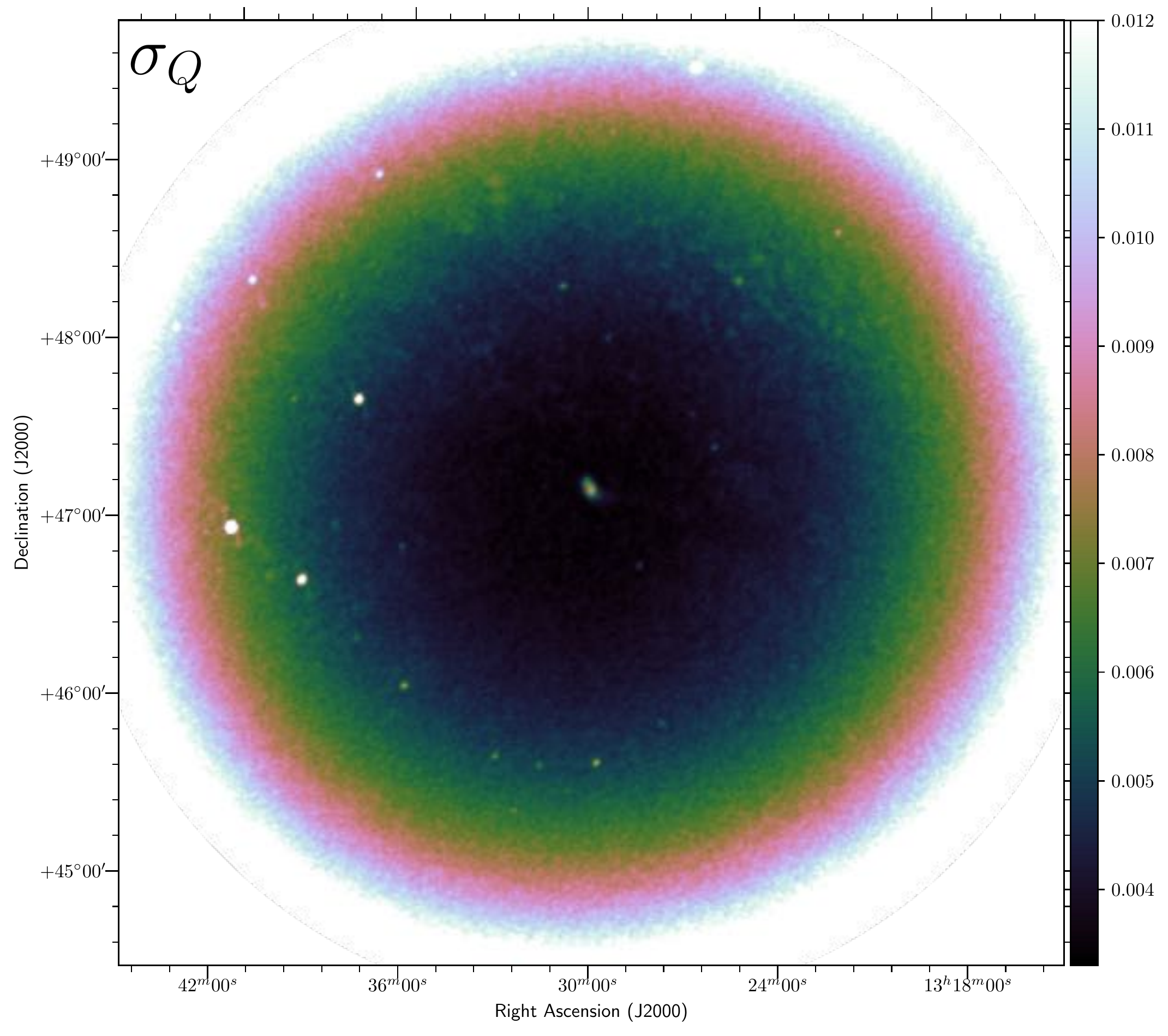}
   \includegraphics[trim=0.0cm 0.0cm 0.0cm 0.0cm,clip=false,angle=0,origin=c,width=5.8cm]{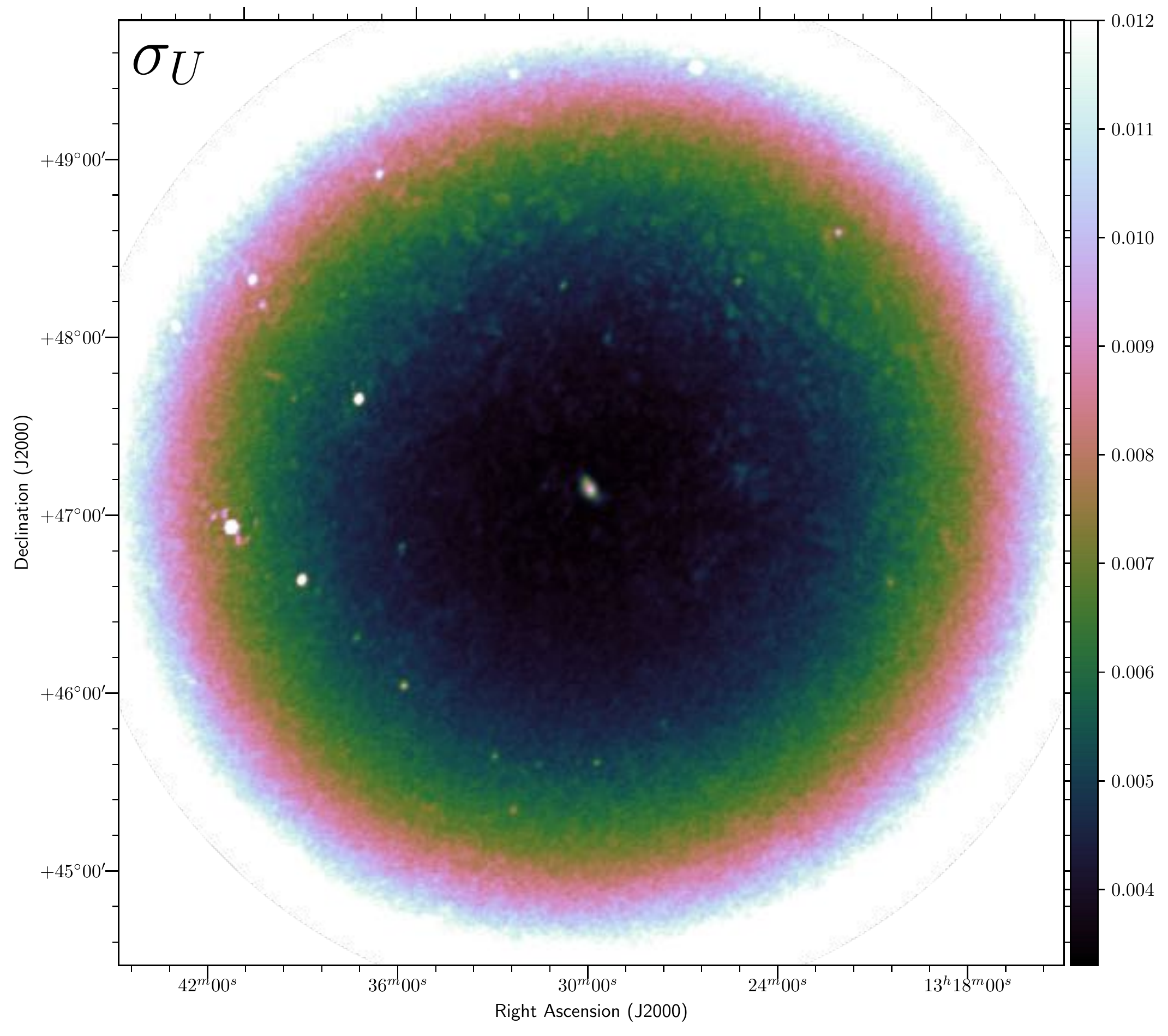}
   \includegraphics[trim=0.0cm 0.0cm 0.0cm 0.0cm,clip=false,angle=0,origin=c,width=5.8cm]{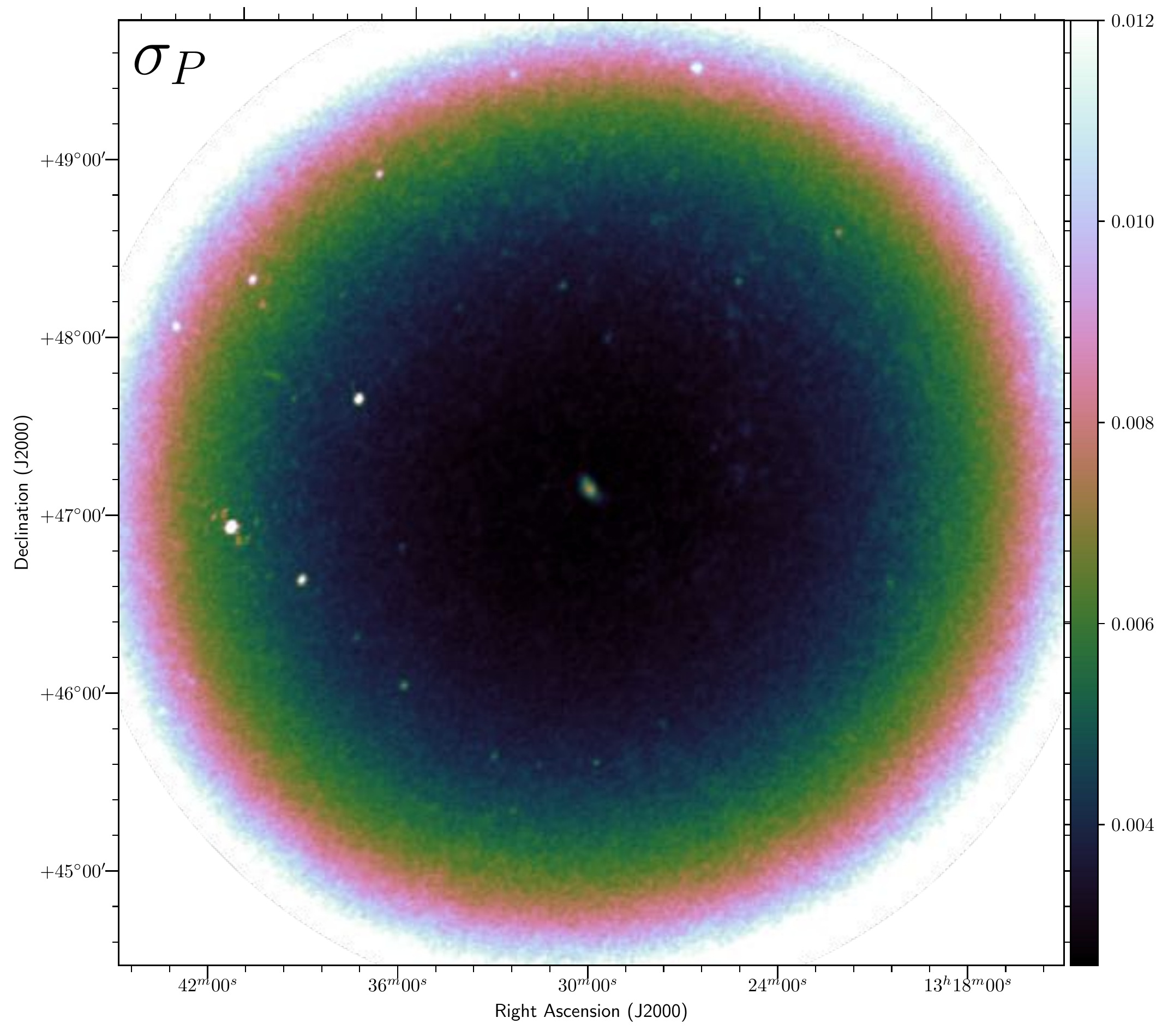}\\
      \includegraphics[trim=0.0cm 0.0cm 0.0cm 0.0cm,clip=false,angle=0,origin=c,width=5.8cm]{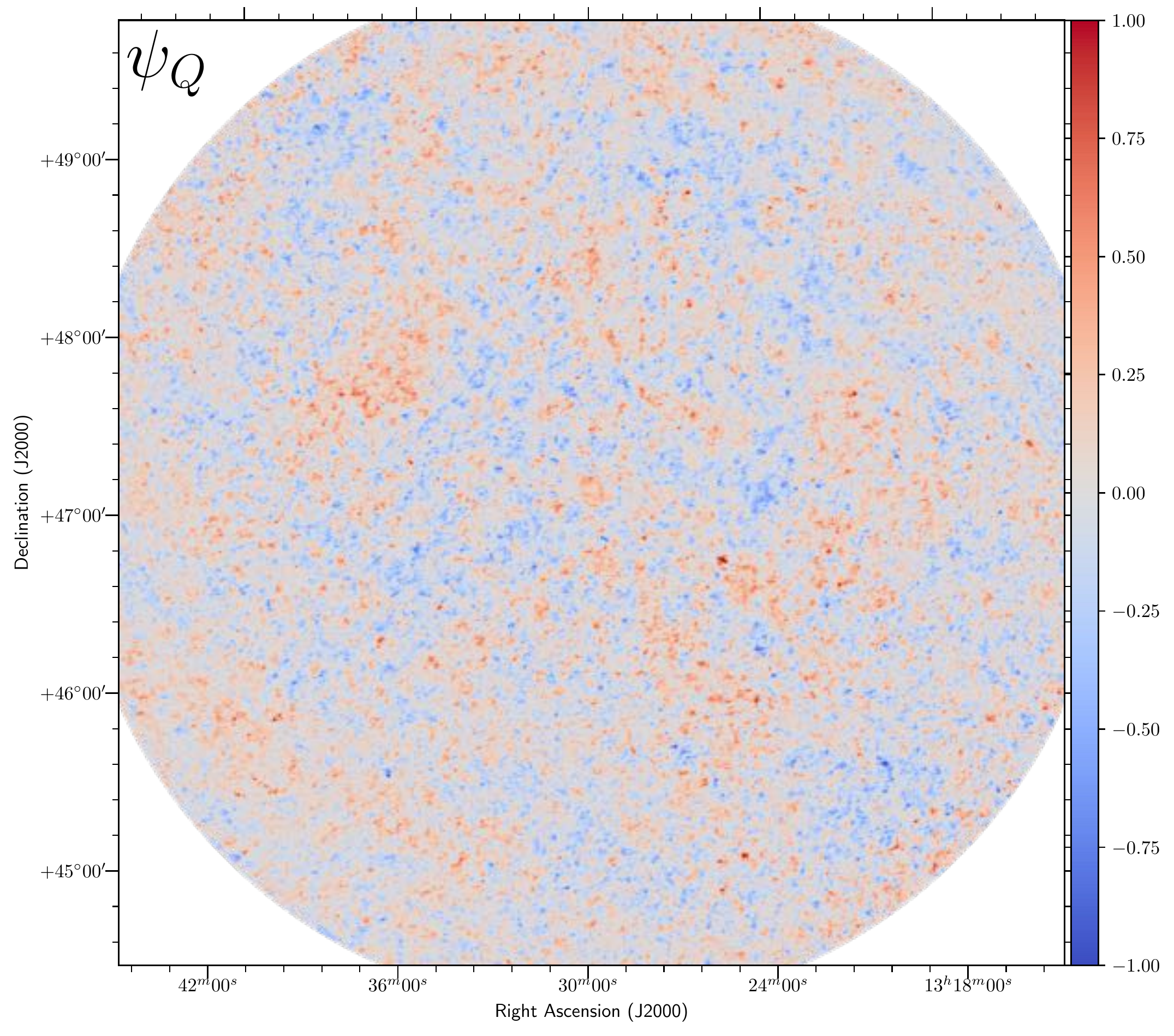}
   \includegraphics[trim=0.0cm 0.0cm 0.0cm 0.0cm,clip=false,angle=0,origin=c,width=5.8cm]{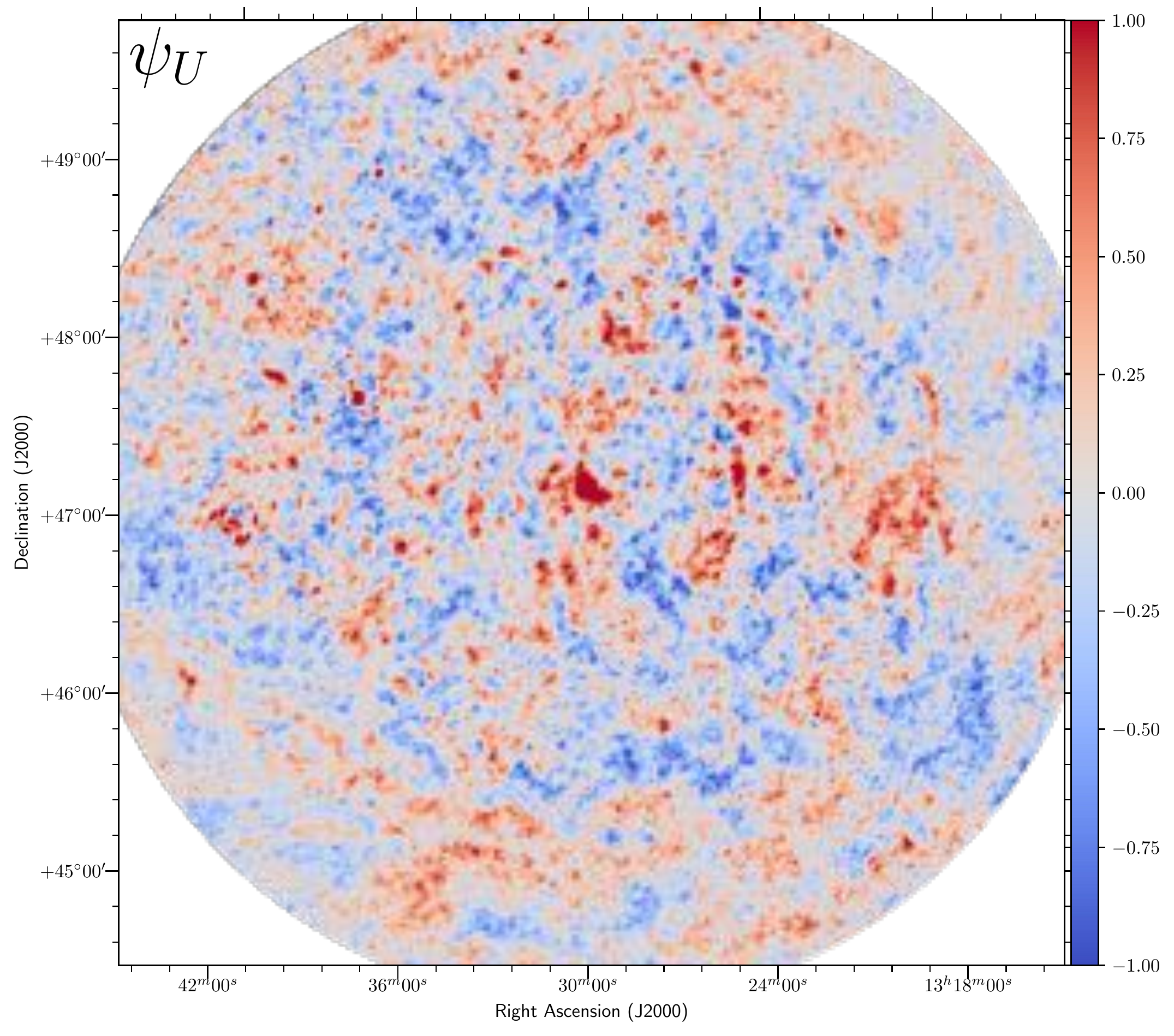}
   \includegraphics[trim=0.0cm 0.0cm 0.0cm 0.0cm,clip=false,angle=0,origin=c,width=5.8cm]{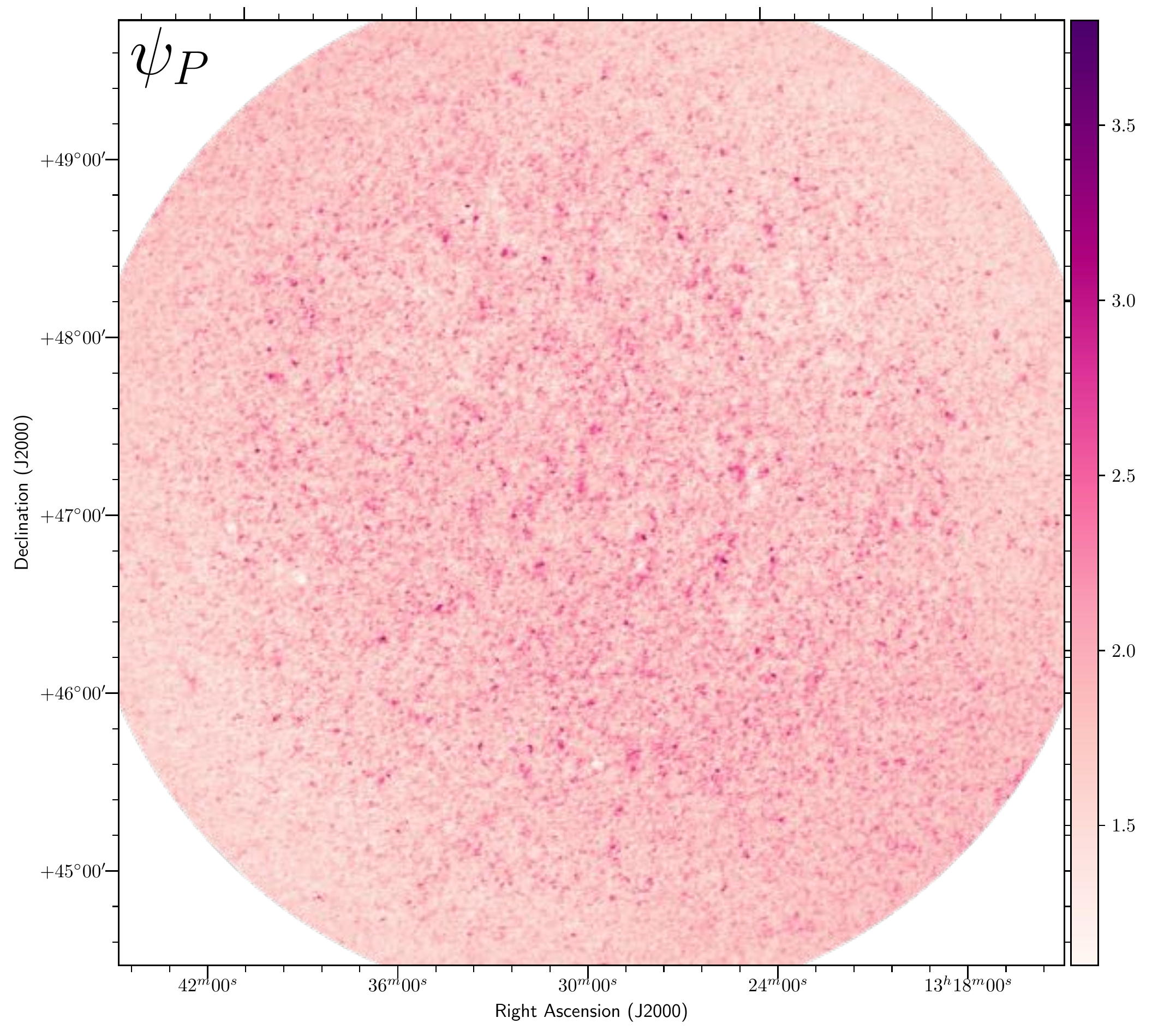}\\
      \includegraphics[trim=0.0cm 0.0cm 0.0cm 0.0cm,clip=false,angle=0,origin=c,width=5.8cm]{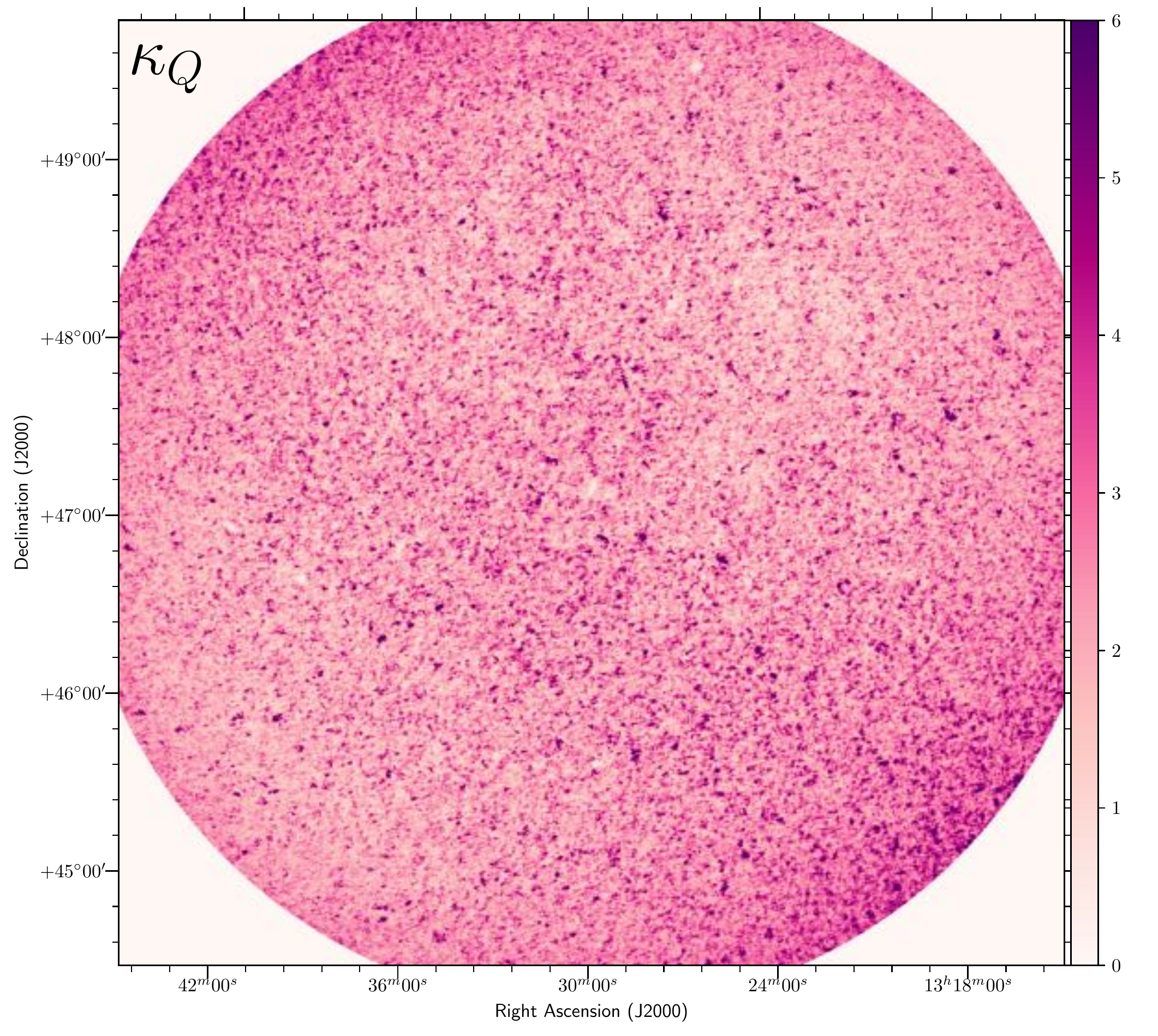}
   \includegraphics[trim=0.0cm 0.0cm 0.0cm 0.0cm,clip=false,angle=0,origin=c,width=5.8cm]{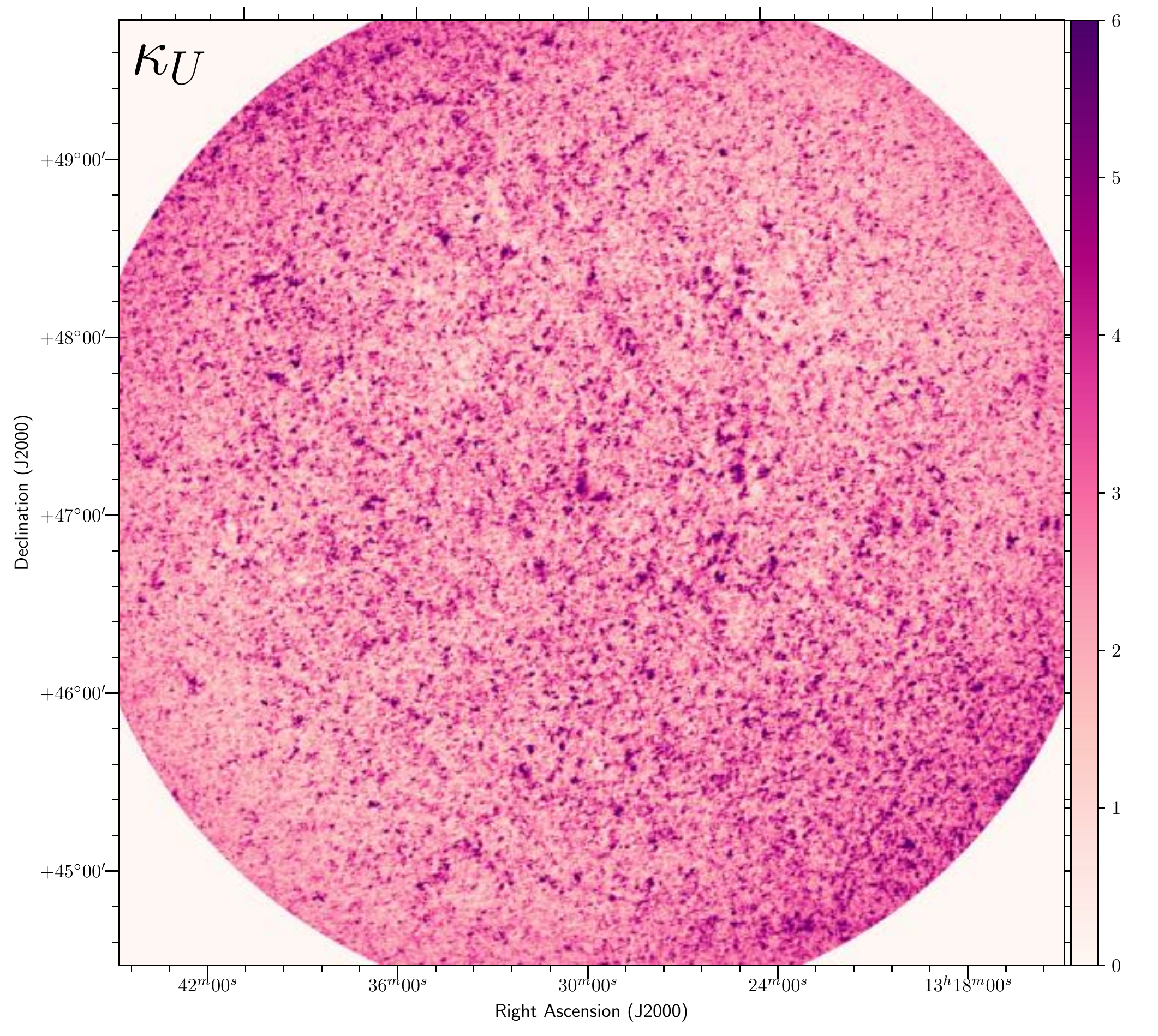}
   \includegraphics[trim=0.0cm 0.0cm 0.0cm 0.0cm,clip=false,angle=0,origin=c,width=5.8cm]{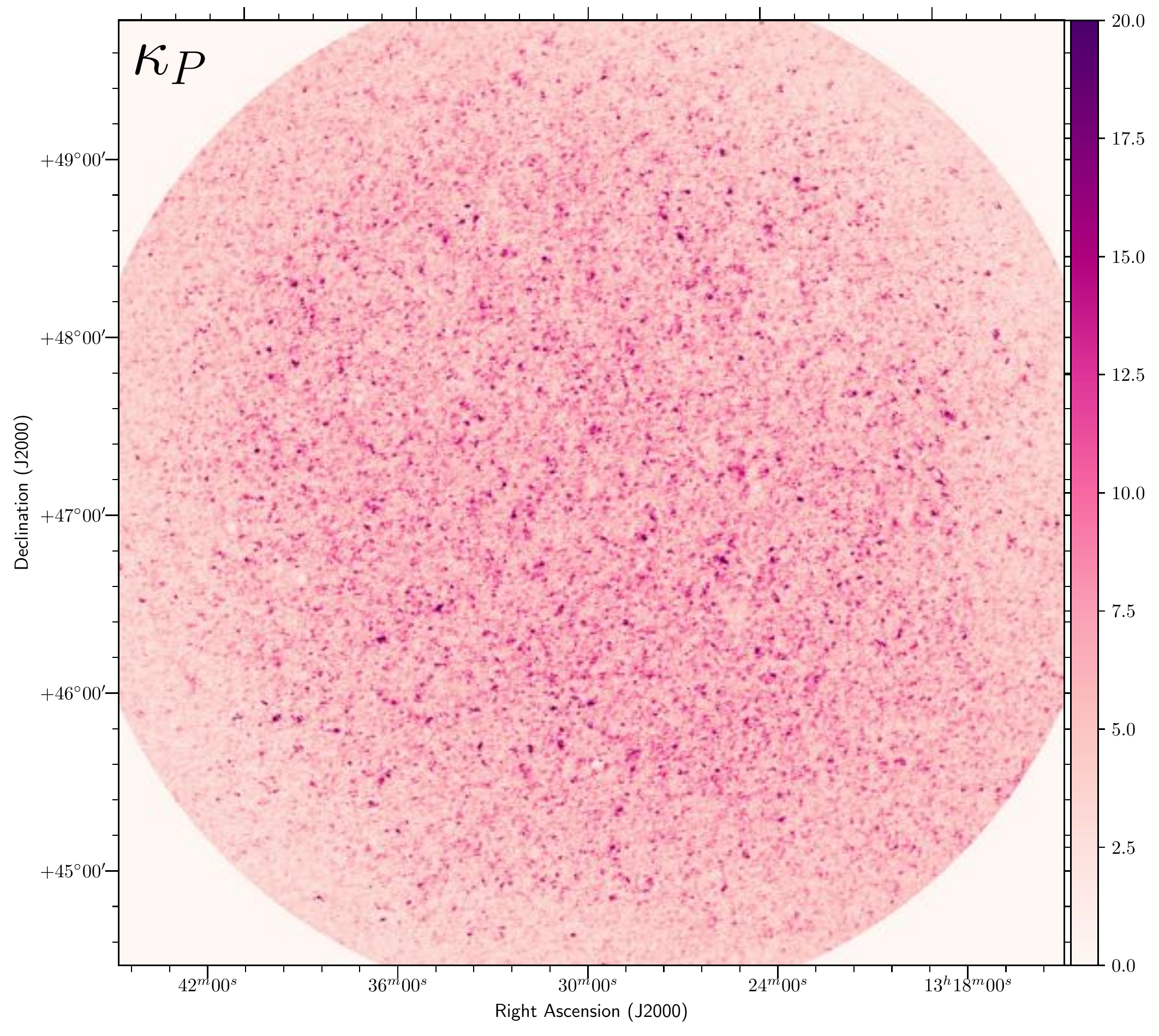}
   \caption{The moment images, as derived from the M51 datacubes presented in \citet{2014A&A...568A..74M}. The data here have been reimaged to 2~arcmin resolution. The moment images for the mean $\mu$ (top row), standard deviation $\sigma$ (2nd row), skewness $\psi$ (3rd row), and excess kurtosis $\kappa$ (bottom row), are all shown. The images shown are derived from the $Q$ data (left column), $U$ data (middle column), and $P$ data (right column). The source seen at the centre of the field in some of the moment images is due to polarization leakage from the M51 galaxy. The pseudo-colour scales are chosen to provide contrast to each specific moment image.}
              \label{M51momentimages}%
    \end{figure*}

To ensure this is not the case, we have applied the Faraday Moments technique to real LOFAR data, the results of which have been presented elsewhere \citep{2014A&A...568A..74M}, and which were taken towards the nearby galaxy M51. The data have been reimaged at 2~arcmin resolution in order to optimise sensitivity to diffuse Galactic polarized emission in the field. For a full analysis of the identified emission and sources in this field, please see the scientific study of these data in \citet{2014A&A...568A..74M}. For the investigation of Faraday Moments using these data, the moment images were produced using the same method detailed in Section~\ref{calcmoments}. The resulting Faraday Moments are shown in Fig.~\ref{M51momentimages}.

There are several key ways in which the moments are similar to, and differ from, the results of the simulations. Very many sources can be seen in the moment images, compared to the six sources reported by the careful analysis in \citet{2014A&A...568A..74M}. This is the result of instrumental polarization leakage, which leads to unpolarized sources that are bright in total intensity `leaking' into Stokes $Q$ and $U$. Furthermore, there is also clearly a diffuse background across the field of view surrounding M51. This diffuse background is likely partially diffuse Galactic polarized emission, as is frequently observed with LOFAR \citep[e.g.][]{2013A&A...558A..72I,2014A&A...568A.101J,2017A&A...597A..98V}, and also partially the result of polarization leakage. For example, in our simulations, the $Q$ and $U$ skew images did not work well for detecting polarized sources. For these real data, many sources are visible in the $Q$ and $U$ skew images, which are displayed on the same scales, and show a larger skew in Stokes $U$. This skew is also associated with M51 itself, which \citet{2013arXiv1309.4646F} and \citet{2014A&A...568A..74M} showed to be unpolarized at low radio frequencies. The increased skew is therefore most likely the result of polarization leakage. This is unusual, as leakage typically manifests at an RM$=0$~rad~m$^{-2}$, which would suggest that the skewness should be that of the noise (i.e.\ zero skew). However, in this case, an RM correction was applied to the data in order to correct for ionospheric Faraday rotation (see \citealt{2014A&A...568A..74M} for further details). This shifts the leakage by approximately 1 to 3~rad~m$^{-2}$, and consequently can be expected to affect the measured skew.

Similarly, from our simulations we defined an alternative source-finding method for identifying sources in the skew and excess kurtosis images (see Section~\ref{alternatives}). In particular, we defined an excess kurtosis $p$-value cut-off of $\le0.001$ in order to ensure reliability. In the real data, no pixels have a $p$-value at this level. The lowest $p$-values reach $\approx0.006$ in the $Q$ images, for the leakage from M51 itself and from two sources detected to the north and south east of the field-of-view. This demonstrates that leakage and other artefacts have a strong effect on the real LOFAR data, and that this inhibits the use of these higher moments. These effects are not expected to have a strong dependence on the flux density of the sources, indeed the brightest sources are probably the most affected, as these sources are more likely to have associated image artefacts and leakage that is substantially above the noise. We therefore do not currently recommend the use of $p$-value cut-offs in order to improve the source-finding in LOFAR data. In the future, and with further instrumental and algorithmic development, these higher moments will eventually help to provide increased completeness to even lower s/n ratios, as improved calibration models and techniques become available. We do not discourage use of the skew and excess kurtosis moments altogether, but rather highlight that these moments do not appear to be a useful addition to our method given the present data constraints. These higher moments will have increased applicability to other SKA pathfinder and precursor instruments, particularly those with reduced leakage levels.

The mean and standard deviation moment images also show the presence of diffuse foregrounds, whether due to leakage or real polarization. In our simulations, we found that conventional source-finders such as \textsc{aegean} could be used on these moment images (see Section~\ref{conventional}). In the real data, we initially found that source-finding was inhibited due to these diffuse backgrounds. One approach to overcome this would be to incorporate the much higher resolutions available with LOFAR, which would in essence filter out the Galactic foreground. However, within \textsc{aegean} the associated Background And Noise Estimation software (\textsc{bane}) can be used to calculate the background emission across the image. By running \textsc{bane} prior to running \textsc{aegean}, we find that all of the polarized sources that are visible by eye in Fig.~\ref{M51momentimages} are detected. Crucially, this includes the six real polarized sources reported by \citet{2014A&A...568A..74M}. In future, it may even be possible to develop additional techniques that use the moments (rather than RM Synthesis) to separate these real sources from the leakage-dominated sources.

\begin{figure*}
   \centering
  \includegraphics[trim=0.5cm 0.0cm 0.666cm 0cm,clip=false,angle=-90,origin=c,width=0.666\hsize]{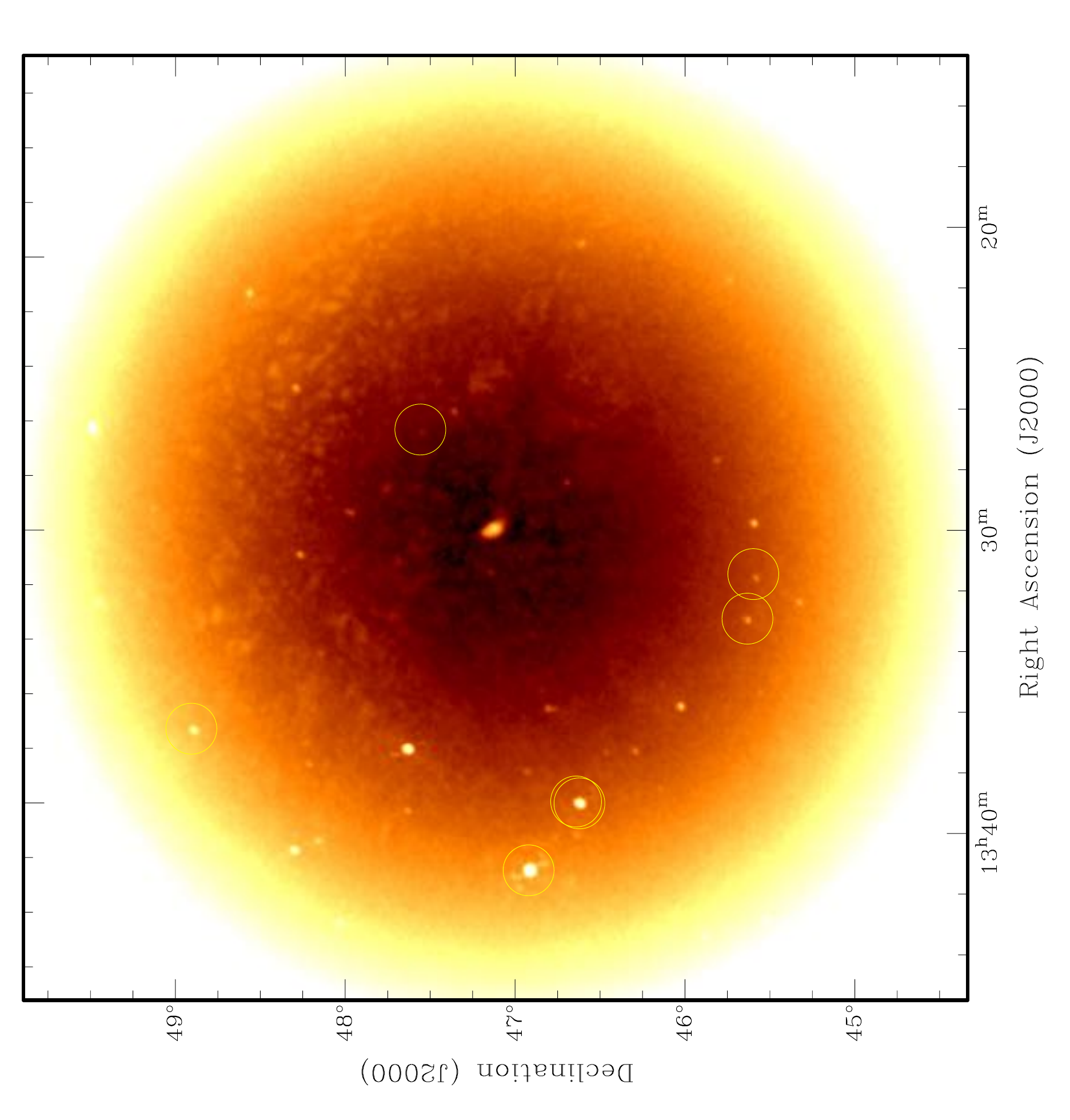}\\
   \includegraphics[trim=0.3cm 0.0cm 0.5cm 0cm,clip=true,angle=-90,origin=c,width=0.666\hsize]{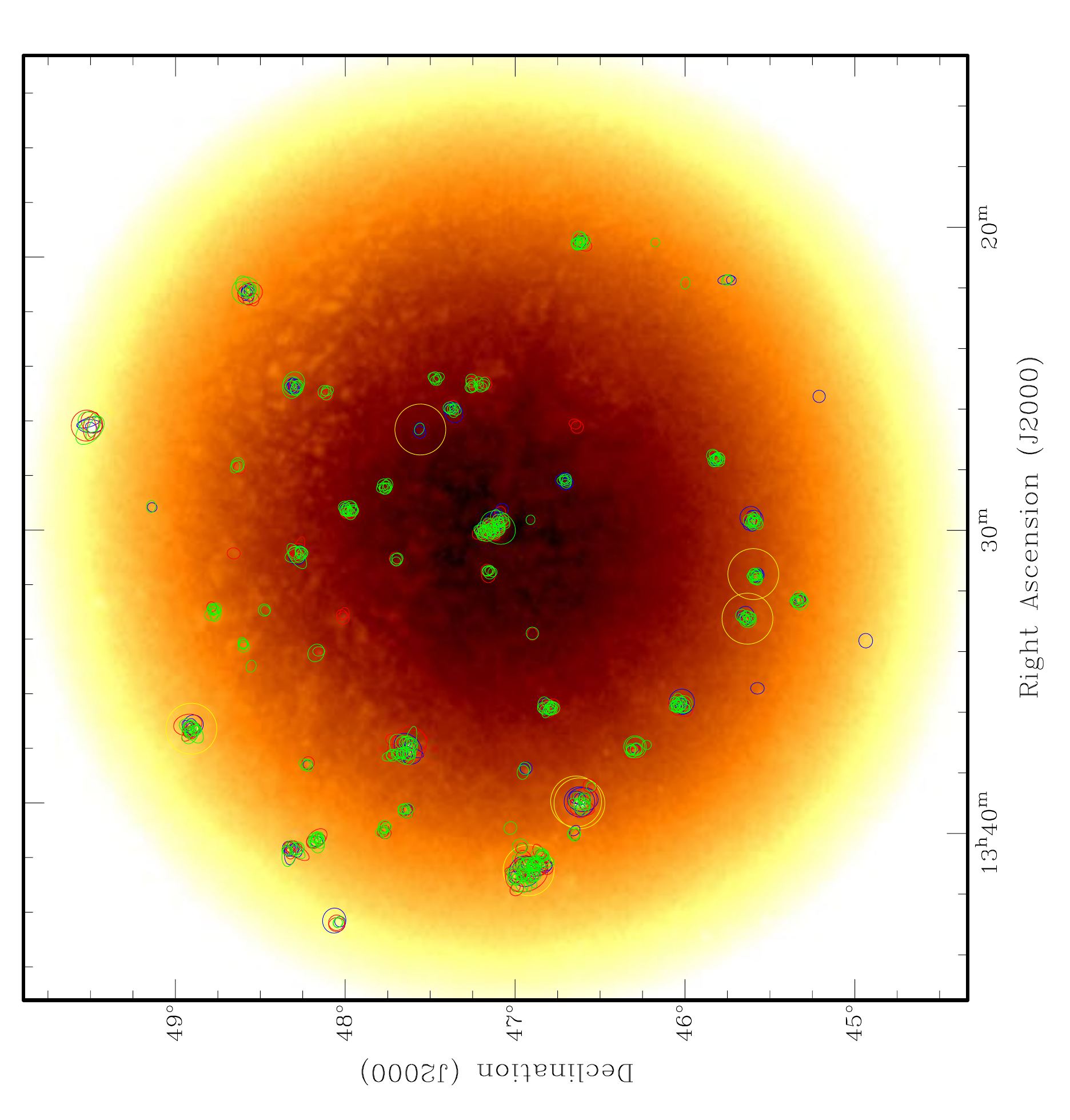}
   \caption{\textbf{Top:} The mean polarized intensity moment image, $\mu_{P}$, from the M51 observation presented in \citet{2014A&A...568A..74M}. The locations of the six polarized sources detected by \citet{2014A&A...568A..74M} using RM Synthesis are shown with yellow circles. \textbf{Bottom:} The same image as above, with annotations showing the sources found using Faraday Moments source-finding. Sources found using the $\mu$ or $\sigma$ of (i) $P$ are in green, of (ii) $Q$ are in blue, and (iii) $U$ are in red. All of the six previously identified sources are detected.}
              \label{M51-detections}%
    \end{figure*}


\section{A Full Formalism for Polarized Source-Finding}
\label{prescription}
\begin{figure}
   \centering
  \includegraphics[trim=0.0cm 0.0cm 0cm 0cm,clip=false,angle=0,origin=c,width=\hsize]{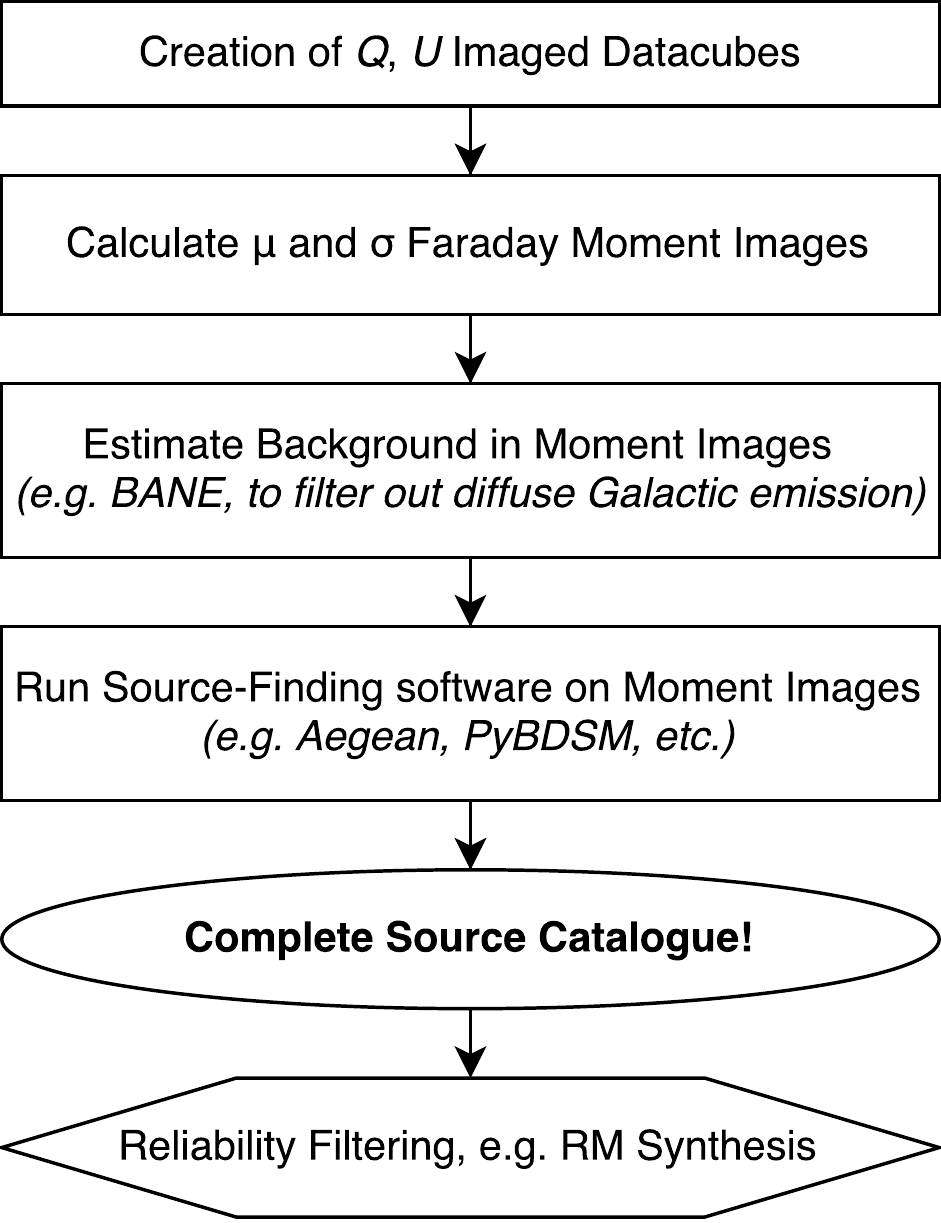}\\
   \caption{A logical flow-diagram, showing the typical work-flow for polarized source-finding. We recommend not using the skew and kurtosis moments. The source-finding can be run as deeply as one desires, given the personal appetite for false-detections. With \textsc{aegean}, we recommend a clipping value for seeding islands (`seedclip') of 5$\sigma$ in the majority of moment images, and slightly deeper to 3$\sigma$ in the $\mu_{Q,U}$ images. In our simulations, we find this provides the essential completeness, whilst minimising the number of false detections. Future stages may use RM Synthesis or alternative filtering techniques in order to convert the complete source catalogue into a complete and reliable catalogue. All of these stages are discussed further in the main text.} 
              \label{logicaldiagram}%
    \end{figure}

Following our investigations, we suggest a formalism for polarized source-finding using the technique of Faraday Moments. The overall logical flow of this procedure is shown in Fig.~\ref{logicaldiagram}. To minimise computational expense, we suggest that only the $\mu_{Q,U,P}$ images, and the $\sigma_{Q,U,P}$ images are calculated. On the basis of the data that we have considered in this paper, we do not recommend calculation of the skew, kurtosis, or the $p$-value images.

Given that we have limited the analysis to just the $\mu$ and $\sigma$ images, in these images, the background should be estimated using an appropriate software such as \textsc{bane} in order to remove the impact of diffuse polarization, whether Galactic emission or instrumental leakage, on the subsequent source-finding. Source peaks should then be found using a conventional source-finding software such as \textsc{aegean}, which can also simultaneously subtract the background emission identified by \textsc{bane}. In the $\mu_{Q,U}$ and $\sigma_{Q,U}$ images, the noise statistics are approximately Gaussian. However, in the $\mu_{P}$ and $\sigma_{P}$ images, the noise is more non-Gaussian, and the source-finding cannot be run as deeply without more false detections. Nevertheless, the source-finding can be run on each moment image, and the lists of identified source candidates can be combined. We recommend source-finding in $\mu_{Q,U}$ down to $3\times$, and in $\mu_{P}$ and $\sigma_{Q,U}$ down to $5\times$ the respective local noise-levels. We find that running $\sigma_{P}$ down to $5\times$ the local noise provides reasonable results, although $8\times$ may be a more conservative approach for some specific datasets.

The depth of source-finding can be tuned in order to ensure maximum completeness, while also allowing for a \emph{large number} of false detections. The key concept here is to provide a highly complete catalogue of polarized radio sources and to greatly reduce the number of pixels on which RM Synthesis needs to be performed. This number may be able to be reduced even further by analysing and comparing the moments themselves, and will be able to be explored in future works. Note that source-finding in Stokes $I$ alone does not suffice in order to achieve this, as it is known that at sub-arcminute resolution, the peak in $P$ is often offset from the peak in $I$, and this is one of the key motivations for developing this alternative source-finding technique (see Section~\ref{intro}). As a secondary step, RM Synthesis can be used on the source candidates to identify any false detections (e.g.\ via a s/n cut-off on the Faraday spectra, such as the $8\sigma$ cut-off proposed by \citealt{2012PASA...29..214G}) and to identify pseudo-sources that originate due to instrumental leakage (e.g.\ by finding peaks in the Faraday depth spectra located at 0.0~rad~m$^{-2}$). The result will be a highly complete and reliable list of polarized sources, that uses a minimal amount of computational resources.

\begin{figure}
   \centering
  \includegraphics[trim=0.0cm 0.0cm 0cm 0cm,clip=false,angle=0,origin=c,width=\hsize]{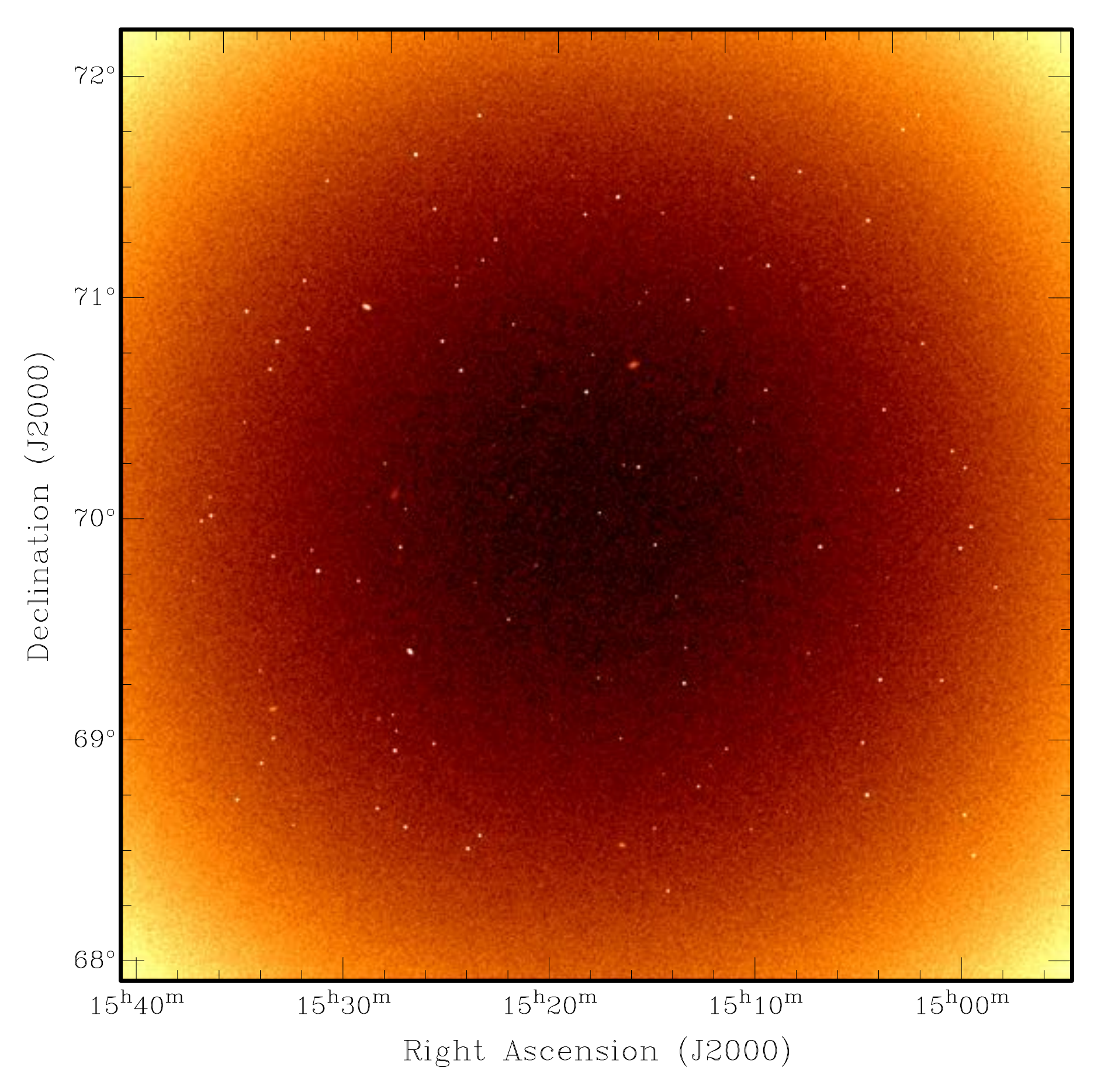}\\
   \includegraphics[trim=0.0cm 0.0cm 0cm 0cm,clip=false,angle=0,origin=c,width=\hsize]{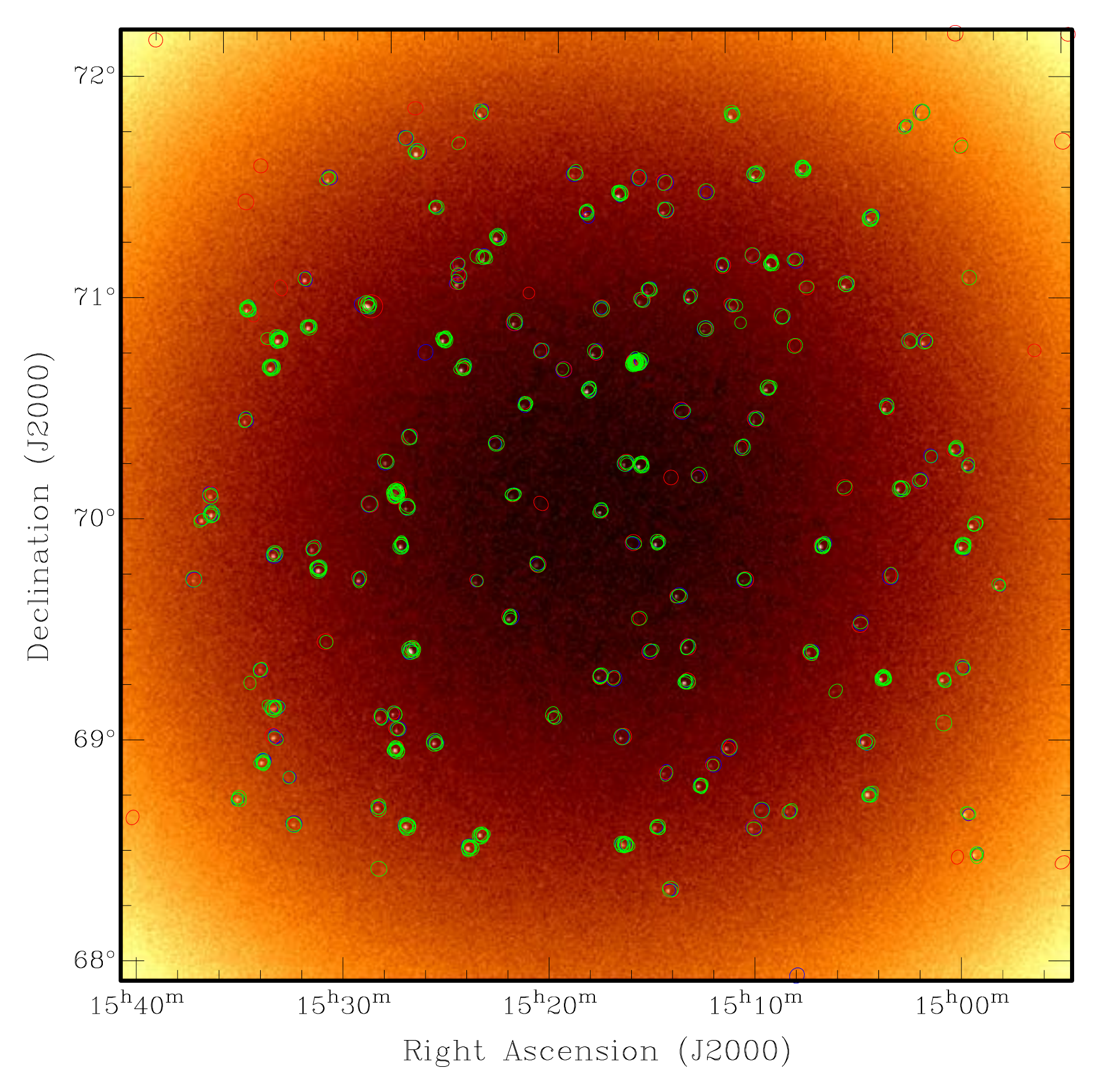}
   \caption{\textbf{Top:} The mean polarized intensity moment image, $\mu_{P}$, from the LOFAR simulations for the entire field-of-view. \textbf{Bottom:} The same image as above, with annotations showing sources found using Faraday Moments following Section~\ref{prescription}. Sources found using the $\mu$ or $\sigma$ of (i) $P$ are in green, of (ii) $Q$ are in blue, and (iii) $U$ are in red. Sources found using excess kurtosis are not shown. The method has good completeness and recovers $\ge98.5$\% of sources at a s/n of 5, although it is unreliable with many false detections. The false detections can be removed and filtered using RM Synthesis, which no longer needs to be performed on every image pixel.}
              \label{sources-simulated}%
    \end{figure}

\begin{figure}
   \centering
   \includegraphics[trim=2.3cm 1.3cm 0.2cm 0.4cm,clip=true,angle=0,origin=c,width=\hsize]{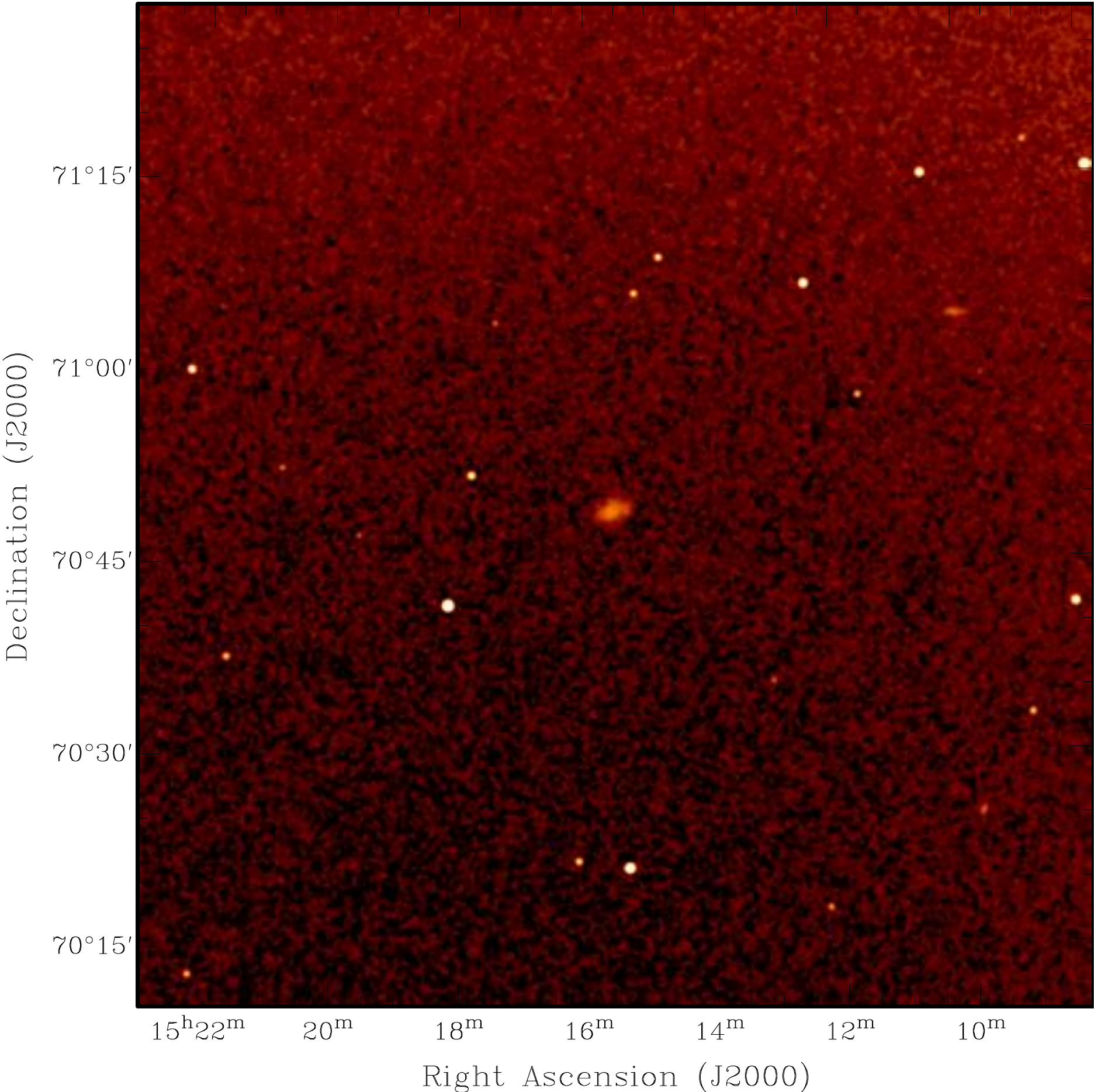}
   \includegraphics[trim=2.3cm 1.3cm 0.2cm 0.4cm,clip=true,angle=0,origin=c,width=\hsize]{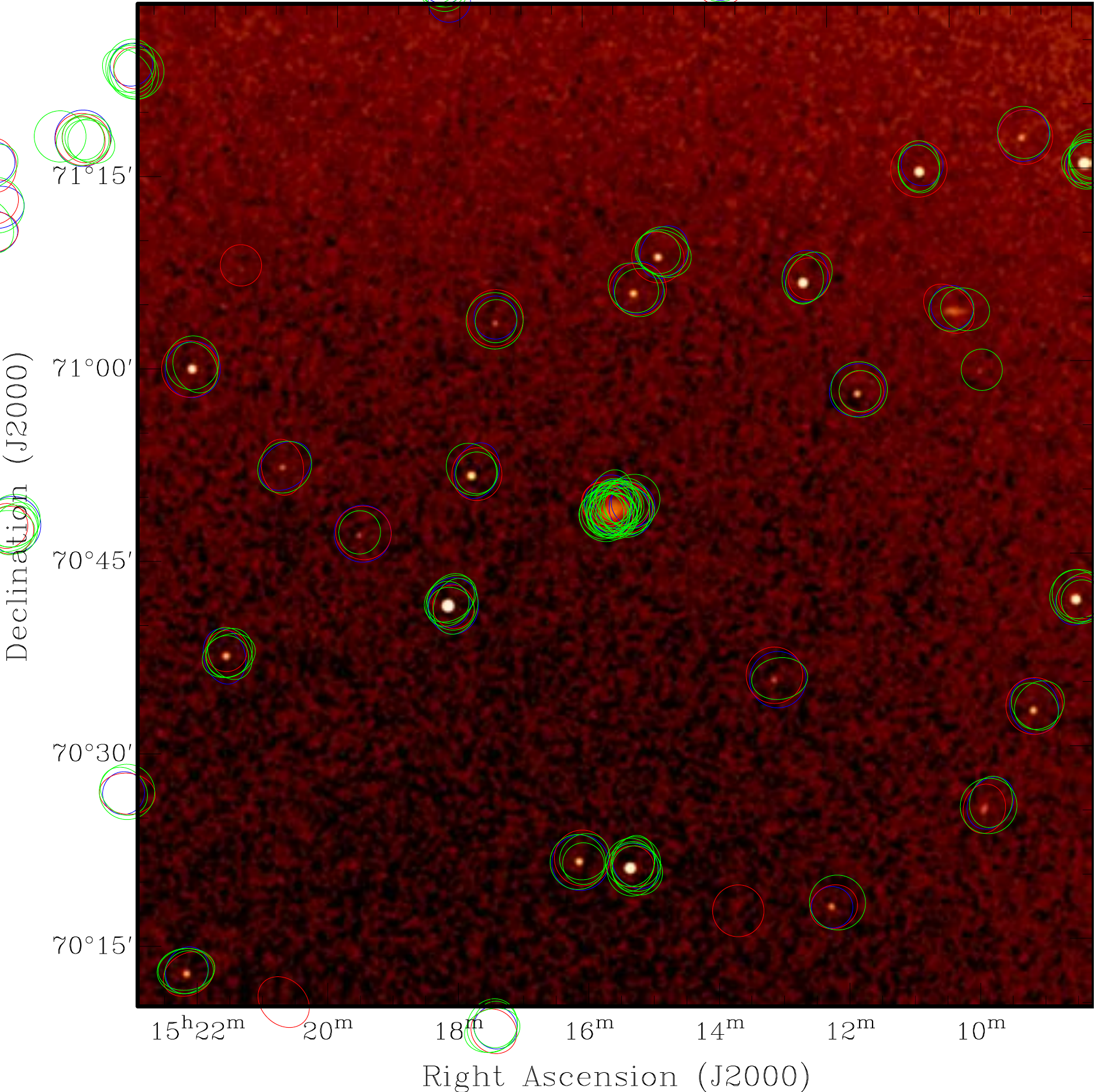}
   \caption{Similar to Fig.~\ref{sources-simulated}, but zoomed-in on a small region to highlight the sources themselves. The Figure is shown in order to allow better detail of the sources themselves to be seen, and the coordinates are therefore not shown.}
              \label{sources-zoomed}%
    \end{figure}

\begin{figure*}
   \centering
   \includegraphics[trim=0.5cm 0.5cm 0.5cm 0.0cm,clip=true,angle=0,origin=c,width=0.49\hsize]{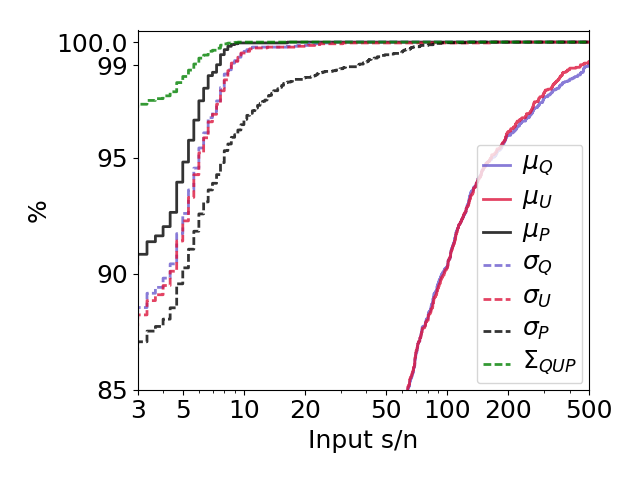}
   \includegraphics[trim=0.5cm 0.5cm 0.5cm 0.0cm,clip=true,angle=0,origin=c,width=0.49\hsize]{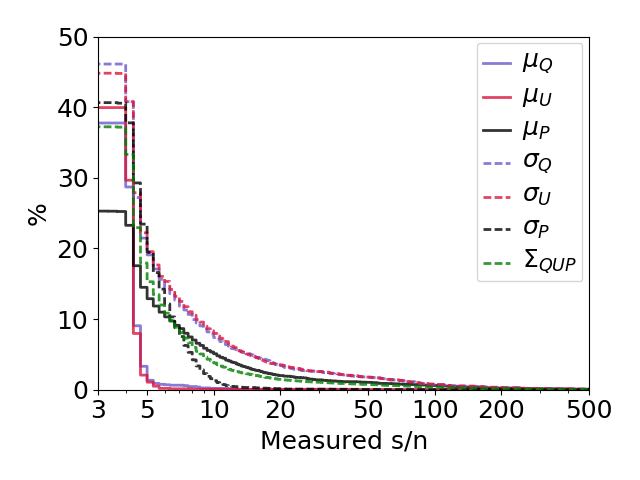}
   \caption{\textbf{Left:} The completeness for each Faraday Moment as compared to the input catalogue, as calculated using 2250 compact sources from 12 separately simulated fields of view. The s/n uses the band-averaged noise. The lines show the completeness for each individual Faraday Moment, as indicated by the legend. The combination of all moments is shown by $\Sigma_{QUP}$. There is significant overlap, with many sources being found in more than one moment. The subsequent degeneracy in finding sources leads to an overall completeness of 98.5\% at s/n=5. The benefit of using multiple moments is also visible: as $\mu_{Q}$ and $\mu_{U}$ add little to the completeness, the difference between $\Sigma_{QUP}$ and $\mu_{P}$ is entirely due to the added completeness obtained via the $\sigma$ moments. \textbf{Right:} The false detection rate (FDR) for each of the individual Faraday Moments, as indicated by the legend. The combination of all moments is again shown by $\Sigma_{QUP}$. The FDR is quite significant at a s/n$\le10$, and becomes substantial below s/n$\le5$ but can be reduced further by the application of RM Synthesis.
}
              \label{complete-reliable}%
    \end{figure*}

The results of the Faraday Moments source-finding from our tests on simulated data are shown in Fig.~\ref{sources-simulated}. This simulated LOFAR dataset shows the distribution of sources across the field-of-view, and also shows the simulated field-of-view overlaid with annotations from the source-finding. A zoom-in towards the simulated sources, also showing the annotations, is shown in Figure~\ref{sources-zoomed}. The method is very complete -- 98.5\% of injected sources above the noise are found down to a s/n ratio of $5$. This is an improvement over conventional source-finders such as \citet{2012MNRAS.422.1812H}, which reaches a completeness of 93.87\% at a s/n ratio of $5$, although we are aided by the degeneracy provided from searching multiple moment images. Nevertheless, the completeness of the Faraday Moments method is alongside a significant number of false detections. The measured completeness and reliability from our simulated dataset are shown in Fig.~\ref{complete-reliable}. This Figure demonstrates the performance of our method across a range of s/n ratios between 3 and 500. As we use simulated data, we can determine if a detected source is ``real'' based upon whether a source was injected into the simulation at that location. The completeness and reliability were therefore assessed using cross-matches in the proximity of each identified source. This also makes the completeness and reliability measurements independent of any RM Synthesis step. All of the displayed $\mu$ and $\sigma$ moments appear to be useful, with the less complete moments also often tending to be more reliable. The less efficient $\mu_{Q}$ and $\mu_{U}$ moments are well suited for low-RM sources, so we do not recommend their removal. We note that the main aim of our technique is to provide high completeness and to reduce the computational overhead from RM Synthesis. Ultimately, the user will need to confirm the real nature (or otherwise) of each source independently. This is a substantial improvement on the typical work scheme, in which a full-Stokes datacube has been required to be searched pixel-by-pixel for real sources and emission. Future work and the development of additional techniques will enable the production of a reliable catalogue from the complete catalogue provided by Faraday Moments.

Ultimately, each identified source can be further investigated using RM Synthesis, which allows a robust way to filter out instrumental or noise peaks in the data. This drastically reduces the number of pixels on which RM Synthesis needs to be performed, and in our tests (which uses a source distribution anticipated for modern radio telescopes) results in $5.184\times10^7$ pixels ($7200^2$ pixels), reducing to 590 pixels. This constitutes a reduction by a multiplicative factor of 1/87,800, or $\approx1\times10^{-5}$. The resulting 590 pixels also include duplicates that are found in multiple Faraday Moment images. These pixels can then be used for full RM Synthesis. Most importantly however, we have now generated a complete catalogue of candidate polarized radio sources from Stokes $Q$ and $U$ datacubes. Further inspection of these data, and the implementation and development of other techniques, will improve the reliability and computing time when producing such a catalogue even further.

\section{Conclusions}\label{conclusions}
We have identified a new source-finding strategy for finding a complete sample of linearly polarized sources in radio astronomy data. This resulting source-list is suitable for efficient application of RM Synthesis. 

We have shown that the technique can reduce the number of pixels on which RM Synthesis needs to be performed by a factor of $\approx1\times10^{5}$ for source distributions anticipated with modern radio telescopes. Due to the computationally efficient implementation of moment calculations, relative to implementations of RM Synthesis, this corresponds to a significant improvement in source-finding speed. It also provides the only known way to obtain a complete sample of polarized sources from $Q$ and $U$ datacubes as a function of frequency. Together with regular source-finding in Stokes $I$, using Faraday Moments for source-finding is therefore capable of providing a complete heuristic of the polarized emission throughout a field-of-view. Note that averaging polarized intensity alone cannot provide the same advantages as using many individual Faraday Moments. This is as the drop-off in completeness as a function of s/n for $\mu_{P}$ is considerably more rapid than for the combination of all moments, as shown in Fig.~\ref{complete-reliable}.

It will be of significant interest to compare these results in the future with more advanced simulations. The results presented here make only the assumption that the noise in $Q$ and $U$ is normally-distributed, and does not at any stage treat a Rician/Rayleigh distribution as approximately normally-distributed. Such assumptions have previously been shown to strongly affect polarized intensity statistics \citep[e.g.][]{2012PASA...29..214G,2012ApJ...750..139M}. The high completeness of our technique is undoubtedly assisted by the modelling of sources by the \textsc{aegean} source-finder as Gaussians. While beyond the scope of this paper, it is likely possible to use Faraday Moments together with source-finders that do not parameterise the shape of the emission a priori (such as e.g.\ \textsc{blobcat}) in order to identify diffuse polarized emission in an automated way. Such an investigation would also allow us to test to what extent the assumption of a Gaussian-shaped source, as made with \textsc{aegean}, improves the identification of sources and hence our completeness. The Faraday Moments technique also has advantages over alternative approaches, such as for example concentrating on small fields around sources seen in total intensity emission, as our method provides a fully automated polarization source-finder, rather than requiring manual inspection of many small fields. Our method is also advantageous over source-finding on Stokes $I$ only, or clipping based on Stokes $I$, as at the angular resolution accessible with LOFAR the peak in linearly polarized intensity is known to be often offset from the peak in total intensity (see Section~\ref{intro}).

Through tests on real LOFAR data of the M51 field, we have found that by using \textsc{bane} or a similar appropriate background estimation software, the technique continues to operate even in the presence of diffuse polarized Galactic emission. The Faraday Moments method was able to find all of the sources that were previously identified via careful manual analysis. We here only focussed on providing a complete catalogue of sources. The real nature of each source in Faraday space still requires manual inspection. However, we have considerably reduced the number of pixels that require such an inspection. Future investigations into automated techniques that allow the filtering of sources based upon their instrumental properties and the significance of the RM peak in Faraday space, while considerably beyond the scope of this current work, could lead to a fully-automated source-finding procedure that provides reliability as well as completeness. Furthermore, differences between the moments may possibly allow for the identification of instrumental sources based upon their moments alone, although leakage may be indistinguishable from emission with a low RM. Moreover, the technique may even enable the easy creation of catalogues containing different physical source classifications, by separating based upon the magnetic properties as revealed in the Faraday Moments. Further investigations will be able to study such possibilities.

Such a new source-finding technique could also enforce a set of selection effects upon any resulting source catalogue. In this respect, the most likely sources to be missed are those with a combination of both a low RM and low depolarization. However, as we shown via our tests of the completeness, this can only affect a very small population of sources. This may be a more significant problem for higher-frequency surveys, in which a low RM source could be less distinguishable via its moments if also combined with a polarization angle that leads to low signal in both Stokes $Q$ and $U$. In the case of LOFAR, as shown by the simulated SEDs in Figure~\ref{SEDs}, it is doubtful that any RM can in reality be low enough to not be detected with moments.

The source-finding we have performed was done using \textsc{aegean}, although this could in principle also be tested using PyBDSF, PySE, or similar software. Future work could expand our analysis to include and optimise this technique to work with other source-finders, and to investigate second-order effects that influence source-finding robustness such as correlated noise. Further iteration and development of this technique has the potential to provide a fully automated source-finding algorithm for full-polarization radio data -- a tool that is currently completely absent from an astronomer's toolkit and yet is much needed. The new technique in this paper now reduces the overhead required for manual data inspection. However, given the expected data deluge with upcoming SKA pathfinder and precursor surveys, we hope in future studies to explore techniques to automate these processes even further. The Faraday Moments technique is likely also applicable at higher frequencies. While at high frequency, typical sources may no longer exhibit full-cycles of $Q$ and $U$ rotation across the observing band, this would have the simple outcome of moving sources in Fig.~\ref{SEDs} from case (ii) to case (i). Extensions of this method and further testing will be useful for LOFAR, as demonstrated here, while the same principles can also be expanded for ASKAP, GALFACTS, MeerKAT \citep{2009arXiv0910.2935B}, the SKA, and other upcoming polarization surveys with radio telescopes.
%
\section*{Acknowledgements}
We are very grateful to both Justin Bray and Rainer Beck, for providing helpful comments that improved the paper. We also thank the anonymous referee for useful comments that enhanced the paper. DDM gratefully acknowledges support from ERCStG 307215 (LODESTONE). LOFAR, designed and constructed by ASTRON, has facilities in several countries, that are owned by various parties (each with their own funding sources), and that are collectively operated by the International LOFAR Telescope (ILT) foundation under a joint scientific policy.


\bsp
\label{lastpage}
\end{document}